\documentclass[]{mn2e}
\usepackage{psfig}
\usepackage{graphicx}  
\usepackage{mnras_cite}

\usepackage{dcolumn}  
\usepackage{bm}       
\usepackage{amssymb,amsmath}  
\usepackage{color}

\newcommand{\bea}{\begin{eqnarray}}
\newcommand{\eea}{\end{eqnarray}}

\newcommand{\xips}{\xi(\pi,\sigma)}

\newcommand{\beq}{\begin{equation}}
\newcommand{\eeq}{\end{equation}}

\newcommand{\beqa}{\begin{eqnarray}}
\newcommand{\eeqa}{\end{eqnarray}}

\def\kk{\kappa}
\def\calP{{\cal P}}
\def\calC{{\Theta}}

\def\vecx{{\vec x}}

\def\kvecMpc{\, h \, {\rm Mpc}^{-1}}

\def\la{\mathrel{\mathpalette\fun <}}
 \def\ga{\mathrel{\mathpalette\fun >}}
\def\fun#1#2{\lower3.6pt\vbox{\baselineskip0pt\lineskip.9pt
\ialign{$\mathsurround=0pt#1\hfil##\hfil$\crcr#2\crcr\sim\crcr}}}

\begin{document}

\title[Cosmology from lensing \& redshift distortions]
{Cross-Correlation of spectroscopic and photometric galaxy surveys: 
cosmology from lensing and redshift distortions}
\author[Enrique Gaztanaga, Martin Eriksen, Martin Crocce etal.]
{\parbox{\textwidth}{Enrique
    Gazta\~naga$^1$, Martin Eriksen$^1$, 
Martin Crocce$^1$, Francisco J. Castander$^1$, 
Pablo Fosalba$^1$,  Pol Marti$^2$, Ramon Miquel$^{2,3}$, 
Anna Cabr\'e$^4$}\vspace{0.4cm}\\
$^1$Institut de Ci\`encies de l'Espai (IEEC-CSIC),  E-08193 Bellaterra (Barcelona), Spain \\
 $^2$Institut de F\'{\i}sica d’Altes Energies (IFAE), E-08193 Bellaterra
 (Barcelona), Spain \\
$^3$Instituci\'o Catalana de Recerca i Estudis Avan\c{c}ats (ICREA), E-08010
Barcelona, Spain \\
$^4$University of Pennsylvania, Philadelphia, PA 19104, USA\\
}

\date{\today}
\pagerange{1--10} \pubyear{2011}
\maketitle 

\begin{abstract}

Cosmological galaxy surveys 
aim at mapping  the largest volumes to test models
with techniques such as cluster abundance, cosmic shear correlations
or baryon  acoustic  oscillations (BAO), which are designed to
be independent of galaxy bias.
Here we explore an alternative route to constrain cosmology:
sampling more moderate volumes with the cross-correlation of 
photometric and spectroscopic surveys.
We consider the angular galaxy-galaxy autocorrelation 
in narrow redshift bins and its combination with different probes of
weak gravitational lensing (WL) and redshift space distortions (RSD). 
Including the cross-correlation of these surveys improves by
factors of a few the constraints on both the dark energy equation of state $w(z)$ 
and the cosmic growth history, parametrized by $\gamma$.
The additional information comes from using many narrow
redshift bins and from measurement of
galaxy bias with both WL and RSD,  
breaking degeneracies that are present when using
each method separately. We show forecasts for a  joint
$w(z)$ and $\gamma$ figure of merit (FoM$_{w\gamma}$)
using linear scales over a deep ($i_{AB}<24$) photometric
survey and a brighter ($i_{AB}<22.5$) spectroscopic 
or very accurate ($0.3\%$) photometric redshift survey.
Magnification or shear
in the photometric sample produce 
FoM$_{w\gamma}$ that are of the same order of magnitude of those
of RSD or BAO over  the spectroscopic sample.
However, the cross-correlation of these probes over the same area yields
a FoM$_{w\gamma}$ that is up to a factor 100 times larger.
Magnification alone, without shape measurements, 
can also be used for these cross-correlations and can produce
better results than using shear alone.
For a spectroscopic follow-up survey strategy, 
measuring the spectra of the 
foreground lenses to perform this cross-correlation provides 5 times better
FoM$_{w\gamma}$ than targeting 
the higher redshift tail of the galaxy distribution to study BAO over a 2.5 times  
larger volume.

\end{abstract}

\maketitle

\section{Introduction}\label{sec:intro}

Weak gravitational lensing is a unique tool to study the large scale 
dark matter distribution. The correlated
shear measurements (i.e. shear-shear correlation function) 
provide direct information on the power spectrum of matter fluctuations
(see Bartelmann \& Schneider 2000, Refregier 2003,
Bernstein 2009  for a review),  avoiding the problem of bias, i.e. of how 
light traces the mass.
The drawbacks are that the information is projected on the
sky over a very  broad radial kernel, and that the signal-to-noise is quite
poor on small scales. For the former, shear-shear tomography can be of
some help, but the radial precision is still typically limited to a few redshift
bins (e.g. see Hu 1999,  Hu \& Jain 2004, Refregier et al. 2011).
For the latter we would need to measure shear for fainter magnitudes
and, from the ground,  this is limited by the atmospheric PSF distortion.
Cosmic shear also suffers from systematic difficulties, such as
shape measurement errors, the galaxy intrinsic alignment
and photo-z errors.  Here we will explore the combination
and cross-correlation of weak-lensing probes with the 3D information in 
the (foreground) large scale galaxy distribution 
in order to complement and extend the constraining power of 
weak lensing and clustering measurements.
The galaxy-shear cross-correlation can also be used to measure 
galaxy bias. Once bias is known, we can recover the full 3D matter clustering
from the 3D galaxy clustering.

We  pay special attention to cosmic magnification as an
alternative or complementary probe to shear. Lensing changes
the area of the background image, which
can result in density fluctuations in the background sources
that are correlated with the density fluctuations in the foreground
lenses. The resulting galaxy-galaxy cross-correlation is called
cosmic magnification and contains very similar cosmological information
to that provided by shear. 
Magnification will be compared with galaxy-shear (e.g. Johnston et al. 2007)
and shear-shear as a reference  (see e.g. Van Waerbeke 2009,
Bernstein 2009, Van Waerbeke et al.  2010 and references therein).
 Cosmic magnification has long been investigated as a potential source
 for the puzzling  galaxy-QSO
cross-correlation (see Gaztanaga 2003 and references therein). 
Thanks to the improvements in photometric homogeneity
in the SDSS sample, magnification has 
been detected in the cross-correlation of foreground photometric
SDSS galaxies with both spectroscopic quasars (Scranton et al. 2005) 
and Lyman break galaxies  (Hildebrandt, van Waerbeke \& Erben 2009).
 In our approach, we envision that, in contrast to these current
detections,  in the near future the lensing sample should be the
one with the highest radial resolution, i.e. spectroscopic or with very good (0.3\%)
photo-z error. For the background sources we do not need such good radial
resolution to predict the lensing response. This will be ideal for
3D power spectrum reconstruction, as will be shown here.

Jain \& Taylor (2003) and Bernstein \& Jain (2004) 
proposed the use of photometric redshift data to
produce a template map of foreground galaxies to cross-correlate
with the  induced shear as a function of the background (or source) galaxy redshift.
They considered the ratios of this cross-correlation to
recover the geometrical ratios rather
than trying to do a full 3D tomography. Bernstein \& Jain (2004) 
called this cross-correlation cosmography. This is related to 
our approach here. The common advantage in both cross-correlation
techniques is that galaxy-shear cross-correlation 
is less sensitive than shear-shear to systematic errors in the PSF correction.
The main differences with what we do here
is that  we consider direct
cross-correlations, rather than using a template, 
over spectroscopic or very good
photometric  foreground galaxies
using narrow bins. We also include
cosmic magnification, restrict to linear scales (to avoid
the uncertainties in galaxy power spectrum), and use the full 
signal (rather than just ratios). This allows to recover cosmological
information in the 3D power spectrum as well as recovery of the geometrical
information in lensing efficiency (see \S3).

As a new ingredient, we also consider the information
contained in redshift space distortions (RSD). On linear scales,
galaxy (peculiar) velocities can be measured by comparing radial to
transverse correlations and this provides a measurement of the linear growth
rate of dark matter fluctuations.
A challenge and opportunity in our approach is the need to model
galaxy biasing which affects both RSD and WL cross-correlations,
but in  different ways. In real space, or for transverse modes,
 on linear scales, the galaxy auto-correlation depends on the square
 of bias, the galaxy-shear cross-correlation only depends linearly on bias
and shear-shear is independent of bias. 
This means that their combination can be used 
to measure bias as well as the growth
information (e.g. see  Hoekstra et al. 2002 and \S\ref{sec:bias} below).
In redshift space, i.e.  for radial modes, bias only affects the
density growth but not the velocity growth (e.g. see 
Eq.\ref{eq:kaiserdeltag} below), which provides an
alternative route to separate growth from bias.
We will show here that the combination of WL probes and RSD results 
in a very accurate determination of bias, also breaking the
degeneracy of growth with cosmic expansion. 
Thus, we will combine three different types of probes here: 
\begin{itemize}

\item Angular clustering from galaxy-galaxy autocorrelation 
in narrow redshift bins (which we call GG) 

\item WL from shear-shear (SS),
galaxy-shear (GS) or magnification (MAG, i.e. galaxy-galaxy
cross-correlation)

\item RSD, from the ratio of transverse to radial modes.
\end{itemize}

The importance of combining redshift space distortions and
weak lensing information to test cosmological models and
modified gravity has been highlighted by several authors (e.g.
Zhang et al. 2007, Guzik, Jain \& Takada 2010, 
Song et al. 2011 and references therein).
Recently, Reyes et al. (2010) used this combination
to measure the growth of structure from a ratio of galaxy-galaxy
auto-correlation with
galaxy-shear cross-correlation in combination with RSD.
Here we will see how this can be generalized when we use all the information,
not only the ratios, and have many redshifts bins with much 
better radial resolution. Potentially RSD could benefit from 
the combination of samples with different biases over the same
region of sky (McDonald \& Seljak 2009). This effect is also
included in our analysis as we will consider bright and faint galaxy
populations with different biases. But this has little
effect on our forecast for the particular surveys we are modeling.
It might also be possible to exploit this further by 
further division of the galaxy population or direct lensing
calibration of biasing (Bernstein \& Cai 2011). 

Cross-correlation of photometric and spectroscopic samples in the same
redshift bin can also be used to estimate photometric redshift contamination
 and determining the redshift distribution of the photometric sample (e.g. Newman
2008). This technique will be generalized here to the cross-correlation between
separate bins by inclusion of transition probabilities $r_{ij}$.

A motivation for our study has been the science case
for the PAU Survey (www.pausurvey.org) based on
the PAU Camera (PAUCam,  Casas et al. 2010) 
a photometric camera with 40 narrow (100\AA) filters and 6 broad band filters
that will be commissioned at WHT Telescope in La Palma at the end
of  2012. PAUCam can map about 2 deg$^2$ of the sky each night 
with all these filters producing an imaging survey to a depth of about $i_{AB} \simeq
24$. Because of the narrow band filters, galaxies with $i_{AB} <22.5$ 
 will have a photo-z accuracy of about 0.35\% (i.e. about 10 Mpc/h), while the
accuracy for the rest of the galaxies is about 3\% (both with around 
50\% completeness for all type of galaxies). 
The bright sample  will be much denser ( $\sim$ 15000 galaxies per
deg$^2$)  than any spectroscopic surveys
to the same depth.
Unlike most spectroscopic surveys, which target predetermined galaxies, 
PAU will contain all the objects in the
surveyed area  to a given brightness.
In our initial studies for PAUCam, we only explored BAO with  $i_{AB} <22.5$ 
LRG galaxies  (Benitez et al. 2009).
However, as we discuss next, this is a rather 
restrictive use for the PAUCam data.
A similar approach has been taken by several spectroscopic surveys
which plan to measure BAO at high redshifts by selecting appropriate
spectroscopic targets out of a given parent photometric catalog.
Here we will consider forecasts using both the 
faint photometric and bright (quasi) spectroscopic samples and we will 
show that one can get  significantly better cosmological constraints when 
using probes based on the combination of WL and 
RSD. 

We will consider both a modest 200 deg$^2$ PAU-like
Survey and a more ambitious 5000 deg$^2$ survey.
Such a survey could also result from the Dark Energy 
Survey (DES), which will image 5000 deg$^2$ in
five passbands to ~24th mag over 5 years starting in 2012,
producing weak lensing shape measurements for about 200 million galaxies. A
massive follow-up spectroscopic survey over a substantial part of the
DES footprint, e.g., using multi-fiber spectrographs such as BigBOSS
or DESpec, would enable RSD measurements of the lensing population and
the implementation of this cross-correlation technique.
 Unlike BAO or supernovae, 
WL and RSD can also provide very valuable  information
on cosmic growth history and this is a key ingredient
to understand the physics of the accelerating Universe.

This paper is organized as follows. In \S2 we present the
modeling, methodology and different assumptions used.
We include a subsection on how we model galaxy bias
and another one on modeling of redshift space distortions.
 Section 3 describes the approximations we make for
weak gravitational lensing. The reader familiar with
these techniques can directly jump to
 \S4, where we present our fiducial surveys.
In  \S5 we present results of our forecast as a
function of the different ingredients ,
to disentangle the different contributions to the FoM.  
This section could also be skipped if the reader is not
interested in such details.
In \S6 we present the main result in this paper, i.e.
the comparison of forecasts for different
surveys. We finish in \S7 with some conclusions and
summary of the main results.
Table\ref{table:notation} summarizes the notation that
will be used in this paper.

\begin{table}
\begin{tabular}{|c|c|}
\hline
RSD &  Redshift Space Distortions \\
WL & Weak Lensing \\
BAO & Baryon Acoustic Oscillations \\
GG &  Galaxy-Galaxy auto-correlation \\
MAG & Magnification from GG auto and cross-correlation \\
SS  & Shear-Shear correlations \\
GS & Galaxy-Shear correlations \\
WL-all & combination of  SS, GS, GG and MAG \\
F & Fain sample of galaxies $22.5 < i_{AB} < 24$ \\
B & Bright sample of galaxies $i_{AB}< 22.5$  \\
F+B & Combination of independent F and B samples \\
FxB & Cross-correlation of F and B over same area \\
\hline
\end{tabular}
\caption{ Notation and acronyms used in this paper.}
\label{table:notation}
\end{table}

\section{Modeling}

In this section we will introduce the different assumptions
and modeling used in this paper. Section 2.1 and 2.2 introduce
the cosmological model and the parameters that we want
to study. In Section 2.3 we present the Fisher Matrix
approach and the figures of merit that will be used to compare
experiments. Sections 2.4 and 2.5
present and justify the models for galaxy bias
and redshift space distortions.
 
\subsection{Growth and Cosmic History}

The cosmic expansion  history, $a=a(t)$ or $H=H(t)$,
 in a flat Friedman Lemaitre Robertson Walker
(FLRW)  background
with matter density $\rho_m$ and dark energy (DE)
equation of state $w=p_{DE}/\rho_{DE}$, can be
written as:
\bea
H^2 \equiv \left(\frac{\dot a}{a}\right)^2 &=&\frac{8\pi
  G}{3}\left(\rho_m +\rho_{DE}\right) - {k\over{a^2}} \label{fe2} \\ 
& = &\nonumber
H^2_0\left[\Omega_{m}a^{-3}+\Omega_k a^{-2}+\Omega_{DE}a^{-3(1+w)}\right]
\eea
where $\Omega_k=1-\Omega_m-\Omega_{DE}$ measures
deviations from flat curvature $k=0$ and we have neglected radiation. 
Our fiducial model corresponds to  $\Lambda CDM$:
a flat universe with $w=-1$, so that the DE density is constant
with redshift $z$. We will explore how well
our different observational probes can 
 constraint $w$ and its variation $w=w(z)$.\footnote{When $w=w(z)$ we
   need replace $w$ in  $\Omega_{DE}a^{-3(1+w)}$ in Eq.\ref{fe2} by the
corresponding  integral over redshift.}
 Parameters $w_0$ and $w_a$ are used to
characterize the evolution of DE equation of state (Chevallier \&
Polarski 2001, Linder 2003):
\beq
w(z)= w_0 + w_a (1-a) =
w_0 + w_a z/(1+z)
\eeq

According to General Relativity (GR), given this cosmic history, 
the equations that determine the growth history, i.e. the
cosmic evolution of the linear density contrast $\delta$,
are of the form \cite{peebles1980,bernardeau02}

\beq
\ddot {\delta} +2 H  {\dot \delta} = 4\pi G \rho_m \delta 
\label{eq:linear1}
\eeq
with the solution
\beq
\delta= D(a) \delta(0)
\label{eq:linear}
\eeq
where the growth factor $D(a)$ depends on the expansion history $H(a)$
(through $w(a)$) and on $\Omega_m(a)$. Any discrepancy found between the
observed growth and the growth $D$
predicted for a given expansion history $H$
can be use as a test for modifications to GR or
variations on the cosmological model.
This linear growth can also be characterized by its derivative,
the velocity growth factor:

\begin{equation}
f \equiv \frac{d\;ln\;D}{d\;ln\;a}= {\dot{\delta}\over{\delta}} \equiv \Omega_m^\gamma(a)
\label{eq:f(z)}
\end{equation}
where $\gamma$ is the gravitational growth index  (see Linder 2005).
So when normalized to $D=1$ today, then
\begin{equation}
 D(a)= \exp{\left[-\int_a^1 d\ln   a  ~f(a) \right]}
 \label{eq:dgamma} 
\end{equation} 
 For GR with  DE equation of state $w$, Linder (2005) finds that to 
a good approximation:
\begin{equation} 
\gamma \simeq \frac{3(w-1)}{6w-5}
\label{eq:GR}
\end{equation}
This reduces to the well known result $\gamma \simeq 0.55$ for
$\Lambda CDM$  ($w=-1$).
 Thus for DE models in GR, measurements of $f(z)$ can be used to estimate
 $w(z)$, independently from measurements of $H(z)$.
Other cosmological models have a different  relation between $D(z)$ and $H(z)$  (e.g. see
Gaztanaga \& Lobo 2001)  which results in different effective values for $\gamma$.
 For example, in the DGP model $\gamma \simeq 0.68$
(Lue, Scoccimarro \& Starkman 2004).
More generically, an independent measurement of
$\gamma$ and $w$ can be related to a measurement of time variations of
Newton's constant $G$  (e.g. see Eq.32 in Pogosian et al. 2011), 
which provides a direct test of GR. This will be our
approach here: we will check how well we can measure separately $w$ and
$\gamma$ with different probes, assuming they are independent of each
other. We will not  make use of GR to relate cosmic growth to cosmic
history (i.e. we will not use Eq.\ref{eq:GR} or Eq.\ref{eq:linear1}),
but rather we will use a parametric description in terms of
independent parameters: $w_0$, $w_a$  and $\gamma$ (plus the standard
cosmological parameters in Eq.\ref{eq:parameters} below). This
can be used to separate GR from other theories of gravity and
learn about the nature of the accelerated expansion.

\subsection{Fiducial Cosmological model}
\label{sec:priors}

To characterize cosmic history and the linear matter power spectrum,
we use the standard set of 8 cosmological parameters 
(e.g. see Komatsu et al. 2011):
\beq
w_0, w_a, h, n_s, \Omega_m,\Omega_B,\Omega_{DE}, \sigma_8
\label{eq:parameters}
\eeq
In addition, the growth index in Eq.\ref{eq:f(z)} is allowed to vary
independently of $w(z)$ to  characterize the growth history and 
to test Modified Gravity separately from DE evolution. 
To model bias we include an additional set of independent
bias parameters as described later on in \S\ref{sec:bias}.
Thus we have a total of 9 cosmological parameters together with 
some nuisance bias parameters $b_i$:
\beq
p_\nu = \left(w_0, w_a, h, n_s, \Omega_m,\Omega_B,
\Omega_{DE}, \sigma_8, b_i, \gamma \right).
\label{eq:parameters}
\eeq
Note that in general $\Omega_k=1-\Omega_m-\Omega_{DE}$ differs from zero.
For the fiducial values we use 
$w_0=-1, w_a=0, h=0.7, n_s=0.95, \Omega_m=0.25, \Omega_B=0.044,
\Omega_{DE}=0.75, \sigma_8=0.8$, $\gamma = 0.55$ (corresponding to its
value in GR).
Unless stated otherwise, results are always presented  
with priors from Planck and Stage-II
Supernovae (SN-II)\footnote{We use the Dark Energy Task Force (DETF)  Planck and
SN-II priors Fisher Matrices given in
http://www.physics.ucdavis.edu/DETFast/, with file names {\tt planckfish}
and {\tt SN-II}.  The Planck prior on $\sigma_8$ assumes a GR model for
linear growth but there are several other ways to measure
$\sigma_8$ that could produce similar priors without such assumption.
The actual priors and how they were obtained are
somewhat irrelevant here, they just represent a baseline to compare
different experiments.}, but with no priors in $\gamma$ or bias $b_i$.

Note that for completeness we include  $\sigma_8\equiv \sigma_8(0)$, 
the global normalization of
the amplitude of density fluctuations at $z=0$. Our goal is not
to measure $\sigma_8$  but rather the evolution of this amplitude
 with time, i.e. $D(z)$, which is given by $\gamma$. In our analysis, 
$\sigma_8$ is dominated by
the priors above (around $1\%$) as our probes are more sensitive to
relative than to absolute variations.
One could also try to measure $\sigma_8$ and other parameters,
such as neutrino masses, by including the amplitude 
of CMB temperature fluctuations at z=1100  rather than using priors of
$\sigma_8$ (i.e. at $z=0)$.
This is a very different approach and involves other complications.
 In such case $\gamma$ and $\sigma_8$
could be strongly correlated. But this is not our approach here. Our
goal is to test the best way to  measure cosmic and growth evolution 
starting from some priors at $z=0$.

\subsection{Fisher Matrix and Figures of Merit}
\label{sec:FM}

Throughout this paper we will employ the 
formalism based on the Fisher Matrix (FM,
e.g. Tegmark et al. 1998):

\beq
F_{\mu\nu} = \sum_{\ell \, {\rm or} \, k}  \ \sum_{ij,mn} {\partial
  C_{ij}\over{\partial p_\mu}}  ~\calC^{-1}_{ij;mn} {\partial
  C_{mn}\over{\partial p_\nu}}
\label{eq:fishermatrix}
\eeq
where $p_\nu$ are the cosmological parameters, nuisance parameters
 or additional quantities we want to
forecast and $C_{ij}$ are the observables with $\calC$
covariance. The indices $ij$ refer here to each pair of
redshift bins and $\ell$ or $k$ refers to angular scales: 
multipole or Fourier modes. 
The elements of $F^{-1}_{\mu\nu}$ are the covariance of  $p_\mu$
and $p_\nu$. For a given experiment,
the FM approach is a mapping (or transformation) of the covariance in the observables
to the covariance in the quantities we want to measure.

Here we will model the weak lensing (WL) probes, i.e. shear-shear,
galaxy-shear  and galaxy-galaxy angular
correlations, in real space  and using the Limber approximation (see \S\ref{sec:corr}).
In this case the observables $C_{ij}$ will be the power spectrum of
the angular cross-correlations
between two redshift bins.  Redshift bins will be taken to be
independent, as they are separated and non overlapping in space. 
This means that there are no intrinsic radial correlations and
we only consider transverse modes for angular clustering and WL probes.
For redshift space distortions (RSD) we will use the ratio of the
amplitudes of the 3D power spectrum as we change from transverse to 
radial modes within a single redshift bin, i.e. $\mu$ in Eq.\ref{eq:kaiserdeltag}.
We neglect the covariance between these ratios and the transverse 
angular power spectrum.
When RSD is combined with BAO we also neglect the covariance between 
the RSD ratios and the BAO wiggles.
Hence the combination of RSD with the WL probes or BAO is given by the
addition of the corresponding Fisher matrices.

\subsubsection{Non-linear scales}

The FM  depend on scales \footnote{We employ spherical coordinates in the
  formalism of weak lensing
  (i.e. $\ell$-modes) and f modes for RSD
  (i.e. $k$-modes).} as the summation is over all scales that
contribute to the analysis in Eq.~(\ref{eq:fishermatrix}). 
If we fix a minimum scale $R_{min}$ 
defined by the validity of ``linear'' theory today,  the redshift evolution
of $R_{min}$ can be obtained through this relation: 
\beq
\sigma(R_{min},z) = 1 
\label{eq:Rmin}
\eeq
where $\sigma(R,z) = \int P_{\rm L}(k,z) W^2(k R)d^3k$ is the
amplitude of fluctuations on scale $R$, $W(x)=\exp(-x^2/2)$ is a Gaussian
smoothing kernel and $P_{\rm L}(k,z)$ is the linear matter power spectrum. 
These are the same criteria adopted by \pcite{white2009}.
The maximum $k$-mode is then given by
\beq
k_{max}(z) = k_{max} \frac{R_{min}(0)}{R_{min}(z)} \kvecMpc
\label{eq:kmax}
\eeq
with $k_{max} = 0.1$
to  ``normalize'' this scale to $0.1\kvecMpc$ at $z=0$. The
corresponding maximum $\ell$-mode is given by
\beq
\ell_{max} + 1/2 = k_{max}(z_i) r(z_i) 
\label{eq:lmax}
\eeq
where $r$ is the transverse comoving distance to the redshift bin $i$ under
consideration.  In the forecast, we consider 
cross-correlations to scales up to
the $\ell_{max}$ of the closest bin (i.e. the smallest
$\ell_{max}$ of the two bins). For the maximum scale we use $k_{min}=0.001 \kvecMpc$
for all $z$ and $l_{min}=50$ to avoid inaccuracies in the Limber approximation
 \footnote{Notice that error-bars are larger at large
  scales, hence this choice has little impact on FM constraints.}.

These conditions, in Eq.(\ref{eq:Rmin})-(\ref{eq:kmax}), 
yield very similar values for $k_{max}$ to those found by
studying the limitations of perturbation theory by \pcite{PTreloadedI} (see also
\pcite{PTreloadedII}). 
In this analysis we will only use linear theory for $P(k)$ and Gaussian initial conditions.

\subsubsection{Figures of Merit}

In the FM approach 
the marginalized 1-sigma variance in parameter $p_\nu$ is given by the
diagonal element of the inverse of the FM: 
$\sigma^2(p_\nu)=[F^{-1}]_{\nu\nu}$. 
 When optimizing an experiment with more than one parameter, 
you need to take into account their covariance.
We will then cast our results in terms of:
\beq
{\rm FoM}_S= \sqrt{{1\over{{\rm det}[F^{-1}]_{S}}}},
\label{eq:FoM}
\eeq
where $S$ is the sub-space of parameters we are
interested on. For just one parameter, this is the inverse
error, for two this is proportional to the inverse area included
within  the 1-sigma error ellipse \cite{detf}, for three
this is  the inverse
volume within the 1-sigma error ellipsoid, and so on. 
In this paper, we will focus on three different figures of merit:

\begin{itemize}

\item{${\rm FoM}_w$} : Constraints on $(w_0,w_a)$ marginalizing over the
  remaining cosmological and bias parameters, similar to the DETF FoM
(Albrecht et al. 2006). 
 But note how here we allow $\gamma$ to vary, while
in Albrecht et al. (2006)
this figure of merit is quoted for a fixed value of $\gamma \simeq 0.55$ 
corresponding  to GR. 

\parskip .2cm
 
\item{${\rm FoM}_\gamma$} : the inverse of the achievable 1-sigma error in $\gamma$ after
  marginalization over the remaining parameter space.

\parskip .2cm

\item{${\rm FoM}_{w\gamma}$} : Joint constraint on
  $(w_0,w_a,\gamma)$, a natural extension to the dark energy FoM of
  the DETF.

\end{itemize} 

 We find that there is little covariance between
$\gamma$ and $(w_0,w_a)$, which means that FoM$_{w\gamma} \simeq$  
FoM$_{\gamma} \times$ FoM$_w$.
Note that the above definitions have different dimensions since
they involve different numbers of parameters. To have a direct comparison of 
the (geometrical) mean error in the different cases we should compare
FoM$_w$ to FoM$_\gamma^{2}$ and FoM$_{w\gamma}^{2/3}$.

\subsection{Galaxy Bias and its cross-correlation}
\label{sec:bias}

In this section we will justify how we model galaxy
bias. This is a key ingredient in our approach
as both RSD and WL cross-correlations depend on 
bias. We will show that we do not need to include
a correlation coefficient between galaxy and
matter cross-correlations, and argue that
we only need to allow bias to vary on time scales
larger than $\Delta a\simeq 0.1$.

\subsubsection{Bias stochasticity and scale dependence}

In the local bias model (Fry \& Gaztanaga 1993),
the smoothed galaxy fluctuation $\delta_g$ at a point $\vec x$
is a function of the matter field $\delta_m$ at the same point.
For small fluctuations we can Taylor-expand this function and
approximate the local bias model as:

\beq
\delta_g(\vec x,z)= b(z) ~\delta_m(\vec x,z)
\label{eq:bias0}
\eeq
The galaxy-galaxy (auto) correlation $\xi_{gg}=<\delta_g\delta_g>$
and galaxy-mass (cross) correlation $\xi_{gm}=<\delta_g\delta_m>$ are then:
\bea
\xi_{gg}(s,z) &=& b^2(z)  ~\xi_{mm}(s,z) 
~~\propto  ~~ b^2(z) D^2(z) 
  \nonumber \\
\xi_{gm}(s,z) &=& b(z)  ~\xi_{mm}(s,z)
~~\propto ~~ b(z) D^2(z) 
\label{eq:bias}
\eea
where $\xi_{mm}(r)$ is the matter-matter correlation and
$s\equiv |\vecx_2-\vecx_1|$ is the separation between pairs and
the second step uses the linear gravitational growth.
Under these linear and local assumptions one can
combine $\xi_{gg}$ and $\xi_{gm}$ to break the degeneracy between $b$ and
$D$. A more general parametrization of bias is based
on the cross-correlation coefficient, r, defined as:
\beq
r \equiv {\xi_{gm}\over{\sqrt{\xi_{gg}\xi_{mm}}}} 
\label{eq:r}
\eeq
If instead of using the local bias model we assume
an effective bias $\bar{b}^2 \equiv \xi_{gg}/\xi_{mm}$ that is constant as
a function of scale we find:
\bea
\sqrt{\xi_{gg}/{\xi_{mm}}} &=& \bar{b}(z) \nonumber \\
{\xi_{gm}/{\xi_{mm}}} &=& \bar{b}(z) ~ r(z)  \label{eq:bratios} \\
{\xi_{gg}/{\xi_{gm}}} &=& {\bar{b}(z)\over{ r(z)}} \nonumber
\eea
where $r(z)$ can differ from unity
(e.g. see Tegmark \& Peebles 1998, Pen 1998, Dekel \& Lahav 1999,
Casas-Miranda etal 2002, Seljak \& Warren 2004, 
Bonoli \& Pen 2009). In the local model $r(z)=1$
and all these ratios are just equal. It is still possible
to measure $D(z)$, $b(z)$ and $r(z)$ using more observables, such as $\xi_{mm}$
from shear-shear or $P(k,\mu)$ from redshift space distortions. 

We will show next that on large scales, 
when $\bar{b}$ is constant as a function of scale,  $r$ has to be close
to unity and there is no need to introduce this parameter.
Let us add non-linearities and a
stochastic component $\epsilon(\vec x, z)$ which could be caused
by shot-noise or non-local perturbations:
\beq
\delta_g(\vec x,z)= b_1(z) ~\delta_m(\vec x,z) + b_2(z)~\delta^2_m(\vec
x,z) ~+ \epsilon(\vec x, z)
\label{eq:epsilon}
\eeq
By definition, $\epsilon$ is not correlated to $\delta_m$. Otherwise,
the correlated part will just renormalize  $b_1$ and $b_2$. Then
\bea
\xi_{gg} &=& b_1^2~\xi_{mm} + \xi_\epsilon+ 2b_1 b_2 \xi_3 + b_2^2\xi_4
\nonumber \\
\xi_{gm} &=& b_1 ~\xi_{mm} + b_2 \xi_3
\eea
where we have introduced
$\xi_\epsilon \equiv <\epsilon\epsilon>$, 
$\xi_3 \equiv <\delta_m\delta_m^2>$ and $\xi_4\equiv
<\delta_m^2\delta_m^2>$ for the  3 and 4 point correlations. 
All these new correlations have a different scale dependence
on the separation $s$ than $\xi_{gg}(s)$.
 For Gaussian initial conditions:
 $\xi_3 \simeq \xi_{mm}^2$ and $\xi_4 \simeq \xi_{mm}^3$ (see
Bernardeau et al. 2002).
On large scales, $s>10 Mpc/h$,  where $\xi_{mm}<1$:

\bea
\xi_{gg} &\simeq & b_1^2 ~\xi_{mm} + \xi_\epsilon ~=~ b_1^2 ~\xi_{mm}
(1+ {\xi_\epsilon\over{b_1^2\xi_{mm}}}) 
\label{eq:xigg}
\\ \nonumber 
\xi_{gm} &\simeq& b_1 ~\xi_{mm} \\
r & \simeq &{1\over{\sqrt{1+{\xi_\epsilon/b_1^2\xi_{mm}}}}}
\simeq 1 - {\xi_\epsilon\over{2b_1^2\xi_{mm}}}
\label{eq:r1}
\eea
In general $\xi_\epsilon(s) <<\xi_{mm}(s)$ and the above equations
reproduce the local bias results in Eq.\ref{eq:bias} with $b=b_1$, 
even when the stochastic $\epsilon$
can be shown to be quite large, even larger that $\delta_m$
 (see Fig.1 in Manera \& Gaztanaga 2011).
When $\xi_\epsilon(s)/\xi_{mm}(s)$ is not negligible then we have that
$r$ is scale dependent, since both $\xi_\epsilon$ and $\xi_{mm}$
depend on the pair separation $s$.  But, in such a case the effective bias, defined
as $\bar{b}^2=\xi_{gg}/\xi_{mm}$,  will also depend on scale. 
In other words, whenever the effective bias $\bar{b}$ is independent of 
scale,  Eq.\ref{eq:xigg}
indicates that we can neglect the scale dependent 
terms, i.e. $\xi_\epsilon/b^2\xi_{mm}$ and then,  according to Eq.\ref{eq:r1},
we force $r \simeq 1$. This result 
is well reproduced in halo bias simulations
on scales $s>20 Mpc/h$  (see Fig.4 and Fig. 13 in Manera \& Gaztanaga
2011 and next sub-section).
 
Seljak \& Warren 2004 and Bonoli \& Pen 2009 found  that $r$ tends 
to unity on large scales in halos in numerical simulations but deviations
from $r=1$ are significant even on the largest scales.  It should be noted that these references
studied $r(k)$ in Fourier space for halos and not in configuration space. In Fourier space $r$ is also
subject to shot-noise, because $r$ is build from the power spectra
 $<\delta(k)^2>$, while shot-noise cancels for the correlation $<\delta(r)\delta(r')>$.
In our forecast we use fourier and harmonic space, but we want to
model galaxies and not halos. For galaxies or optimaly weighted halos
shot-noise is less important and $r$ is closer to unity (see Scherrer
\& Weinberg 1998, Matsubara 1999, Hamaus etal 2010, Cai \& Bernstein 2011).
Correcting for shot-noise of halos is possible but complicated as
halos have a exclusion region which results in a sub-Poisson
correction (see Casas-Miranda etal 2002, Manera \& Gaztanaga 2011).
Below we will show that indeed $r$ is very close to unity in numerical
simulations for all redshift. In \S\ref{sec:sto} we will test
 how a stochastic bias impact our result.

\subsubsection{Bias evolution}

\begin{figure}
\includegraphics[width=3.4in]{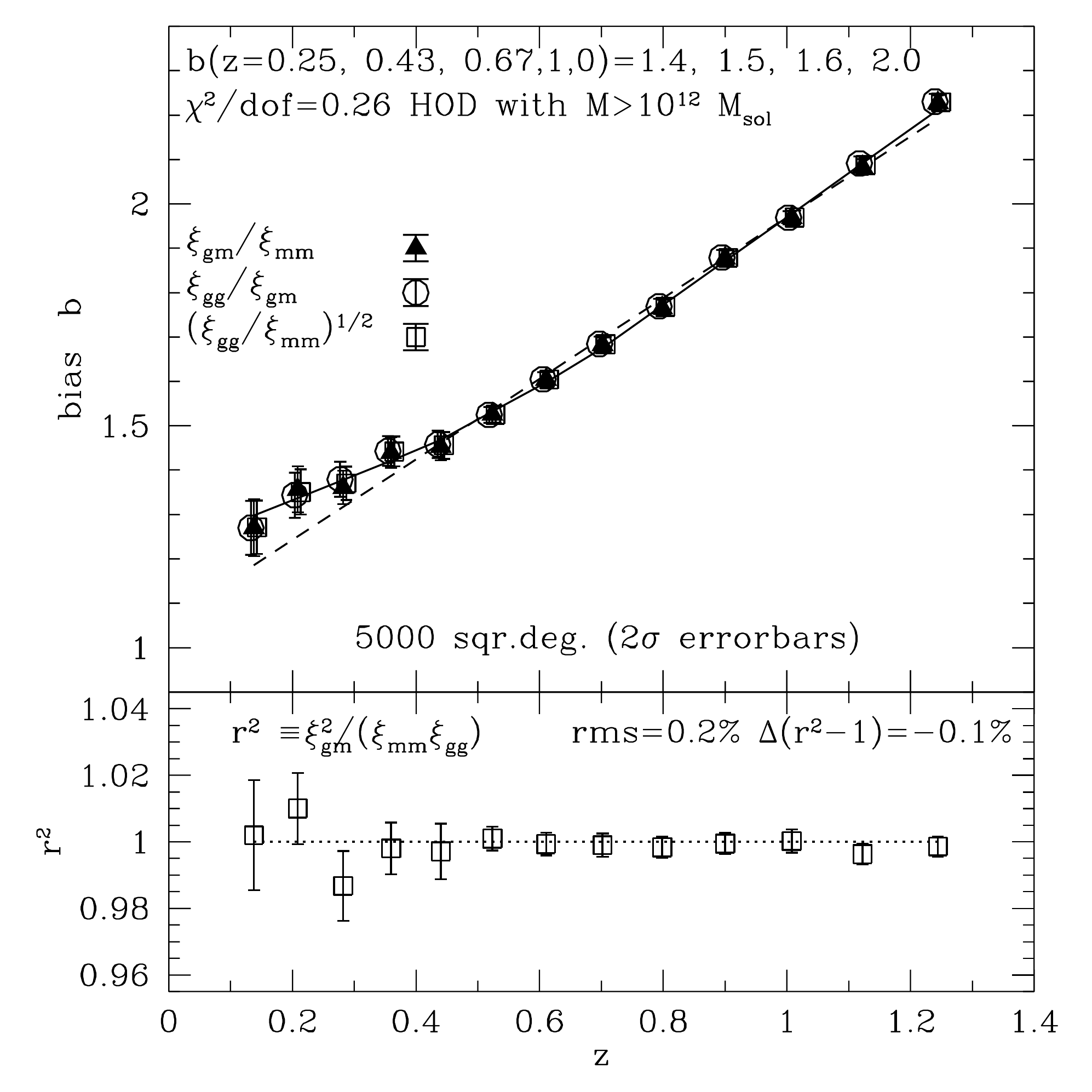}
\caption[rfitb]{Bias in mock MICE galaxy simulations. Mean and (2-sigma) errors correspond
to  a 5000 deg$^2$ survey.
Top panel shows the evolution of the effective bias defined in different ways:
$\sqrt{\xi_{gg}/\xi_{mm}}$ (squares), $\xi_{gg}/\xi_{gm}$ (circles)
or $\xi_{gm}/\xi_{mm}$  (triangles). The agreement is excellent: we include a small horizontal
displacement to be able to see the different symbols. Continuous (or dashed) line shows the
best fit  bias evolution model: $b(z)$ based on linear inter/extrapolation
between 4 (or 2) free bias parameters $b(z_i)$ spaced at regular
scale factor values $a=0.8,0.7,0.6, 0.5$ (or $a=0.7, 0.5$).
Bottom panel shows the cross-correlation coefficient
$r^2=\xi_{gm}^2/(\xi_{gg}\xi_{mm})$.
}\label{fig:rfitb}
\end{figure}

 Figure \ref{fig:rfitb} illustrates
the above modeling as a function of redshift.
Galaxy mocks are built from halos in the light-cone of a MICE simulation
of side $L=3072 Mpc/h$ and $N=2048^3$ particles
 (see Fosalba, Gaztanaga, Castander \& Manera 2008,  
Crocce, Fosalba, Castander \& Gaztanaga 2011 for more details on the
 simulation). 
The large size of this MICE run allows to build a light-cone 
sampling one octant of sky without repetition to $z=1.5$.
Halos are identified
 and weighted  with a simple 
HOD (Halo Occupation Distribution)
prescription. The number of galaxies in a halo
of mass $M>M_{min}$ is 
$1+(\frac{M}{20M_{Min}})^\alpha$ with $M_{min} \simeq  10^{12}
M_{sun}/h$ and $\alpha\simeq 1$
(e.g. see Scoccimarro et al. 2001).\footnote{This assignment can also
  introduce additional stochasticity but its contribution to
  $\xi_\epsilon$ vanishes if this stochasticity is not spatially
correlated between halos} Evolution with redshift is all given
by the halo mass evolution.
We divide the light-cone
simulation in redshift  bins of width $\Delta z \simeq 0.1$ and
estimate the correlations $\xi_{mm}$, $\xi_{gm}$ and $\xi_{gg}$ 
in each bin as a function of separation $s$.  Three different
bias parameters are then fitted 
to the ratios in Eq.\ref{eq:bratios} on scales  $s>30 Mpc/h$ 
where we find that these ratios are constant within the errors.
Results for the different  ratios agree well as shown in
the top panel of Fig.\ref{fig:rfitb}. These results
 indicates that $r \simeq 1$, as confirmed by the bottom panel.
Errors are from a 100 jackknife angular patches
 which we have checked agree well with subsample
errors (e.g. see Norberg et al. 2009 and
references therein). The error-bars are similar for the different cases
because sampling variance cancels to a good extend in the ratios.
The bottom panel shows that $r^2$
is indeed very close to unity in agreement with the arguments above. 
The mean rms error in $r$ is only 0.2\% and the rms deviation
from unity is less than $-0.1\%$. Similar results are also found
for different choices of
HOD weights. 

Note that because of the large volume and density of our simulation,
the evolution of bias is known here to better than 1\%.
Halo model predictions do not yet reach these levels of accuracy
\cite{Manera2010,Manera2011}, but could potentially be calibrated
with simulations to achieve similar accuracies. As shown in the figure,
the values at different redshift are strongly correlated, indicating that only a
few parameters are needed to characterize the evolution of bias. 
The characteristic time scales for bias evolution in our samples seem to
be $\Delta a>0.1$, corresponding to $t>1 Gyr$, which is typical of
galaxy evolution. This is also in agreement with recent bias measurements (e.g. Coupon etal
2011), which find a smooth dependence of
observed bias with redshift for different galaxy luminosities and
color, in good agreement with HOD. These results 
corresponds to  $i_{AB}<22.5$, similar to
the  sample that will be studied here. 
Systematics and selection effects do not seem to 
introduce higher frequency variations in the recovered biases. 

\begin{figure*}
\includegraphics[width=6in]{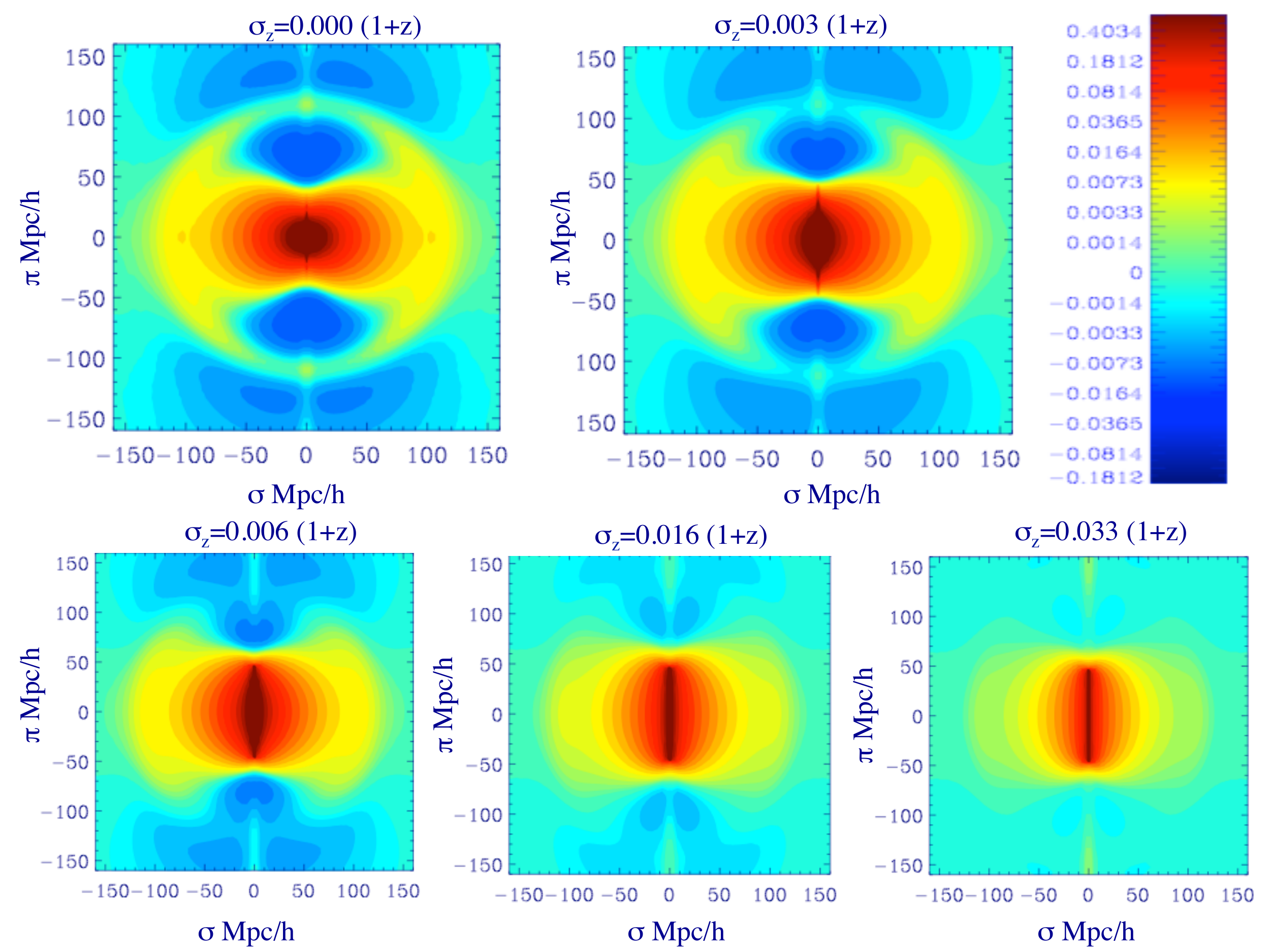}
\caption[xsp]{
Top left panel shows the $\xips$ correlation in the Kaiser model with 
no photo-z error, i.e. $\sigma_z=0$ (from \pcite{paper4}). 
The correlation is clearly squashed in the radial direction with
a region of negative correlation (in blue) between $\pi=50-100$ Mpc/h. Top right panel 
shows the same model but with a photo-z degradation of $\sigma_z=0.003(1+z)$, corresponding
to the PAU Survey. The difference is small and is mostly confined to small radial scales. Bottom panels
show how the results are degraded as we increase $\sigma_z$ to 0.006 (left), 0.016 (center)
and 0.033 (right panel). As the photo-z increases the radial squashing disappears, turning 
instead into a radial elongation. Note also how the region of negative correlation vanishes
as we increase the photo-z error.}
\label{xsp}
\end{figure*}


Following these indications, we
model $b(z)$ by including free bias parameters $b_i$
at some fixed redshifts,  with  linear interpolation to other redshifts.  
We choose a constant spacing in scale factor $\Delta a_i=0.1$ to fix the
interpolation positions in $b(z)$. Given the 
redshift distribution of galaxies in our sample we choose
4 points as the interpolation locations: $a_i=0.8, 0,7, 0.6, 0.5$, 
corresponding to redshift $z_i=0.25, 0.43, 0.67, 1.0$.
Fig.\ref{fig:rfitb} shows, as continuous line, a fit to such bias
model. The fit has a $\chi^2$ per degree of freedom of 0.26
which is ``too good''. This indicates that this model has too many free
parameters. If we use 2 bias parameters, instead of 4, at $a_i=0.7, 0.5$, we
find a $\chi^2$ per degree of freedom of 1.2 (dashed line in the
Figure). This indicates that 2 parameters is probably not enough,
given the small error-bars. So 3 parameters seems to be the right number for large
samples (5000 deg$^2$) and 2 parameters could be enough for smaller area surveys (ie
200 deg$^2$) which have larger errors. Here we will always use 
4 parameters to be on the safe side. 

In summary, we will use 4 biasing parameters with
no priors as default, but also show how results change as a 
function of priors.
Both from HOD modeling and from its comparison with observations 
we already have strong priors on what is the variation of  
bias for a given galaxy sample. These priors can also be estimated
from the same data we are considering here.  Recall that we will only
be using linear scales in our forecast, while HOD modeling can take advantage of the
data on smaller non-linear scales to constraint  halo mass and HOD. Moreover,
biasing could also be measured and constrained with other techniques,
such as higher order correlations 
in the same galaxy sample (e.g. see Gaztanaga, Norberg, Baugh \&
Croton 2005, Sefusatti, Crocce, Pueblas \& Scoccimarro 2006 and
references therein).


\subsection{Redshift Space distortions (RSD)}
  \label{sec:z-dist}

The measured redshift distance $s$ to a galaxy differs from the cosmological
distance $r$ by its peculiar velocity $\vec{v}$ along the line-of-sight.
These displacements lead to redshift distortions. On large (linear) scales
the dominant effect is due to coherent bulk motions induced by
gravity. Mass conservation implies that  the velocity divergence is
 $\theta \equiv  {\vec{\nabla} .  \vec{ v}} =-\dot{ \delta}$,
 which on linear scales results in
$\theta=-f \delta$ (see Eq.\ref{eq:f(z)}). Thus in the line-of-sight
direction fluctuations are distorted  by a factor $(1+f)$
resulting in a squashing effect in the 2-point redshift
correlation function \cite{Kai}.
 At small scales, random
velocities inside clusters of galaxies produce a radial stretching
pointing towards the observer, known as fingers of God (FOG).
This have very little effect on linear scales considered here
(Hikage, Takada \& Spergel 2011).

The top left panel of Fig.\ref{xsp} shows the squashing effect
in the amplitude of the 2-point correlation
as a function of radial (vertical axes) and perpendicular (horizontal axis) separation
\cite{paper4}. 
The squashing effect is quite clear in the inner regions of 20-50
Mpc/h and it also produces a  large region with
negative correlation between 50 and 100 Mpc/h which is 
characteristic of RSD (it is not present in the transverse direction).
This effect allow us to measure $f$ \cite{paper1}. 
One can also use this information to measure $H(z)$ from the BAO
ring position at $\simeq 100$ Mpc/h (see Gaztanaga, Cabre \& Hui 2009,
Matsubara 2004) and put constraints on DE equation of state
(Gaztanaga, Miquel \& Sanchez 2009).

The remaining panels display the same measurements done over a
distribution of objects with {\it photometric} redshifts of increasing
photo-z error (as labeled). The distortion pattern due to coherent
infall for the case of $\sigma_z=0.003(1+z)$ (top right panel) 
is remarkably similar to the one in the original redshift sample.
And it is precisely this anisotropic signal that we will employ to
disentangle bias and growth of perturbations. 
For most purposes $\sigma_z \sim 0.003(1+z)$ is almost
equivalent to having a (spectroscopic) redshift sample. This
will be quantified better later. Notice how this agreement
degrades rather quickly with $\sigma_z$.

In the large-scale linear regime and in the plane-parallel 
approximation (where galaxies are taken to be sufficiently faraway
from the observer that the displacements induced by peculiar 
velocities are effectively parallel), the distortion caused by
coherent  infall velocities takes a particularly simple form 
in Fourier space \cite{Kai}:
\begin{equation}
\label{eq:kaiserdelta}\delta_s(k,\mu) = (1 + f \mu^2) \delta(k)
\end{equation}
where $\mu$ is the cosine of the angle between $k$ and the line-of-sight, the subscript $s$ indicates redshift space, and $f(z)$ is  given by Eq.~(\ref{eq:f(z)}).
If we assume that galaxy fluctuations
are linearly biased by a factor $b$ relative to the underlying matter density 
$\delta$ (i.e. $\delta_g = b \, \delta$)  but velocities are unbiased,
then 
\begin{equation}
\label{eq:kaiserdeltag}\delta_g(k,\mu) = (b + f \mu^2) \delta(k)
\end{equation}
where $\delta_g$ are the measured galaxy fluctuations in redshift
space. However in practice what we measure is the rms value
$\sigma (b+\mu^2f)$, where
$\sigma$ is the r.m.s. amplitude of fluctuations at some scale. 
Hence by fitting the $\mu$ dependence
we can have independent measurements of $b \, \sigma$ and $f \, \sigma$ as a
function of redshift (or the ratio $\beta=f/b$, but we can not
separate $f$  from both $\sigma$ or $b$).
This means that $f(z)$ can not be measured unless we fix the normalization of
$\delta(z)$. Combining RSD with weak lensing 
 (see below)  or higher order correlations 
in the same galaxy sample (e.g. see Gaztanaga, Norberg, Baugh \&
Croton 2005, Sefusatti, Crocce, Pueblas \& Scoccimarro 2006 and
references therein)  we can break this degeneracy.
Nonetheless having an estimate for $f(z)\sigma(z)$ (independent of
bias) can be as valuable  as just having $f$, as both are ways to
constrain $D(z)$ and therefore $\gamma$.

To implement RSD constraints on the growth rate $f$ and bias we follow
the approach of White et al. (2009). First we write
the cross power spectrum of
galaxy samples A and B at {\it the same} redshift bin $i$ as, 
\beq
P^{(i)}_{AB}(k,\mu) = (b_A+f \mu^2)(b_B+f \mu^2) \,{\rm G}_{\sigma_z}(k,\mu)\,P_{\rm L}(k)  
\label{eq:pAB}
\eeq
where A and B are indices  for different galaxy
types and $G_z$ 
\beq
{\rm G}_z(k,\mu) = \exp\left[-(1/2)k^2 \mu^2 (\sigma^2_A+\sigma^2_B)\right]  
\eeq
accounts for the radial smearing due to photometric
redshift uncertainties (and possibly the small scale peculiar velocity dispersion).
The factor $\sigma$ relates to the photo-z error 
$\sigma_z$ by $\sigma=\sigma_z c /H(z)$.
In Eq.~(\ref{eq:pAB}) the bias $b$, the growth rate $f$ and the photometric uncertainty $\sigma$
are functions of redshift $z_i$.
In turn, the linear power spectrum $P_{\rm
  L}$ is determined up to an overall normalization, say $\sigma_8(z) =
D(z) \sigma_8(0)$, leading to the well known degeneracy between $f(z)$ or $b(z)$
and $\sigma_8(z)$. 
The FM corresponding to a given redshift bin 
$i$ is (see Eq.~(\ref{eq:fishermatrix})), 
\begin{equation}
F^i_{\mu\nu}=\sum_{XY} \int \frac{V_0 \,d^3k}{(2\pi)^3} \left(\frac{\partial
    P^{(i)}_X}{\partial q_\mu}\right) {\rm C}^{-1}_{XY} \left(\frac{\partial
    P^{(i)}_Y}{\partial q_\nu}\right)
\label{eq:fisher}
\end{equation}
where $V_0$ is the total volume of the redshift bin and $X,Y = (g_1g_1,g_1g_2,g_2g_2)$ in the case of two galaxy types
 or just $X=g1$ for only one type. The derivatives in
 Eq.~(\ref{eq:fisher}) for just one population are simply given by,
\beq
\frac{\partial\,P}{\partial\,b} =\frac{2}{b+f \mu^2} P(k) \ \ {\rm and} \ \
\frac{\partial\,P}{\partial\,f} = \frac{2\mu^2}{b+f\mu^2} P(k), \nonumber 
\eeq
while the covariance matrix is,
\begin{equation}
{\rm Cov}[P(k),P(k)] = 2 P^2(k) N^2(k) 
\label{eq:covRSD}
\end{equation}
with $N(k)=1+(\bar{n} P(k))^{-1}$ and ${\bar n}$ the galaxy number density.
It is in this error term that the damping due to photometric
uncertainty has an impact by increasing the shot-noise contribution to $N(k)$
and therefore the total error in $P(k)$.
Expressions for the derivatives and covariance
of two or more galaxy types can be readily obtained
(see White et al. 2009).

The combination of parameters that are
constrained with RSD are $q_\nu = b(z)\sigma_8(0) D(z)$
and $f(z)\sigma_8(0) D(z)$.  In turn these depend
on the set of cosmological parameters $p_\mu$ 
defined in Eq.~(\ref{eq:parameters})
through Eqs.~(\ref{fe2},\ref{eq:f(z)},\ref{eq:dgamma}).
To be able to join the constraints from RSD and WL
we transform the variables by:
\beq
F_{p_{\mu}p_{\nu}} = M F_{q_\mu q_\nu} M^{\rm T}
\eeq
where $M = \partial p / \partial q$.

As explained in \S\ref{sec:priors},  the value
of $\sigma_8(0)$ is dominated by priors so that we only measure
relative amplitudes here. As we do not measure 
the shape of $P(k)$ either, in practice our forecast
is equivalent to a measurement proportional to the ratio of the radial
and transverse directions. 

When we combine RSD with WL measurements we assume that these
measurements are not correlated. This is not quite correct when the
measurements are over the same area (FxB case below). To avoid the complication of
estimating  the  full covariance of these two observables (i.e. $P(k)$
with $C_\ell$) we will
assume  here that they are 100\% correlated and make sure that we 
only count once the same modes (i.e. we use only independent measurements). 
This is a conservative approach as the true covariance will always be
smaller. The total number ($N_T$) of $k$-modes in the RSD case is:
\beq
dN_T(k)= \left({L\over{2\pi}}\right)^3 2\pi k^2 dk d\mu
\eeq
where $L^3$ is the volume of the survey region (e.g. within a redshift bin).
This can be written as the product of  purely transverse $N_t$ and 
purely radial $N_r$  modes:

\bea
dN_T(k) &=& dN_t(k) ~ dN_r(k) \\ \nonumber
dN_r(k) &=&  \left({L\over{2\pi}}\right) k d\mu \\ \nonumber
dN_t(k) &=& \left({L\over{2\pi}}\right)^2 2\pi k dk   ~~~~(\mu=0)
\eea

In WL (or angular clustering) we will use the $C_\ell$ angular power spectrum
which only measure purely transverse modes. This is easy
to verify. If we use $k= \ell / \chi$ (where $\chi$ is radial distance and
$\ell$ is the angular multipole) we find that $N_t(k) \simeq N(\ell) \simeq 4\ell$, as in
the $C_l$  calculation in Eq.\ref{eq:ecl}).
So we can avoid double counting by subtracting the purely transverse
modes from the RSD calculation:
\beq
N'_T(k)= N_T(k)-N_t(k) = N_T(k) \left( 1 - {1\over{N_r(k)}} \right)
\eeq
where $N'_T$ is the number of independent modes that can be used
for RSD forecast when combined with WL (ie for the FxB case).
This correction is small as $N_t(k)>>1$. In our case this can only
affect the FxB combination and  we have checked
that this correction is always smaller than 30\% in the final FoM.

\section{Weak Lensing}

Weak lensing produces both a coherent distortion of galaxy shapes (known as shear distortion)
and coherent magnification (MAG) of area
that can be a source of density fluctuations. The latter is known
as magnification bias  or cosmic magnification (see e.g. Gunn 1967, Narayan
1989, Broadhurst et al. 1996 and references therein) which introduces both a correction to 
the angular galaxy auto correlation function \cite{VFC97,loverde08}
and also to the cross-correlation of galaxies at different (including
disjoint)  redshifts \cite{MJ98,HVWE09}. 
Magnification bias also  changes the shape of the 
angular auto correlation function in RSD  \cite{matsubara2004,hui07,hui08},  
but these are relatively small effects, so we neglect then here from
now on.

Measurements of cosmic magnification can provide comparable S/N to that of
cosmic shear without the need to measure galaxy shapes (see e.g. Van
Waerbeke 2009 and references therein).
With spectroscopic or good photometric redshifts, such
as those in PAU, one can further explore 3D lensing tomography
by means of the angular cross correlation between galaxies in different
narrow redshift bins, as will be explored here.

 We follow the approach of Hu \& Jain (2004 and references
 therein) who presented forecasts for shear and galaxy cross-correlations. 
As new ingredients, we will also consider 
 the case for weak lensing MAG and the use
of narrower redshift bins.

 As we will show below, i.e. Eq.\ref{eq:cijgg} and Eq.\ref{eq:cijgk}, 
both galaxy-galaxy and galaxy-shear cross-correlations of
narrow redshift bins produce a direct measurement of the 
power spectrum of the foreground (lensing) distribution. This allows
for a 3D reconstruction of the power spectrum, with a radial
resolution that is much better than possible 
with the shear-shear reconstruction, i.e. Eq.\ref{eq:clkk}.
 The smaller the bin width, the 
better the reconstruction.  In our forecast we are limited
radially by the accuracy of the Limber approximation, which 
breaks for very narrow bins. To simply the analysis
we also want to be able to neglect the intrinsic
cross-correlations of separate bins.  To achieve this,
we find that we should use redshift bins which
are larger than about 0.014(1+z) (i.e. $>60$ Mpc/h) for which these
approximations are adequate.
Better results could in principle be obtained by 
using smaller bin widths, but this is left for  a future analysis.
 In this sense our results are conservative.

\subsection{Magnification \& Shear} 

Magnification  $\mu$ is defined as:
\beq
\mu = {1\over{det A}} = {1\over{(1-\kappa)^2-|\gamma|^2}}
\label{eq:mu}
\eeq
where $A$ is the Jacobian matrix for the lensing transformation
(see e.g. Bartelmann \& Schneider 2000). 
In the weak lensing limit, fluctuations in magnification $\mu$, 
convergence $\kappa$ and shear $\gamma$ are closely related. 
Fluctuations in (E-field) shear $\epsilon$ and convergence are 
equal, $\delta_\kappa=\delta_\epsilon$ (see e.g. Hu \& Jain 2004)
and, according to Eq.\ref{eq:mu},  they are half as large as magnification
$\delta_\mu=2\delta_\kappa$.
As $\kappa$ can be obtained from shear measurements, we will
also use $\kappa$ to refer directly to shear. But
we should bear in mind that some shear estimators
are given in terms of $\delta\kappa/2$ (e.g. the iCosmos software of
Refregier et al. 2011) and other combinations.
In our FM approach this will only be relevant when
introducing the scale for the intrinsic noise ellipticity.

We will focus here in convergence, $\delta_\kappa$, as reconstructed from 
galaxy shapes measurements (that we will call shear), 
and magnification, $\delta_\mu$, as estimated
from fluctuations in galaxy number density counts. 
Magnification  changes the area of the background
sources behind lenses, this induces a background fluctuation
 $\delta_g \simeq -\delta_\mu$ which is correlated
with the foreground galaxy  population.
 Additionally, background
magnitudes are also affected 
inducing additional galaxy density fluctuations in $\delta_g$ 
across the sample magnitude limit. Adding both contributions gives
\beq
\delta_g =  (2.5s-1) \delta_\mu \simeq (5s-2) \delta_\kappa
\eeq
where $s$ here is the slope of the galaxy number counts at the flux limit. 
Dust extinction in the lenses can also produce significant
fluctuations. Menard et al. (2010) have shown that at sufficient large
wavelengths, i.e. $I$ and $Z$ bands, dust extinction becomes negligible 
and the change in magnitude is dominated by magnification.
In our  analysis we will use $I$  magnitudes to select galaxies
and neglect dust extinction, although this might not always be
a good approximation (see Fang et al. 2011).

Weak lensing convergence in bin $j$
is given by the projected matter density $\delta_{m_i}$ in all
foreground redshifts $i<j$:
\beq
{\delta}_{k_j}(\vec{\theta}) =  \sum_{i<j} \bar{p}_{ij}\delta_{m_i}(\vec{\theta}) 
\label{eq:deltak}
\eeq
where $\bar{p}_{ij}<1$ is a geometrical WL weight that will be
introduced later on and  $\vec{\theta}$ are sky positions.
We then have the following relation between observed galaxy fluctuation
at  background bin $j$, i.e. $\hat{\delta}_{g_j}$, and the foreground
DM distribution: 
\beq
\hat{\delta}_{g_j}(\vec{\theta})  
\simeq
b_j \delta_{m_j}(\vec{\theta})  + \epsilon_j (\vec{\theta}) + \sum_{i<j} p_{ij} \delta_{m_i}(\vec{\theta})  
\label{eq:deltahat}
\eeq
where $p_{ij}=(5s-2) \bar{p}_{ij}$ 
and we have used the linear local bias with
stochasticity $\epsilon$, defined in Eq.\ref{eq:epsilon}. If we neglect
$\epsilon$ we can use the observed $\hat{\delta}_g(\vec{\theta})$  and $\delta_k(\vec{\theta})$
maps and the above Eq.\ref{eq:deltahat} and Eq.\ref{eq:deltak} to reconstruct both
$\delta_m(\vec{\theta})$ and also $b_i$ and $p_{ij}$
(see Pen 2004).  In general it is not clear to what extend we can 
 neglect $\epsilon$. What we do here instead is to consider cross-correlations.
The galaxy-galaxy and galaxy-shear cross-correlations then relates directly to
the matter auto correlations
\bea
<\hat{\delta}_{g_i}\hat{\delta}_{g_i}> &\simeq& 
b_i^2 <\delta_{m_i}\delta_{m_i}> \\ \nonumber
<\hat{\delta}_{g_i}\hat{\delta}_{g_j}> 
&\simeq&  b_i p_{ij}  <\delta_{m_i}\delta_{m_i}> ~~~~~~i<j\\
<\hat{\delta}_{g_i}\hat{\delta}_{k_j}> 
&\simeq&  b_i \bar{p}_{ij} <\delta_{m_i}\delta_{m_i}>
~~~~~~i< j \nonumber 
\eea
where we  keep only 
leading order in $p_{ij}<<1$. We
have neglected $<\epsilon_i\delta_i>$ and  
 $<\epsilon_i\epsilon_j>$ terms, which seems to be
a good approximation on large linear scales as we find that
the cross-correlation $r$ defined in Eq.\ref{eq:r} is very close to unity
(see discussion in \S2.4 and Fig,\ref{fig:rfitb}). More generally,
when $r \neq 1$ we need to replace $b_i$ by $b_i r_i$ in the second
and thrid equations. This has little impact in our predictions
as shown in \S\ref{sec:sto}.

Galaxy-galaxy cross-correlations from pairs of different redshift bins
can  be combined with the galaxy-galaxy auto-correlation
to measure bias $b_i$ and $p_{ij}$ and $<{\delta}_{m_i}\delta_{m_i}>$
with a radial precision given by the number of independent
redshift bins. The same information can also be obtained 
from combining galaxy-shear cross-correlation and 
galaxy-galaxy autocorrelation.

Note that in our approach we ignore all the radial modes as  
the intrinsic $<\delta_{m_i}\delta_{m_j}>$ correlation is negligible for disjoined
top-hat bins $i\ne j$ when $\Delta z > 0.02$. It could be possible to
include radial modes by using smaller redshift bins but this requires
going beyond the Limber approximation, a study that we leave for future
analysis (see Challinor \& Lewis 2011).
In this sense our results are conservative, as we do not
include radial modes or very fine radial bins.
As we increase the radial resolution 
(or number of independent redshift bins)
 the number of independent transverse modes  $k_i=\ell/r_i$ 
also increases.

Finally, recall that we do not include redshift space distortions (RSD)
in the modeling of angular correlations. This is a very good
approximation for lensing because it has a very broad radial
window which washes away the effect of radial peculiar velocities.
For the angular galaxy-galaxy autocorrelation in narrow redshift
bins, this is not a good approximation 
 (see Nock et al. 2010, Crocce, Cabr\'e \& Gazta\~naga
2010 and references therein). As mentioned in \S2.3 our approach is to
model angular clustering as transverse modes in real space and 
include the effect of RSD through the ratios of  the
amplitude of clustering as we change from transverse to 
radial modes within a single redshift bin (see \S2.5).

\subsection{Correlations \& Power spectrum}
\label{sec:corr}

Consider measurements of angular galaxy density and shear 
 in a set of z-bins $i=1,..,N_z$. Then the projected measurement of
 $A$ in bin $i$ is:
\beq
\delta_{A_i}(\vec\theta) = \int dz ~p_{A_i}(z) \delta_m(r\vec\theta,z)
\eeq
where $\delta_m$ is the 3D dark matter fluctuation, $\vec\theta$ gives the
angular position, $r=r(z)$ is the comoving transverse (or angular diameter)
distance to redshift $z$ and $p_{A_i}$ is the radial weight defined
later. The label $A$ refers to either the galaxy fluctuation, i.e. $A=g$, or a shear
measurement of convergence, i.e. $A=\kk$.

The observables to measure are the galaxy-galaxy ($gg$), galaxy-shear ($g\kk$)
and shear-shear ($\kk\kk$) cross-correlations of angular fluctuations
between redshift bins $i$ and $j$, which are denoted by:
\bea
\label{eq:wcross}
w_{g_ig_j}(\theta) &\equiv& \langle
\delta_{g_i}(\vec\theta_1)\delta_{g_j}(\vec\theta_2)\rangle \\
w_{\kk_i\kk_j}(\theta) &\equiv& \langle
\delta_{\kk_i}(\vec\theta_1)\delta_{\kk_j}(\vec\theta_2)\rangle \\
w_{\kk_ig_j}(\theta) &\equiv& \langle
\delta_{\kk_i}(\vec\theta_1)\delta_{g_j}(\vec\theta_2)\rangle 
\eea
where $\theta=|\vec\theta_2-\vec\theta_1|$ is the angular separation between 2 measured
fluctuations of galaxy density $\delta_g$ or galaxy shear $\delta_\kk$
(see Bernstein 2009 for generalizations of this).
It is convenient to express $w(\theta)$ in terms of harmonic coefficients $C(\ell)$:
\bea
\label{eq:wcl}
w(\theta) &=& \sum_{\ell} \frac{2\ell+1}{4\pi} C(\ell) L_{\ell}(\cos\theta) \\
C(\ell) &\equiv& 2\pi \int_{-1}^{1} d\cos\theta \, w(\theta)\, L_{\ell}(\cos\theta)
\eea
where $L_{\ell}(cos\theta)$ are Legendre polynomials  of order
$\ell$. Thus we have galaxy-galaxy, $C_{g_ig_j}(\ell)$, shear-shear,
$C_{\kk_i\kk_j}(\ell)$  and galaxy-shear, $C_{g_i\kk_j}(\ell)$  cross
power spectrum between bin $i$ and $j$.
Our galaxy sample is split into different galaxy types (or
 populations) that we label by $gn$, where $n$ can take $N_t$ values
 (we only use $N_t=2$ here, but this can be easily extended), ie
 $n=1,2,..N_t$, for each population with different bias. We do not
 split shear by populations because we assume that shear is independent of bias.
 Thus the total number of observables for  2 populations, $N_t=2$, is then:
\bea
\label{eq:cijall}
C_{ij}(\ell) &\equiv& [ C_{g1_ig1_j}(\ell) , C_{g2_ig2_j}(\ell), C_{g1_ig2_j}(\ell),
 \nonumber \\ & & C_{g1_i\kk_j}(\ell),C_{g2_i\kk_j}(\ell),
C_{\kk_i\kk_j}(\ell) ]
\eea
For a generic cross-correlation  we will use the notation:
\beq
C_{A_iB_j}(\ell)
\eeq
 where $A$ and $B$ can take the values $\kk$ for shear or $g1,g2,..g{N_t}$
 for different galaxy types. Note that while $C_{A_iA_j}=C_{A_jA_i}$, in general:
\bea
C_{A_iB_j}(\ell) &=& C_{B_jA_i}(\ell) 
\\
C_{A_iB_j}(\ell) &\neq&  C_{A_jB_i}(\ell)  ~~for~~ A\neq B ~;~ i \neq j
\eea

In the small angle approximation (Limber 1954, Loverde \& Afshordi
2008), there is a simple relation between $C(\ell)$ 
and the 3D matter power spectrum $P(k)\equiv <\delta^2(k)>$:
\bea
\label{eq:clAB}
C_{A_iB_j}(\ell)  &=& 
\int_0^\infty dz  ~p_{A_i}(z) ~p_{B_j}(z) ~ \calP(k,z) \\
\calP(k,z) &\equiv&  {P(k,z)\over{ r_{H}(z) r^2(z)}} 
\label{eq:calP}
\eea
where $r_{H}(z) \equiv \partial r(z)/\partial z$, 
and
$\calP$ is the a-dimensional power spectrum at $k=(\ell+1/2)/r$.
In linear theory $P(k,z)=D^2(z) P(k)$. 

\subsection{Lensing weights} 
\label{sec:pij}

For shear or convergence
fluctuations, i.e. $A=\kk$ the redshift
distribution $p_A=p_\kk$ in the above equation corresponds 
to the weak lensing efficiency between sources in redshift bin $j$ 
and dark mater (lenses) at z:
\beq
\label{eq:ps}
p_{\kk_j}(z)\equiv  {3\Omega_m H_0 r(z) \over{ 2 H(z) a(z) r_0}}
\int_{z}^\infty dz'  ~{r(z';z)\over{r(z')}}~  \phi(z') 
\eeq
where $\Omega_m$ is the DM density at $z=0$, $r_0=c/H_0$,
$a(z)=1/(1+z)$. Here $r(z';z)$ is
 the angular diameter distance between $z'$ and $z$
and $\phi_j(z)$ is the normalized galaxy probability density
distribution for source galaxies in the redshift bin label $j$.
For a uniform galaxy distribution in a top-hat window of total width $\Delta_j$:

\beq
\label{eq:pg1}
\phi_j(z) = 
\left\{ \begin{array}{ll} 
  {1/\Delta_j} 
 & \textrm{ for $|z-z_j|<\Delta_j/2$} \\
\\
0 & \textrm{otherwise}.
\end{array} \right.
\eeq

\subsubsection{Narrow bin approximation}
\label{sec:narrow}

\begin{figure}
\includegraphics[width=3in]{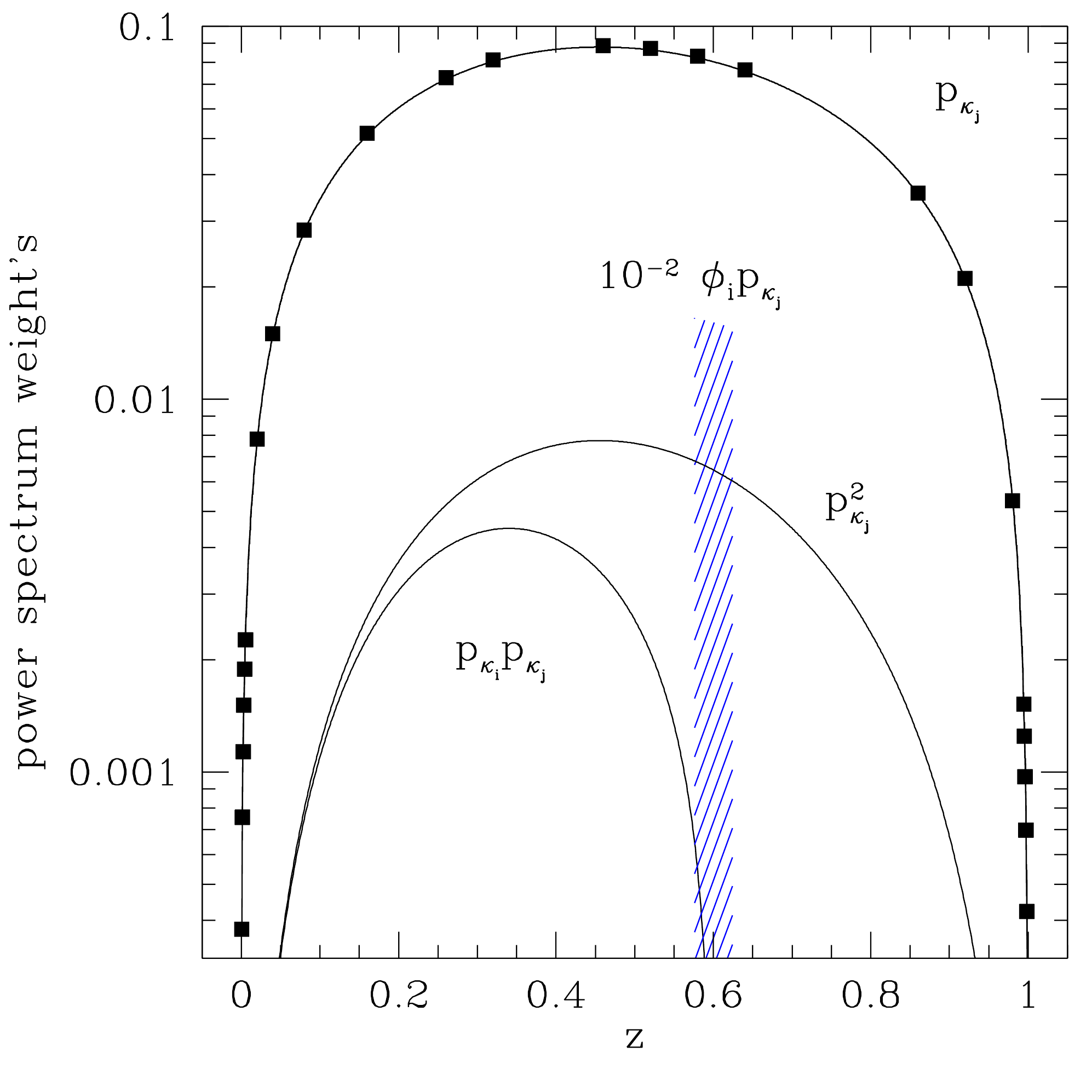}
\caption[]{
Filled squares show the weak lensing efficiency $p_{k_j}(z)$ in
Eq.\ref{eq:ps} as a function of the lensed position $z$
for a source at $z_j=1.0$. The line across the squares corresponds
to the approximation in Eq.\ref{eq:ps2}.
 The other two lines show 
the weights $p^2_{\kappa_j}$ and $p_{\kappa_i}p_{\kappa_j}$ 
 applied to the matter power spectrum $\calP(k,z)$ in the
shear-shear auto and cross-correlation of Eq.\ref{eq:clkk}
for $z_i=0.6$.
These distributions are always very broad,
indicating that shear-shear can only measure the projected
(2D) power. The shaded region shows the corresponding
weights for galaxy-galaxy cross-correlation from MAG (or galaxy-shear), ie
$\phi_{z_i}p_{\kappa_j}$, for $\Delta_i=0.05$. Here the power spectrum is measured
in 3D with a resolution that is only limited by $\Delta_i$.}
\label{fig:pwl}
\end{figure}

For narrow redshift bins we approximate $p_\kk$ in Eq.\ref{eq:ps} by:

\beq
\label{eq:ps2}
p_{ij} \equiv 
p_{\kk_j}(z_i)\simeq
{3\Omega_m H_0 \over{ 2 H(z_i) a_i}}
  ~{r_i r(z_j;z_i)\over{r_0 r_j}}  ~~
\textrm{for}~ i<j
\eeq
where $r_i\equiv r(z_i)$. This is an excellent approximation for
the non-overlapping top-hat bins in our problem. For example for
sources at $z_i=1$, and  a bin width 
of $\Delta z_i=0.1$ (which is 30 times larger than our default value
at $z_i=1$) the accuracy of this approximation is better than 0.7\%
for $z<0.95$, which is all we need for non-overlapping z bins.
This is  illustrated in Fig.\ref{fig:pwl}, which
compares the exact calculation (squares) with the approximation (lines
across the squares) for $\Delta z_i=0.05$.
This approximation  greatly simplifies  Eq.\ref{eq:clAB}, which
becomes a single integral that can be done analytically for the
galaxy-galaxy and the galaxy-shear case.
Also note that $p_\kk$ in this approximation is
independent of the redshift bin width.

\subsection{Galaxy weight} 

For galaxy type ``n'',  i.e. $A=gn$, the redshift
distribution $p_A=p_{gn}$ in the Eq.\ref{eq:clAB}, corresponds 
to the sum of the intrinsic number density  and the lensing
magnification  contribution (see Eq.\ref{eq:deltahat}):
\beq
p_{gn_i}(z) = b_{n_i} \phi_{n_i}(z) + \alpha_{n_i} p_{\kk_i}(z) 
\eeq
where $b_{n_i}\equiv b_n(z_i)$ is the mean linear galaxy bias for galaxy type $n$ at
redshift bin $i$.
The second term is due to weak lensing magnification with
\beq
\alpha_{n_i} \equiv \alpha_n(z_i) \equiv 5s_{n}(z_i)-2 
\eeq
where $s_n(z_i)$ is the slope of the number counts $N_n(<m;z_i)$ of
galaxies of type $n$ with apparent magnitude smaller than the survey
flux limit $m$ (plus whatever other cuts we do to the survey)
at $z=z_i$
\beq
s_n(z_i) \equiv {d {\,\rm log}_{10} N_n(< m;z_i) \over dm}
\label{eq:s}
\eeq
 Its value can be estimated using
the same galaxy sample.  We then have that
$\alpha_n$ modulates the weak lensing magnification effect of
dark matter fluctuations at $z<z_i$. It is clear from this
that both shear and magnification could measure the same information, modulo
$\alpha_n$, noise and systematics, but in practice we need to study
their correlations, which we do next.

\subsection{Galaxy-Galaxy (Magnification)}

For narrow and disjoint redshift bins we have from Eq.\ref{eq:clAB} that for $i<=j$,
\bea
C_{gn_igm_j}(\ell) &\simeq& \left[b_{n_i} b_{m_i} {\delta_{ij}\over{\Delta_i}} +
\alpha_{m_j} b_{n_i} p_{ij} \right] \calP_i
\nonumber \\ &+& \alpha_{n_i}\alpha_{m_j} C_{\kk_i\kk_j}(\ell)
\label{eq:cijgg}
\eea
where $\delta_{ij}$ is the Kronecker delta and
$\calP_i \equiv \calP(k_i,z_i)$ in Eq.\ref{eq:calP}, with $k_i=(\ell+1/2)/r(z_i)$.
When $i=j$ the first term dominates (as $p_{ii}=0$), while for $i\neq j$ the second
term is dominant (as $\delta_{ij}=0$).  The last term is always
sub-dominant and will be neglected here.\footnote{But note that if we can measure very well the two first terms, the last term
contains all the shear-shear information, without need of direct
shear measurements.} So we see here how galaxy cross-correlations can
be used to measure 3D power spectrum without the need of shear
measurements. 

The nuisance
biases $b_{n_i}$, can be measured by comparing $C_{ij}$ with
$C_{ii}$. The main source of contamination here is 
intrinsic or induced correlations of
galaxies in bin $i$ with the ones in bin $j$. 
If these bins are well separated this can
only occur because of photo-z transitions, ie
galaxies moving from one redshift bin to another due to
photo-z errors  (see \S\ref{sec:trans} and \S\ref{sec:out} for details). 

The different pair combinations
can be used to simultaneously measure $b_i$, $\calP_i$ and $p_{ij}$.To
illustrate this claim, consider the ratio: $C_{ij}/C_{ik}$ when
$i\neq j \neq k$ we have:

\beq
\label{eq:rij}
{C_{ij}(\ell)\over C_{ik}(\ell)} = {\alpha_j p_{ij}
  \over{\alpha_k p_{ik}}}
\eeq
which are independent of bias and the power spectrum for any value of $\ell$.
This is also true in the non-linear regime (i.e. with non-linear bias
and non-linear dark matter clustering). This is therefore a direct
geometrical measure. In a similar way one can build up ratios like
$C_{ii}/C_{ij}$ to measure $b_i$ independent of $\calP_i$, or $C^2_{ij}/C_{ii}$
to measure $\calP_i$ with independence of bias $b_i$.

\subsection{Shear-Shear} 

For the shear-shear term
with $i<j$ the cross-correlations are:
\bea
\label{eq:clkk}
C_{\kk_i\kk_j}(\ell) &=& 
\int_0^{z_j} dz ~p_{\kappa_i}(z) ~p_{\kappa_j}(z) ~ \calP(k,z) 
\\ 
&\simeq&
\int_0^{z_i} {dz \over{r_H}}
\left( {3\Omega_m H_0 \over{ 2 H a r_{0}}}\right)^2
  ~{r(z_i;z) r(z_j;z)\over{r_ir_j}}  
~ P(k,z) \nonumber
\eea

and $ C_{\kk_i\kk_j} = C_{\kk_j\kk_i}$ for $j<i$.
It is also possible to combine here different pairs of bins to get
some tomographic (3D) information, but as illustrated in 
 Fig.\ref{fig:pwl} the resulting weights are very broad, 
even if we use narrow galaxy bins. This means that
the tomographic recovery is quite limited.

\subsection{Galaxy-Shear}  

Finally, for narrow bins,  we also find the galaxy-shear
cross-correlation from Eq.\ref{eq:clAB} for $i<j$:

\beq
C_{gn_i\kk_j} \simeq
b_{n_i}  p_{ij} \calP_i
+ \alpha_{n_i} C_{\kk_i\kk_j}
\label{eq:cijgk}
\eeq
and $C_{\kk_ign_j}  = 0$ for $i<j$. Thus the off-diagonal
 information content of $C_{gn_i\kk_j}$ is identical (except for
the global factor $\alpha_{m_i}$) to that of
$C_{gn_igm_j}$, even when the noise and the systematic could be
quite different.

\subsection{3D $P(k)$ recovery}   

A comparison of Eq.\ref{eq:cijgk} or Eq.\ref{eq:cijgg}
with Eq.\ref{eq:clkk} illustrates how cross-correlations of
galaxy-galaxy or galaxy-shear both
recover the full 3D power spectrum, i.e. $\calP(k)$, while the shear-shear 
only provides an integral over very broad redshift distribution.
This is illustrated in Fig.\ref{fig:pwl}, which shows that
$p_{\kappa_i}p_{\kappa_j}$  in Eq.\ref{eq:clkk} 
is always very broad, while $\phi_{z_i}p_{\kappa_j}$ can be as narrow
as needed.

Another point to note
is the fact that the galaxy-galaxy or galaxy-shear
 cross-correlations only depend on $p_k \simeq 10^{-1}$ while shear-shear
depends on $p_k^2 \simeq 10^{-2}$. This means that the amplitude of
shear-shear correlation is much smaller and is therefore sensitive
to smaller amplitudes in  systematic effects. Moreover, the effect
of systematics will tend to cancel in a cross-correlation analysis to
a larger extend than in the auto-correlation. But these are generic
considerations and ultimately the key point is that systematics
are quite different in both measurements.

\subsection{Covariance}
\label{sec:covar}

In the Gaussian limit, the covariance between one  pair of
observables in Eq.\ref{eq:cijall} at redshift bins $(ij)$ and another pair 
at redshift bins $(kl)$ is given by:
\beq
\calC_{ij;kl} \equiv
Cov({C_{ij};C_{kl}}) ={\hat{C}_{ik}\hat{C}_{jl}+
\hat{C}_{il}\hat{C}_{jk}\over{N(\ell)}}
\label{eq:covar}
\eeq
where $N(\ell)$ is the number of modes at a given $\ell$.
In our case we bin the $l$ modes as $\Delta\ell \simeq 2/f_{sky}$ (see
Cabre et al. 2007)
to avoid correlation induced by the limited fraction of sky covered ($f_{sky}$).
The number of modes of each $\ell$ is then 
\beq
N(\ell)=(2\ell+1)f_{sky}\Delta\ell \simeq 2 (2\ell +1) 
\label{eq:ecl}
\eeq
The observables $\hat{C}$  in the covariance include observational noise:
\bea
\hat{C}_{gn_igm_j}&=&C_{gn_igm_j}+~{1\over{\bar{n}_{gn}}} ~ \delta_{ij} ~ \delta_{nm} \\
\hat{C}_{\kk_i\kk_j}&=&C_{\kk_i\kk_j}+ {\sigma_\kk^2\over{\bar{n}_\kk}}  \delta_{ij} \\
\hat{C}_{\kk_ign_j}&=&  {C}_{\kk_ign_j} 
\eea
where $\bar{n}_\kk$ and $\bar{n}_{gn}$ are the surface density of galaxies with
measured shear and galaxies of type $n$ with good photometry
 respectively and $\sigma_\kk^2$  
 is the variance in convergence
from intrinsic ellipticities. Note that $\sigma_\kk<1$ while
$ \bar{n}_{gn} >\bar{n}_\kk$ so that the noise could be larger or
smaller in shear-shear than in galaxy-galaxy, depending on the
depth and quality of data.


\subsection{Signal to Noise}
\label{sec:s2n}

From the above covariance 
we can estimate the signal to noise S/N ratio for the galaxy auto-correlation
$C_{g_ig_i}(\ell)$
\beq
(S/N)_{g_ig_i}^2 = {{N(\ell)}\over{2}}
\label{eq:s2ncii}
\eeq
and compare it to the galaxy-galaxy cross-correlation $C_{g_ig_j}(\ell)$
\beq
(S/N)_{g_ig_j}^2 \simeq {C_{ij}^2 N(\ell)\over{C_{ii}C_{jj}  }}
\propto {N(\ell) \Delta_i \Delta_j }
\propto {N(\ell)\over{N_z^2 }}
\eeq
where in the first step we have used that $C_{ij}^2<C_{jj}C_{ii}$
and $C_{ii} \propto 1/\Delta_i$, where
$\Delta_i \propto 1/N_z$ is the redshift bin width and
$N_z$  is the number of redshift bins.
The total S/N is the sum over all the $N_z$ redshift  bins.
In the case of the cross-correlation there are $N_z(N_z-1)/2$ pairs
and therefore the total S/N
is independent of the number of bins for large $N_z$. Something similar happens
when considering RSD or in general 3D $P(k)$  measurements.
This means that there is no S/N gain in using many redshift
bins. We could gain information in cases where the signal 
varies on radial scales comparable to the redshift bin width.
Both cosmological parameters and galaxy formation (i.e. biasing)
vary on scales $\Delta a=0.1$. This means that there is no information
gain in using smaller bins in this case. 

In the case of galaxy-shear:
\beq
(S/N)_{g_i\kappa_j}^2 \simeq {C_{g_ig_j}^2 N(\ell)\over{C_{g_ig_i}C_{k_jk_j}  }}
\propto {N(\ell) \Delta_i }
\propto {N(\ell)\over{N_z}}
\eeq
Here there are only $N_z$ galaxy-shear pairs per shear bin, so again
the total signal to noise is quite insensitive to the number of bins.

In the case of the autocorrelation $C_{ii}$ in Eq.\ref{eq:s2ncii},
the total S/N increases with number of
bins and there is a net S/N gain in using more bins. This gain is only  limited
by the increase of shot-noise for small redshift bins and the fact
that very narrow nearby bins are no longer independent. Thus, while
shear-shear and galaxy-shear are useful to recover galaxy bias, the
gain from using narrow bins
comes from galaxy-galaxy auto correlations.
This is a key point to understand the results in this paper.

\subsection{Photo-z error transitions}
\label{sec:trans}

First we study the case of galaxy-galaxy cross-correlations.
Consider the transition probability $T_{ij}$ that a galaxy at  bin $j$ is measured
to be at bin $i$ because of the photo-z  error. The number of
galaxies measured in bin $i$,  $\bar{N}_i$, is then:
\beq
\bar{N}_i= \sum_j T_{ij}~N_j
\eeq
where $N_j$ is the true number of galaxies in bin $j$. To include the
effect of photo-z errors in the cross-correlations we need to
define the relative transition probability , or migration matrix$r_{ij}$,
as  the fraction of the galaxies assigned to z-bin $i$
which really are in bin $j$, ie:
\beq
r_{ij} \equiv ~T_{ij}~ { N_j\over{\bar{N}_i}} = {T_{ij} N_j\over{
    \sum_j T_{ij} N_j}} =
{T_{ij} <N_j>\over{    \sum_j T_{ij}<N_j>}} 
\label{eq:rij}
\eeq
where by construction:
\beq
\sum_j r_{ij} = 1   ~~~\forall i
\eeq
Note that when $r_{ij}$ is a smooth function of the redshift 
distance $z_j-z_i$, $r_{ ij}$ will also be independent of the 
redshift width $\Delta z$ for $|z_j-z_i|>\Delta z$.
These $r_{ij}$ transitions correspond to the contamination
matrix $C_{ps}$ in Bernstein \& Huterer (2010) and give the
probability $P(z_s| z_p)$ for a true redshift is $z_s$ to be
measured in photo-z redshift $z_p$.
The last equality in Eq.\ref{eq:rij} just indicates that the same
probabilities apply for the mean density as for regions with
fluctuations: $N_i\equiv <N_i>(1+\delta_i)$. We then have
that the correlation $\bar{w}_{ij} \equiv
<\bar{\delta}_i\bar{\delta}_j>$ 
in photo-z space can be related to the true correlation
$w_{ij} \equiv <{\delta}_i{\delta}_j>$ as (see also Benjamin et al. 2010)
\bea
\bar{w}_{ij} &=& <(1+\bar{\delta}_i) (1+\bar{\delta}_j)> -1
= {<\bar{N}_i \bar{N}_j>\over{<\bar{N}_i><\bar{N}_j>}} -1
\nonumber \\ &=&
 \sum_{kl} r_{ik} r_{jl} (1+w_{kl}) -1 = 
\sum_{kl}  r_{ik} r_{jl} ~ w_{kl}
\eea
Thus, photo-z errors result in a mixing of the cross-correlation
measurements:
\beq
 \bar{C}_{ij} = \sum_{kl} ~r_{ik} ~r_{jl}~C_{kl}
\label{eq:barcij}
\eeq
where $\bar{C}_{ij} $ is the observed cross-correlation in photo-z
space and ${C}_{kl}$ is the true cross-correlation.
The new covariance matrix
(i.e. Eq.\ref{eq:covar}) will now be:
\beq
\bar{\calC} =  Cov(\bar{C};\bar{C}) 
\label{eq:barcT}
\eeq
Note that noise (e.g. shot-noise) should be added to the
observed (not the true) covariance.
 This could degrade the cosmological parameters correspondingly.
These considerations can be extended trivially to the case of
the cross-correlation of 2 galaxy populations. We will then have
different transitions $r_{ij}$ for each population and assume no
transition between populations. In matrix notation:\beq
 \bar{C}_{g1g2} = r_{g1} ~C_{g1g2} ~r_{g2}^T 
\label{eq:barcij2}
q\eeq
where $r_{g1}$ and $r_{g2}$ represent the transition matrix for
galaxy populations $g1$ and $g2$ and $T$ stands for the transpose
matrix. In our case $q_1$ and $g_2$ can take the values of
Faint, $g_F$, or Bright, $g_B$, galaxies.

For galaxy-shear and for shear-shear we  use the same relations
Eq.\ref{eq:barcij}-\ref{eq:barcT}
to account for photo-z contamination.  In the
case of shear, the observable is given by some mean over the
number of galaxies where shapes are measured.  So photo-z leakage
has a different effect in shear fluctuations than in galaxy counts (see Bernstein
\& Huterer 2010). Nevertheless we note  the shear
correlations are also built from pairs of measurements
and one can use the same relations as for galaxy-galaxy to 
account for photo-z contamination.

Consider a photo-z distribution $P(z_p|z)$ giving the probability that a
galaxy at true redshift $z$ is measured to be at $z_p$.
 For top-hat  bins of width $\Delta_i$ and $\Delta_j$ we have:
\beq
T_{ij} = \int_{\Delta_j} {dz\over{\Delta_j}}  \int_{\Delta_i}
dz_p  P(z_p|z)
\simeq
 \int_{\Delta_i} dz_p   P(z_p|z_j)
\eeq
where the integrals are around the corresponding top-hat 
bins centered in $z_i$ and $z_j$. In the second
equality above we have approximated the integral over bin $j'$ by its 
mean value.
Note that $T_{ij} \neq T_{ji}$ because the distribution  $P(z|z_p)$ is 
in general different for different $z$ (i.e. typically photo-z errors are
a function of $z$).
For a Gaussian photo-z distribution with error
$\sigma_z=\sigma_z(z_j)$

\bea
&T_{ij} & = \int_{z_{ij}-\Delta_j/2}^{z_{ij}+\Delta_j/2}
{dx\over{{2}\Delta_j}}
\left[erf({2x+\Delta_i\over{\sigma_z\sqrt{8}}})-
erf({2x-\Delta_i\over{\sigma_z\sqrt{8}}}) \right] 
\nonumber \\ 
& \simeq &
{1\over{2}} \left[
erf\left({{2z_{ij}+\Delta_i}\over{\sigma_z}\sqrt{8}}\right)
-erf\left({{2z_{ij}-\Delta_i}\over{\sigma_z}\sqrt{8}}\right)
\right] 
\label{eq:Tij}
\eea
where
$z_{ij}\equiv z_j-z_i$, 
and
 $erf(x)$ is the standard error function (integration of a normal from 0 to x).

In general, it is possible to recover $r_{ij}$ from the
observables given some model for the intrinsic correlations,
even without the combination of spectroscopic
and photometric samples. The cross-correlation of 
spectroscopic and photometric samples 
provides another route to measure $r_{ij}$, extending the
proposal by Newman (2008) to the cross-correlation in separate redshift bins.

\subsection{Uncertainties in the Photo-z transitions}
\label{sec:out}

So far we have seem how
it is possible to correct for photo-z errors.
Here we will quantify how the uncertainties in our
knowledge of the photo-z transitions $r_{ij}$, which we will denote by $\Delta_{ij}$, could affect our
predictions.  
From Eq.\ref{eq:barcij} we find that the error in $\bar{C}_{ij}$ is:
\beq
\Delta \bar{C}_{ij} = \sum_\alpha \left(
~R_{\alpha j}~\Delta_{i\alpha}
+L_{i\alpha}~ \Delta_{j\alpha} \right)
\label{eq:barcij2}
\eeq
where we have defined the matrices $R$ and $L$ as:
\beq
L_{i j} \equiv \sum_k r_{ik} C_{kj}   ~~~;~~~
R_{i j} \equiv \sum_k r_{jk} C_{ik} 
\eeq

For a spectroscopic sample in the foreground (background)
we have that $L \simeq C$ ($R \simeq C$).
We will assume that this uncertainty is the same 
 $\Delta_{r}=\Delta_{ij}$ for all pairs $ij$ and that they
are uncorrelated, so
\beq
<\Delta_{ij} \Delta_{kl}> =\delta_{ik}\delta_{jl} \Delta_r^2
\label{eq:sigmar}
\eeq
where  $\delta_{ik}$  and  $\delta_{jm}$ are the Kronecker deltas.
Note that in a more realistic situation errors could be
different for different transitions, i.e. $<\Delta_{ij}^2> \neq
<\Delta_{kl}^2>$, but here we just want to evaluate the overall
size of this effect. One can easily extend this analysis to
a more generic situation.
The covariance caused by this uncertainty is then:
\bea
&\calC_{ij;kl} \equiv  
Cov[\bar{C}_{ij};\bar{C}_{kl}] 
= < \Delta \bar{C}_{ij} \Delta \bar{C}_{kl} >= \Delta_r^2
\label{eq:covarr}
&\\
&
\times [
\delta_{ik} (R^TR)_{jl}+
\delta_{il} (LR)_{kj} +
\delta_{jk} (LR)_{il} +
\delta_{jl} (LL^T)_{ik}]
\nonumber
&
\eea
Note that $\bar{C}_{ij}$ here does not include shot-noise as we
are evaluating  the uncertainty in the $\bar{C}_{ij}$ prediction, for a
given value and error in $r_{ij}$. This covariance is then added
 to the covariance matrix in Eq.\ref{eq:barcT},
that includes the observational noise,
to explore how the FoM degrades as
a function of the uncertainties in the bin transition $\Delta_r$. This will give
us an idea of how well we need to know $r_{ij}$ 
for a given required FoM value.

When we have different populations or cross-correlation of different
observables (ie, galaxy with shear) we can easily generalize the above 
expressions, e.g.:
\beq
\Delta \bar{C}_{A_iB_j} = \sum_\alpha \left(
~R_{A_\alpha B_j}~\Delta^{A}_{i\alpha}
+L_{A_i B_\alpha}~ \Delta^{B}_{j\alpha} 
\right)
\label{eq:barcij3}
\eeq
where $\Delta^A$ is the uncertainty for type $A$. Types $A$ and $B$ 
can refer to density fluctuations (e.g.  $A=g_F$ or 
$A=g_B$ for Faint and Bright) and the same formalism can also
be applied to shear.

\section{Fiducial Galaxy Surveys}
\label{sec:fiducialsurveys}

We compare results for  combinations of two galaxy samples:
a deep ($i_{AB}<24$) sample, that we call Faint (F), and a shallower ($i_{AB}<22.5$)
sample, call Bright (B). We consider four cases: the 
two samples alone and two ways to combine them. In one of the combinations
(call F+B) we assume that the two samples are independent, ie
over different parts of the sky.
In the other combination (call FxB) we assume that they are over
the same region of sky: typically one can think 
the B sample as a spectroscopic follow-up of the F sample.


\subsection{Faint Sample (F)}

The F sample is a photometric survey with a broad band photo-z error:
\newline
\newline
\indent $22.5<i_{AB}<24$   ~~ and~~ $\sigma_z \simeq 0.05(1+z)$ \\ 
\newline
similar to the upcoming DES (Dark Energy Survey, www.darkenergysurvey.org).
We assume that we have very good photometry to be able to measure
magnification and also shear (galaxy shape) information. But we will show results
for magnification alone, shear alone and both combined. 

\begin{figure}
\includegraphics[width=3.5in]{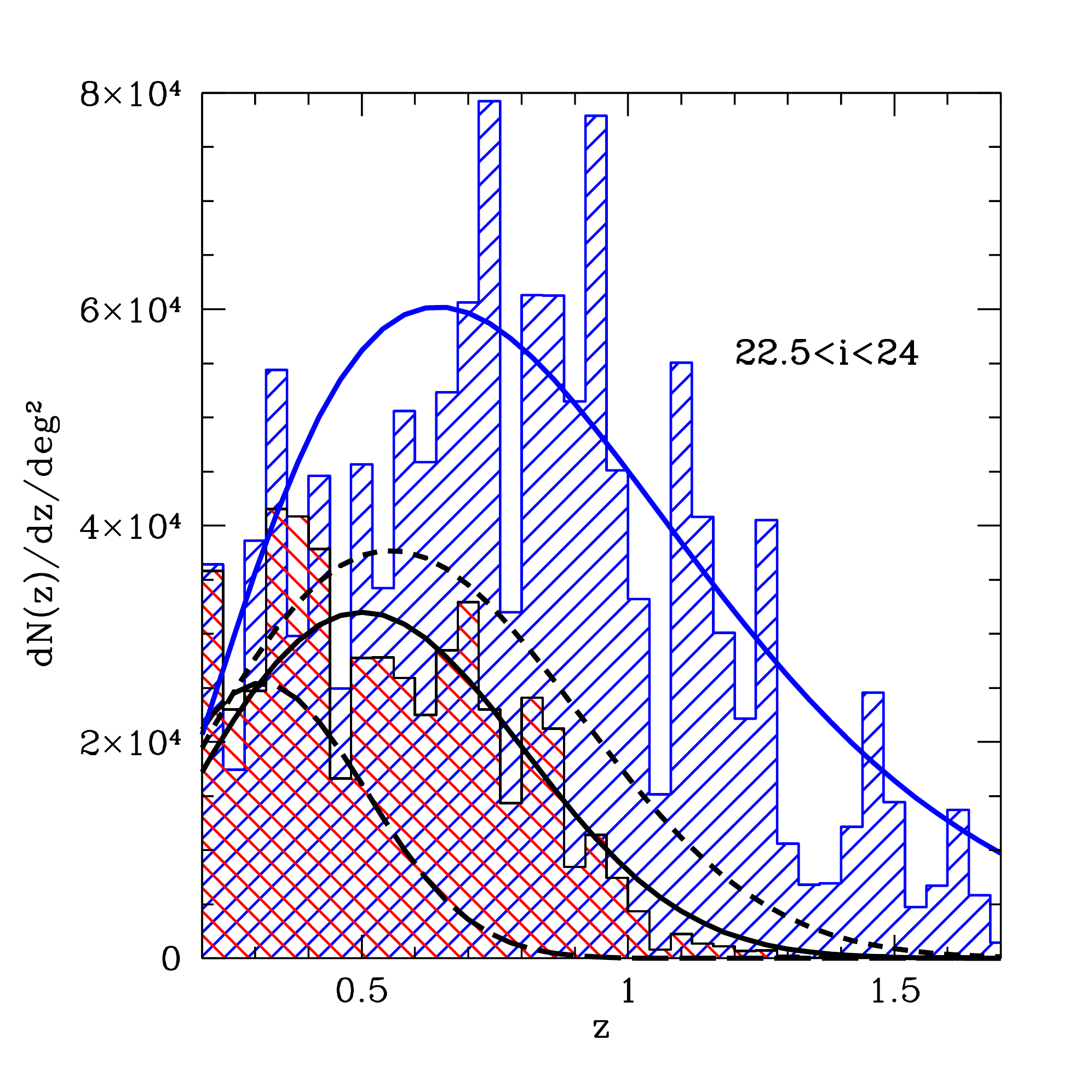}
\caption[]{Upper and lower continuous lines show the
number of galaxies per deg$^2$
in our complete fiducial F (faint) sample with $22.5<i_{AB}<24$ 
 and the B (bright) sample with  $i_{AB}<22.5$
 i.e. Eq.\ref{eq:nz24} and Eq.\ref{eq:nz225} multiplied
by 2 to make then complete. Histograms show the actual measurements
in the COSMOS photo-z survey which has $\bar{z}=0.90$
and $\bar{z}=0.57$ respectively for $z>0.2$.  
The long and the short dashed lines show the
corresponding fits to $i_{AB}<21.5$ and $i_{AB}<23.0$.}
\label{fig:nz}
\end{figure}

The cut at the bright end, $i_{AB}>22.5$ is not an essential
ingredient. It is included  to make this sample separate in magnitude from the B
sample, but all the results to follow are almost identical without
such a cut. This selection results in:
\beq
{dN\over{d\Omega dz}}= 
7.342 \times 10^4 (z/0.467)^{1.913} e^{-(z/0.467)^{1.274}}
\label{eq:nz24}
\eeq
over $0.1 \lesssim z \lesssim 2.0$, 
where $dN/d\Omega dz$ is in units of number of galaxies per ${\rm
  deg}^{2}$. This is a fit to an evolving  luminosity function
that matches Blanton et al. (2003), Dahlen et al. (2005, 2007) and Jouvel
et al. (2009).
The above redshift distribution includes
a completeness reduction of $\sim 50\%$. For a survey of 200 
${\rm deg}^2$ it yields $\sim 6\times 10^6$ galaxies. Note that we 
often refer to this sample briefly as $i_{AB}<24$, but we always use a lower
cut  $i_{AB}>22.5$, so that galaxies are clearly separated from the
Bright  sample. This avoids potential photo-z transitions between 
the two samples. The upper line in Fig.\ref{fig:nz} illustrates the
above model and compares it with the actual data in the public COSMOS
photo-z sample (Ilbert et al.,2010) which has about 100000 galaxies
over 2 deg$^2$ area 
with $22.5<i_{AB}<24$ and  mean photo-z redshift of $\bar{z}=0.9$
for $z>0.2$. Note that this area is quite small and subject to large
sampling fluctuations (see Fosalba et al. 2008).

For the fiducial bias in this sample we use four bias parameters
$b_i^F$ at $z_i=0.25, 0.43, 0.67$ and $1.0$ with linear interpolation to
other redshifts (see \S\ref{sec:bias}). The fiducial values for the
4 bias parameters follow:
\newline
\newline
\indent $b^F_i=1.2 + 0.4 (z_i-0.5)$,\\
\newline
so that the bias approaches unity at $z=0$ and has a linear
scaling, as shown in Fig.\ref{fig:rfitb}. We then allow the 4 bias
to vary without priors around these fiducial values.

Note that we only consider WL and angular galaxy clustering probes
for the Faint sample alone.
It is also possible to measure RSD here even with poor photo-z (e.g. see
Nock et al. 2010, Crocce, Cabr\'e \& Gazta\~naga 2010, 
Ross et al 2011, Crocce et al. 2011)
but we have checked that the additional FoM is 
small compared to the  one in spectroscopic surveys
and decided to restrict to the more traditional use of photometric
surveys.

\subsection{Bright Sample (B)}

The bright sample is a spectroscopic sample or very good photo-z
without lensing information and defined by a flux limit:\\
\newline
\indent $i_{AB}<22.5$ \\
\newline
which results in:\\
\beq
 {dN\over{d\Omega dz}}= 3.481 \times 10^4 (z/0.702)^{1.083}
e^{-(z/0.702)^{2.628}}
\label{eq:nz225}
\eeq
over $0.2 \lesssim z \lesssim 1.25$
also including a completeness of $50\%$, as in the faint sample.
For a survey of 200 
${\rm deg}^2$ it yields $\sim 2\times 10^6$ galaxies.
The lower solid line in Fig.\ref{fig:nz} illustrates the
above model and compares it with the actual
data in the public COSMOS
photo-z sample (Ilbert et al.,2010) which has about 43000 galaxies 
with $i_{AB}<22.5$ with a mean photo-z redshift of $\bar{z}=0.57$
for $z>0.2$.

The fiducial photo-z errors 
will be $\sigma_z \simeq 0.0035(1+z)$, based on the PAU
concept of using narrow band filters (see \S1). On linear
scales, larger than 20 Mpc/h, this is almost
equivalent to spectroscopic accuracy. In \S\ref{sec:photoz} we
will explore the dependence of the results with photo-z error.
Later on, in \S\ref{sec:survey}, we will also  consider a sample named ``Bspec''
with spectroscopic redshifts ($\sigma_z=0$) but with a lower
density of 1000  galaxy/deg$^2$. 
Dependence on density
 will be considered in section \S\ref{sec:shotnoise}.

The four fiducial bias parameters, $b_i^B$, for the B sample follow:
\newline
\newline
\indent $b^B_i=2 + 2 (z_i-0.5)$
\newline
\newline
Note that the bias at $z=0$ is $b=1$ in both F and B samples. This is reasonable
as bias only deviates significantly from unity
when the tracer is dominated by large masses or large luminosities,
and this only occurs at high redshifts.

For the bright sample we focus on RSD, which corresponds to the more traditional
use of a spectroscopic sample. BAO will also be considered in
\S\ref{sec:survey}.  Using a good parent photometric catalog it would
in principle be possible to do WL using shear or magnitudes over the
same spectroscopic objects. The gain in such case
comes from the combination of galaxy positions (with good radial
information)  and the weak lensing information in the parent sample.
This belongs to the combination FxB below and is not
considered  as part of the features of B sample alone. 
The separation will allow for a more clear understanding of
what we gain with the different combinations.

\subsection{Independent Samples (F+B)}

Faint plus Bright: a direct combination (i.e. addition of Fisher Matrix)
of the two cases above, assuming that both samples are independent
and sample different parts of the sky.  For a given area, the number of independent modes
(or total information) available in this case is twice the number of
modes for the individual samples. In this respect one should consider
that F+B has twice the area or cost as F or B. But given that we will be using
different probes in F and B we do not quite have twice the information.

\subsection{Cross-correlation (FxB)}

Faint cross  Bright:  samples are combined over the same
  area. In this case the Bright sample is a subset of the
original Faint sample, but here we consider disjoint magnitude
bins to have independent sets of galaxies.This will avoid
mixing of systematic effects, such as photo-z transitions.
The total area sampled is somehow 1/2 of the case F+B above, but we include 
the cross-correlations of the 2 samples for MAG (or WL) case.  
Here we use the MAG or WL and RSD probes in both the F and B
samples as we have both adequate redshift and photometric data . 
Traditionally one thinks of  WL probes to be applied to deep
photometric samples, such as sample F. As we will show below 
the combination of accurate redshifts and WL information in the B
sample alone produces figures of merit that can be comparable to the ones
in the F sample. This information is only included in the FxB case as 
this involves combining the redshifts information in the B  sample with the WL
information in the F sample over the same objects. This only happens
when we combine both samples and the B sample is a subset of the F
sample over the same area.
This, together with the cross-correlation, are
 key and unique features of this combination.

\subsection{Redshift bins}

As discussed in \S\ref{sec:s2n},
the redshift bin widths have little impact in the RSD analysis
or shear analysis, but they are important for the galaxy-galaxy
auto correlations, i.e. $C_{g_ig_i}$.
 We use  non-overlapping top-hats
with a width  4 times larger than the photo-z error in the Bright sample, i.e. $\Delta=0.014 (1+z)$.
This corresponds to about 60 Mpc/h at $z=0.6$ and is close to the 
smallest width where we can still neglect the intrinsic correlation
between adjacent bins.
For the Bright sample we will show that this width is large enough
(as compared to photo-z error) so that  we can almost neglect photo-z
errors. For the Faint sample,  which has
$\sigma_z =0.05(1+z)$,  adjacent bins are strongly correlated because
of photo-z errors. These correlations are taken into account in our analysis
using  photo-z error transitions  (see \S\ref{sec:trans}), 
so that our results are very similar to those obtained
for larger bins, as
expected. Thus the equivalent effective
width is much larger for the F sample than for the B sample, and is  
comparable to the photo-z error in the F sample. 
We will show in \S\ref{sec:bins}
how our conclusions depend on the redshift bin width.

\section{Results}

In this section we will show the forecast  on FoM produced by different probes
on a particular choice of surveys.
We will also explore how this forecast is affected by applying 
different assumptions on  bias evolution models,
photo-z errors, galaxy densities, survey areas and redshift
bin-width. The goal is to understand the impact of each of the
assumptions and to show how they affect the main results in this paper.
 We focus in MAG and RSD for most of this, but very similar results are
found using other WL probes, as will also be shown.
In \S\ref{sec:survey}  we explore different surveys, including
other WL probes (Shear-Shear, Galaxy-Shear) 
                as well as BAO in addition to MAG and RSD, and compare
                expected constraints for different surveys.

\subsection{Bias and WL}

First we explore the relation between galaxy bias, the clustering
from the galaxy-galaxy autocorrelation and the clustering from the different WL
probes (i.e. shear and magnification). 
In \S\ref{sec:MAGRSD}  RSD is added to this picture.

\subsubsection{Bias fixed}
Fixing the galaxy bias is the same as assuming perfect knowledge
of the bias.
It represents the ideal case for recovery of cosmological
parameters given the galaxy samples. This case is not totally 
uninteresting as there are separate ways to constrain bias (i.e. from map
reconstruction, halo modeling or other probes).
It might also be possible to do direct lensing
calibration of biasing (Bernstein \& Cai 2011). 
 
Results for 200 deg$^2$ with Planck
and SN-II priors are shown as the right value in the pair of numbers 
separated by a dash ($-$) in Table \ref{table:bfix}.
The first two rows compare galaxy-galaxy autocorrelations (GG) for the
Faint (F) and Bright (B) samples. 
The Bright sample gives about a
factor of 2 larger combined FoM than the Faint sample, despite having
2.5 less volume and fewer galaxies. This is because of the higher radial
resolution and illustrates a key point in our approach: better radial
accuracy of the foreground galaxies improve both  growth and
cosmic history reconstruction. 
Bellow we show  (\S\ref{sec:bins}) how this changes
with the number of bins (or the bin width).

The third entry is the results for
galaxy-shear (GS)  in the Bright sample.
Here the combined FoM$_{w\gamma}$ is a factor of
24 times lower than using the galaxy-galaxy autocorrelation over the same
(B) sample. This shows the problem with weak-lensing:
 that has a broad kernel and is intrinsically 2D. 
Having good radial resolution in the foreground galaxies of
the galaxy-shear cross-correlations does not help 
as much as in the galaxy-galaxy case (e.g. see \S\ref{sec:s2n}).
 
The next entry shows the combination
of galaxy-shear with galaxy-galaxy autocorrelation, which is still
dominated by the galaxy-galaxy results, but note how there is
an increase in the FoM$_w$ due to the geometric ratios
$p_{ij}$ which are only measured in galaxy-shear. The last entry
include all weak lensing and galaxy cross-correlations, including
magnification  (we will call this WL-all). Again here we find a significant
improvement, but results seem dominated by the galaxy-galaxy
auto-correlation. Note how there is very little improvement in the
FoM$_\gamma$ in the last two entries with respect to the galaxy-galaxy auto-correlation.
When bias is known, weak lensing does not add new information
about the growth over what we can measure with galaxy clustering alone.

\begin{table}
\begin{tabular}{|c|ccc|}
\hline 
& \multicolumn{3}{c|}{bias: {\bf free}-fixed}  \\
\hline
Probe/Sample &
FoM$_{w}$& FoM$_\gamma$ & FoM$_{w\gamma}$\\
\hline
GG  Faint (F)         & {\bf 24}-34 & {\bf 0.8}-28 &  {\bf 17}-930\\ 
GG  Bright (B)        & {\bf 27}-53 & {\bf 0.1}-31 &   {\bf 4}-1665 \\
GS (B)                 & {\bf 8}-11 & {\bf 0.04}-6.2 & {\bf0.3}-70 \\
GS+GG (B) & {\bf 27}-56 &  {\bf 4}-31 & {\bf 102}-1749 \\
WL-all  (B)                             & {\bf 30}-63 &  {\bf 6}-32 &{\bf 186}-1998 \\
\hline
\end{tabular}
\caption{
 Figures of Merit for DE equation of state
 $w_0- w_a$ (FoM$_w$),  growth $\gamma$ (FoM$_\gamma$)
and joint (i.e. FoM$_{w\gamma}$).  Each line shows  different 
combinations of samples and probes: galaxy-galaxy auto-correlation
(GG), galaxy-shear alone (GS) and their combination
(excluding or including galaxy-galaxy cross-correlations).
The first two entries  correspond to GG over the Faint (F) and
the Bright (B) samples. In each column we show two values separated
by a dash``-''. The first value (bias free) uses a fit for $b(z)$ using 4 free biasing
parameters. The second assumes that bias is known (fixed bias). 
For  200 deg$^2$ with Planck+ SN-II priors.}
\label{table:bfix}
\end{table}

\subsubsection{Bias free}

This picture changes dramatically when we allow bias to vary.
The bold value on the left (in the pair separated by a dash) in
Table \ref{table:bfix} shows how the above results change when
we measure the bias evolution $b(z)$ without priors. As justified in \S\ref{sec:bias} we
use 4 free bias parameters for each population (F or B) spaced by
$\Delta a\simeq 0.1$ (i.e. $\Delta t \simeq 1 Gyr$) and interpolate
linearly in redshift between these 4 points. 

In general, there is significant reduction in all FoMs. Notice how
FoM$_\gamma$ is more affected than FoM$_w$ because bias
is degenerate with growth in the power spectrum amplitude. 
In the case 
of Galaxy-Shear the FoM$_w$ is hardly
affected by the bias. Recall from Eq.\ref {eq:cijgk}
 that in this case the angular  power spectrum
between galaxies at redshift bin $i$ and shear at $j>i$ is
$C_{g_i\kappa_j} = b_i p_{ij} \calP_i$. The information about $w(z)$ is
encoded in $p_{ij}$  but is also degenerate
with the growth information which depends on $w(z) $ through $\Omega(z)$.
This produces a degeneracy in $w(z)$ for GS that is not broken by
knowning the bias.
These effects are less severe for GG because of the better
radial resolution which helps breaking this degeneracy even when bias
is free. Note how in this case FoM$_{w\gamma}$ is now better for the
Faint sample than for the Bright sample in contrast to the 
situation when bias is known. 

Note here how the WL-all combination (last entry) provides a significant
improvement in FoM$_\gamma$ with respect to the GG (B)
or GS+GG cases. In contrast, when bias
is known, there is no improvement. This improvement comes from
bias, which can be measured with the combination of 
galaxy-shear and galaxy-galaxy statistics, eg
$b_i \propto C_{g_ig_i}/C_{g_i\kappa_j}$. This is a key point to notice.
So far these results do not include the cross-correlations between the
Faint and Bright samples which better exploit the complementarity
of deep weak lensing measurements with good radial resolution
in the foreground lenses. They also do not include RSD which provides
another way to measure bias. We will explore this  in the following
section about magnification (MAG) 
and RSD and also later on (\S\ref{sec:survey})
when we combine all the different probes.

\begin{figure*}
\includegraphics[width=5.in]{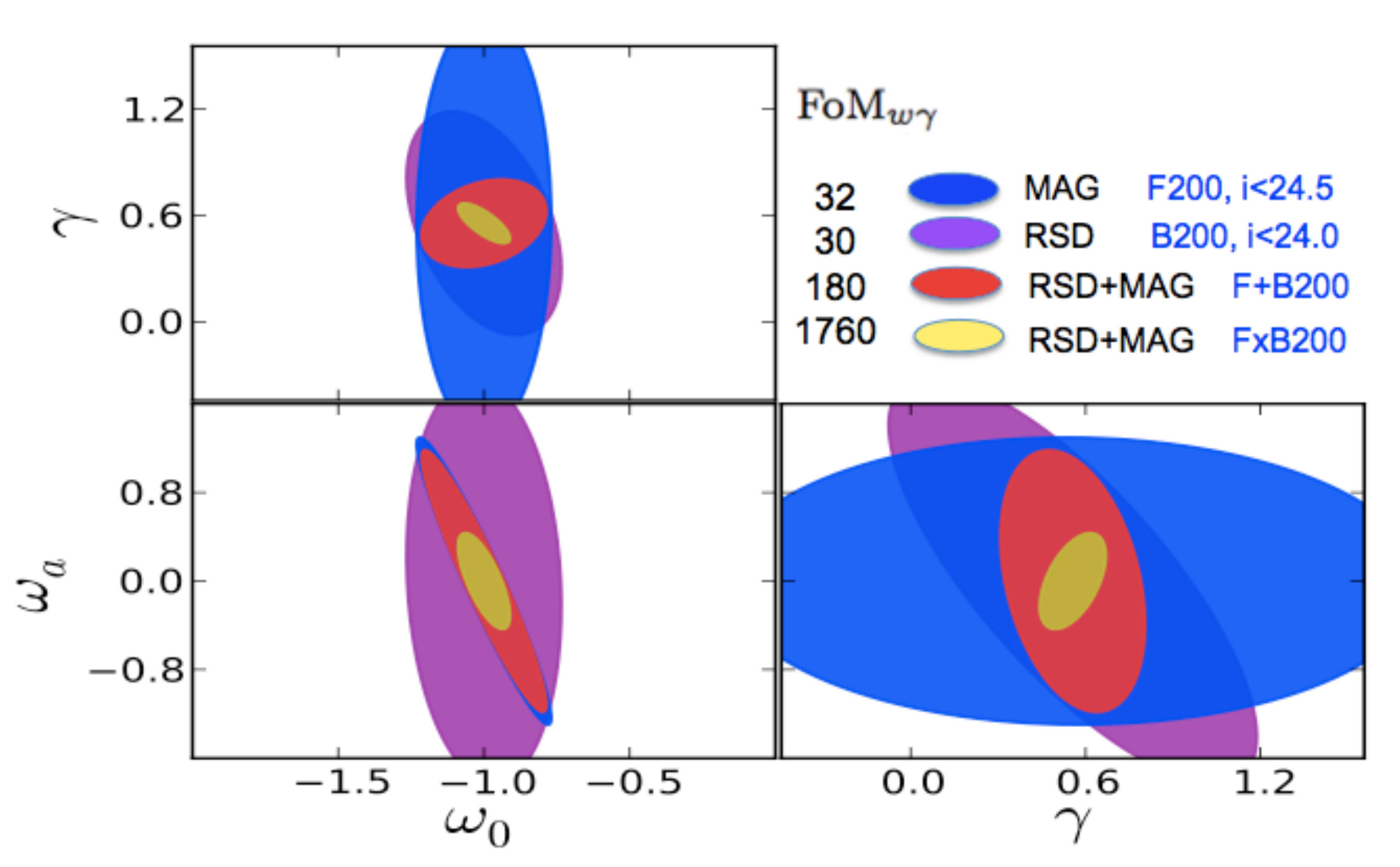}
\caption[]{Contours  ($\Delta \chi^2=2.3$) of $w_0-w_a-\gamma$
 in magnification (MAG)
over a Faint sample with $22.5<i_{AB}<24$ (labeled F200, blue) as compared with redshift
space distortions (RSD) over a bright
subset (labeled B200, purple) with $i_{AB}<22.5$ with very good
photo-z resolution: $\sigma_z=0.0035(1+z)$.  Bias evolution $b(z)$
and other 6 cosmological parameters are marginalized over.
 The red contours correspond to the combination of both
probes over different sky areas  (F+B), while the yellow contours
show the combination over the same area (FxB), including also
all cross-correlations between galaxies in both samples.}
\label{fig:degene2}
\end{figure*}

\subsection{Combining RSD and MAG}
\label{sec:MAGRSD}

\begin{table}
\begin{tabular}{|c|ccc|}
\hline 
& \multicolumn{3}{c|}{bias: {\bf free}-fixed}  \\
\hline
Probe/Sample &
FoM$_{w}$& FoM$_\gamma$ & FoM$_{w\gamma}$\\
\hline
MAG (F)              & {\bf 25}-40 & {\bf 1.3}-30 &  {\bf 32}-1198\\ 
RSD (B)               &  {\bf5}-29 & {\bf6}-16  &   {\bf 30}-463 \\
MAG+RSD  F+B   &  {\bf28}-138 & {\bf 7}-35 & \fbox{{\bf 180}-4859}\\
MAG+RSD  FxB  &  {\bf86}-197 & {\bf 21}-39 & \fbox{{\bf 1760}-7733} \\
\hline
\end{tabular}
\caption{Same as Table\ref{table:bfix} for magnification (MAG)
and Redshift Space Distortions (RSD).
The first two entries  correspond to MAG over the Faint (F) and
RSD over the Bright (B) samples. In the third row
(F+B) the MAG and RSD probes over the F and B samples are
combined assuming to be uncorrelated and to come from different
regions of the sky.  In the last line (FxB) the two samples are combined as
being over the same area of the sky, including its
 cross-correlation and covariance. 
}
\label{table:bfix2}
\end{table}


Table \ref{table:bfix2} shows a comparison of the different FoM
for MAG and RSD probes. As above, we will 
focus first on the results for a fixed bias.

\subsubsection{Bias fixed}

As shown in Table \ref{table:bfix2},  MAG with fixed bias
produces  better results than RSD.
The combined FoM$_{w\gamma}$ is a factor of 2.6 times better for the Faint
sample using MAG.

One of the main points we want to stress in this paper is shown 
when we combine MAG+RSD. We can see a large improvement both for
the separate samples (F+B) and for the cross-correlation analysis
(FxB). This is highlighted by a box in Table \ref{table:bfix2}. 
  Note how the improvement is important for all FoMs.

The F+B combination improves the FoM$_{w\gamma}$ and FoM$_w$ by
factors of 3-8 and 3-4, while the gain in FoM$\gamma$ is more
modest. These results will change dramatically when we allow the bias to be
free.

For the F+B combination with fixed bias,  having two separate samples
over different regions of the sky is worse than doing the
analysis over the same region, even when we are effectively doubling
the area in the first case. The difference between F+B and FxB  is almost
a factor of 2 for fixed bias. This is because FxB includes more
observables than F+B. As explained in Section 4.4, in the FxB case we
use the MAG and RSD probes in both the F and B samples and we also include the
cross-correlation of the F and B samples. For F+B we only consider
RSD in B,  MAG on F and no cross-correlations.

\subsubsection{Bias free}

The above picture changes when we allow the bias
to be free.  In all cases the FoM degrades significantly.
The degradation depends on the sample, probe and FoM considered,
as illustrated by bold numbers in Table \ref{table:bfix2}. 
In MAG  the FoM$_\gamma$ degrades more than FoM$_w$,
while the opposite happens in RSD. Note how fixing bias improves
FoM$_w$ for RSD (B). This is because of the improvement in 
$D(z)$ which allows $w(z)$to be measure through its dependence on
$\Omega(z)$. This is in contrast with GS in 
Table \ref{table:bfix} which shows little improvement of 
 FoM$_w$ when bias is fixed because of further degeneracies with
the lensing geometrical parameters in $p_{ij}$. The combination
of MAG and RSD can break these additional degeneracies.

When we correlate samples over the same 
region of the sky (FxB) we find much better results than 
when the samples are independent  (F+B). The improvement is
about a factor of 3 for FoM$_w$ or FoM$_\gamma$ and 10 for
FoM$_{w\gamma}$.  
Bias is better constrained when both samples are over
the same region as  we can include the cross-correlations between
then and use both MAG and RSD probes in both samples. 
 So even when there is less information in terms of number
of independent modes or total area in FxB  than in F+B, there is more constraining
power in the FoM of cosmological parameters. 
This is illustrated in Fig.\ref{fig:degene2}, which shows the 3
projections ($w_0, w_a. \gamma$) in the FoM$_{w\gamma}$
ellipsoid. Note how in the $w_0-w_a$ plane there is little gain
in the MAG+RSD (F+B) combination as the improvement is 
in $\gamma$. In FxB the degeneracies are strongly reduced thanks
to the bias measurements.

For the F+B comparison there is also a significant
improvement with respect to the separate F and B cases. In this
case the relative improvement is smaller when bias is free (as compared
to the case of known bias).  It comes from having different biases 
for the F and B samples, but no cross measurement to relate them, 
as in the FxB case where we use MAG cross-correlation to relate the bias in F and B.

Allowing the bias to be free reduces 
the combined FoM$_{w\gamma}$ by factors of 37 and 15 in
MAG and RSD. But this reduction is only a factor of 4.4 for the
FxB combination. 
These results assume no priors on bias. We will check
in next sub-section how this changes when we have some priors on bias.

\subsubsection{Bias stochasticity}
\label{sec:sto}

In section \S2.4 and  Fig,\ref{fig:rfitb}
we have argued for $r$ being very close
to unity in numerical simulation. Therefore through the paper,
the bias is assumed deterministic (r=1), but here we in
addition test how a stochastic bias would impact our result.
A stochastic bias is introduced by replacing Eq.\ref{eq:cijgg} by

\bea
C_{gn_ign_i}(\ell) &\simeq& {b_{n_i}^2 \over{\Delta_i}} \calP_i
\\ \nonumber
C_{gn_igm_j}(\ell) &\simeq& 
\alpha_{m_j} r_{n_i} b_{n_i} p_{ij} \calP_i~~~~j>i
\label{eq:cijgg2}
\eea
where $r_{n_i}$ is the galaxy-matter cross-correlation coeﬃcient
for the $n_i$ galaxy population. The variation of $r_{n_i}$ with redhift
is parametriced in the same manner as the bias  $b_{n_i}$, using 4
parameters which are scale independent on linear
scales. We use $r_{n_i}=1$ as the fiducial values and
 allow $r_{n_i}$ to vary without priors.

When marginalizing over these new parameters,  FoM$_{w\gamma}$
changes by less than 4\% for the Faint sample.. The impact on the
Bright sample is even smaller and under 1\%. This small effect
can be understood by comparison to the GG result for the F
sample in top right corner of Table  \ref{table:bfix}:
most of the FoM in MAG comes from the GG contribution (the
first  of the above equations), which is independent of  $r_{n_i}$.
Such a small dependence on $r_{n_i}$  will have  little impact on the 
measurement of bias or on the FxB cases.

\begin{figure}
\includegraphics[width=2.8in]{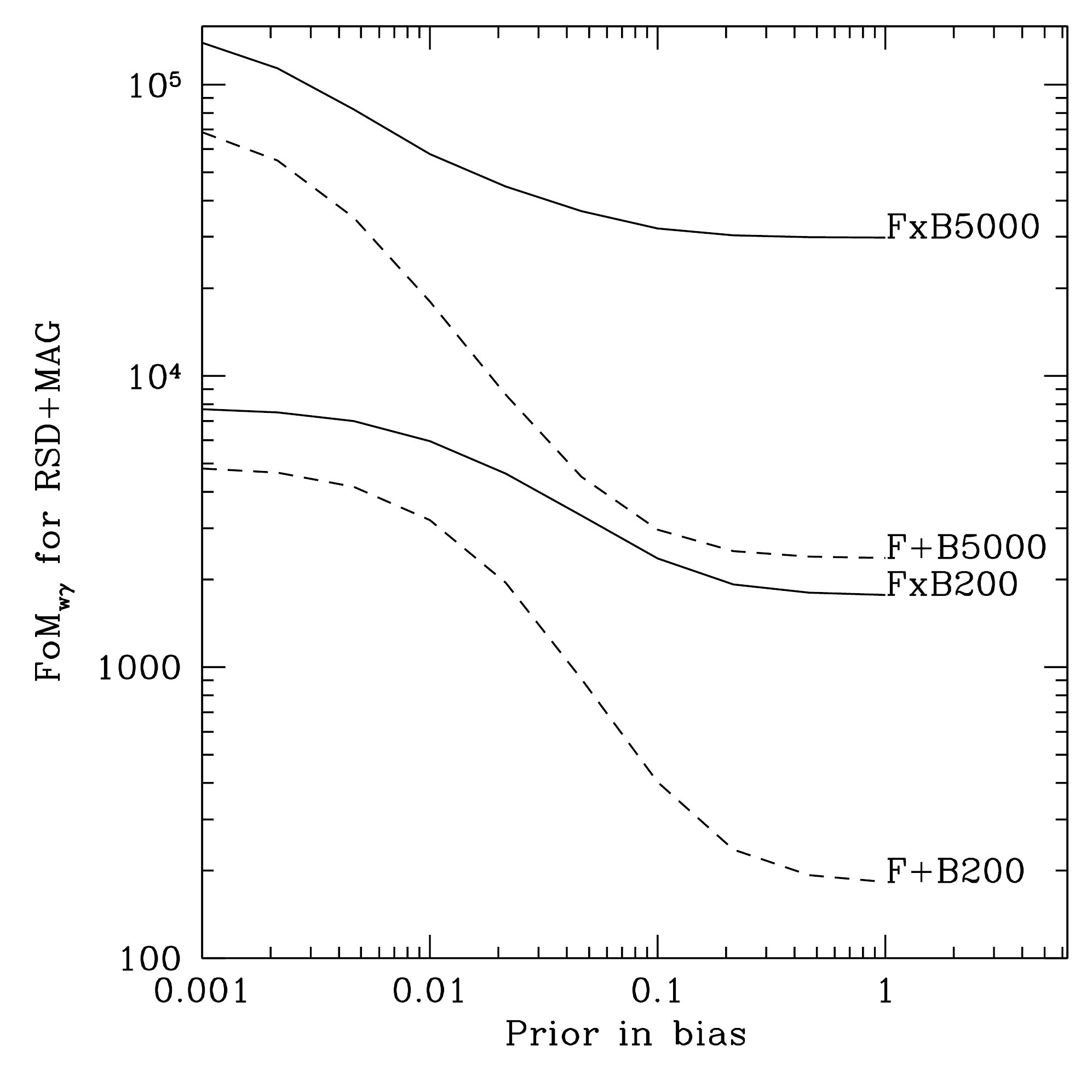}
\caption[]{FoM$_{w\gamma}$ in the MAG+RSD 
combination as a function
of priors in the bias parameters (absolute rms error). 
Dashed lines correspond
to the FoM when we combine samples from different 
regions of the sky (F+B), while continuous lines correspond
to the cross-correlation of samples over the same area (FxB).
Upper/lower lines corresponds to 5000/200 deg$^2$ surveys.} 
\label{fig:fomplotbias}
\end{figure}

\subsection{Priors on bias}

Figure \ref{fig:fomplotbias} shows how the FoM$_{w\gamma}$ increases
as we reduce the priors in the bias parameters. These priors
can come from other observations or from modeling
of bias (e.g. see \S\ref{sec:bias}). The same 
gaussian prior is used for all the bias parameters. 
F+B adds MAG in the F sample
with RSD in the B sample, while FxB include 
all combinations and cross-correlations.
The figure shows that 
the FxB combination is less affected by priors than the other combinations
and reflects the fact that  bias is better measured  by the FxB combination. 
A prior of 0.01 in bias improves the FoM of
FxB and F+B by about a factor of 2.4 and 3.8 respectively 
for the 5000 deg$^2$ sample. With such priors, the FxB combination
still has  FoM$_{w\gamma}$ that is 3.2 times larger than F+B. This is
because in for the FxB combination we use both RSD and WL over
the bright and faint sample, while in F+B we only use RSD over B
and WL over F.

Also note how for no priors the FoM in the small sample combination FxB200 is almost as
good as the F+B5000 combination which has 25 times larger area.

If we use 2 biasing parameters per population instead of 4 to describe
$b(z)$ (i.e. see \S\ref{sec:bias}) we find an increase of about $40\%$ in the FxB combined
FoM$_{w\gamma}$ for both 200 or 5000 deg$^2$ samples.

We conclude  that the impact
of bias evolution is well understood and our results are quite
robust with respect bias modeling.

\begin{figure}
\includegraphics[width=2.8in]{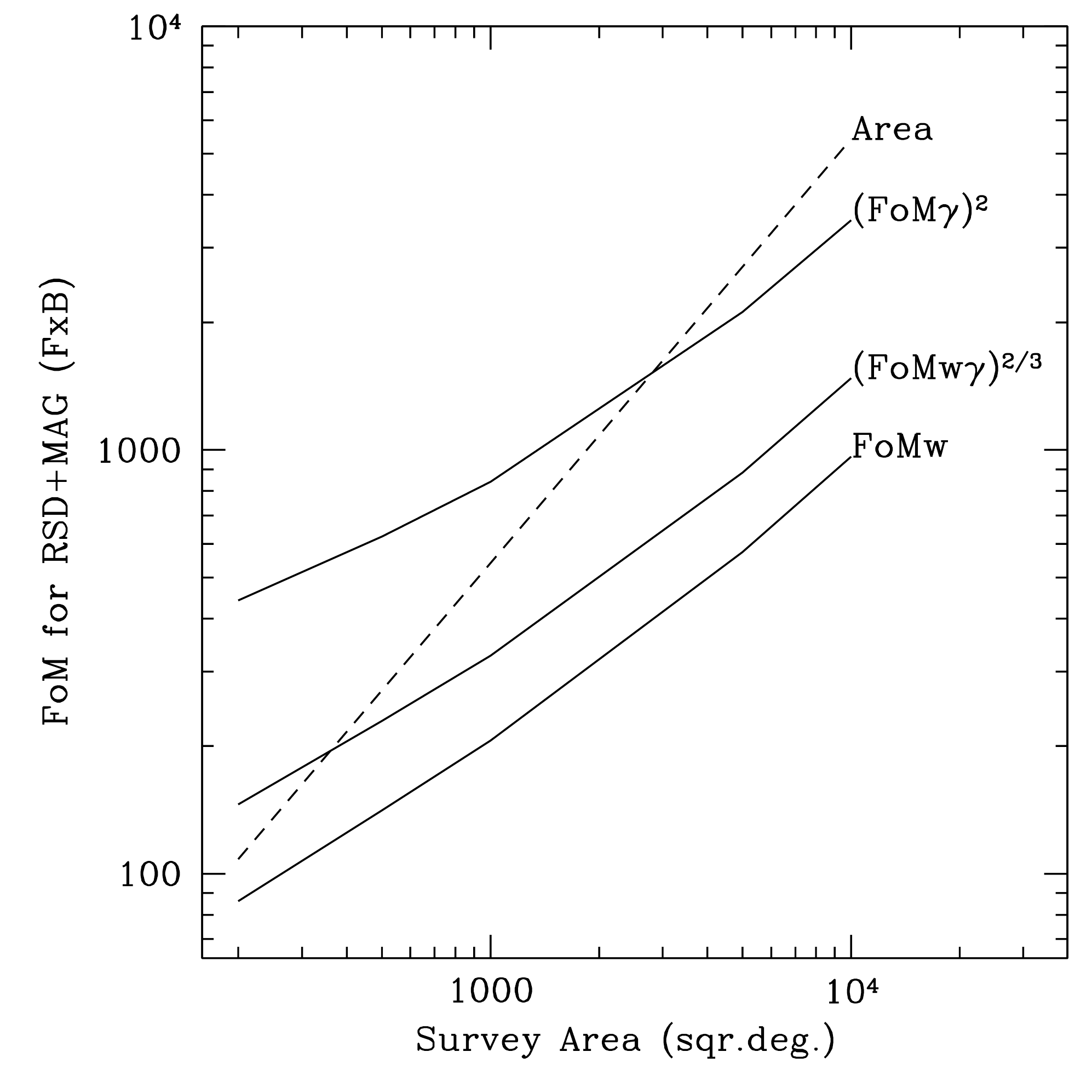}
\caption[]{Variation of the different FoM:
 FoM$_\gamma^2$ (top line),  FoM$_{w\gamma}^{2/3}$   and  FoM$_{w}$ 
(bottom line)
as a function of the Survey Area. This is RSD+MAG combined over the same 
area (FxB). The dashed line shows the scaling with area for reference.}
\label{fig:fomplotA}
\end{figure}

\subsection{Survey Area}
\label{sec:area}

Figure \ref{fig:fomplotA}
shows how the  MAG+RSD combination over the same area (FxB)
changes as a function of the
  survey area.
For the Fisher Matrix
 without priors we expect the following scaling: FoM$_{w}\propto A$,
FoM$_{\gamma}\propto A^{1/2}$ and FoM$_{w\gamma}\propto A^{3/2}$,
according to Eq.\ref{eq:FoM}, as it depends
on the number of parameters in each FoM definition.
The values shown in Fig.\ref{fig:fomplotA} are rescaled accordingly and
show the FoM corresponding to the geometrical mean of the 
parameters involved. After this rescaling, they should all scale
linearly with the area. But we are
using Planck and SN-II priors which tend to dominate the results for smaller area,
resulting in a weaker dependence in $A$. A fit to Fig.\ref{fig:fomplotA}
yields:
\bea
FoM_{w}\propto  A^{0.62} \\
FoM_{\gamma}\propto A^{0.26}\\
 FoM_{w\gamma}\propto A^{0.89}
\label{eq:area}
\eea

Note that FoM$_{\gamma}^2$ is
larger than FoM$_{w}$, which indicates that $\gamma$
is better determined than the geometrical mean of $w_0$ and
$w_a$. This is  despite the fact that we include Planck and SN-II
priors, which include priors in $w_0$ and $w_a$, but no priors
in $\gamma$.


\subsection{Bright Magnitude limit}

\begin{figure}
\includegraphics[width=2.8in]{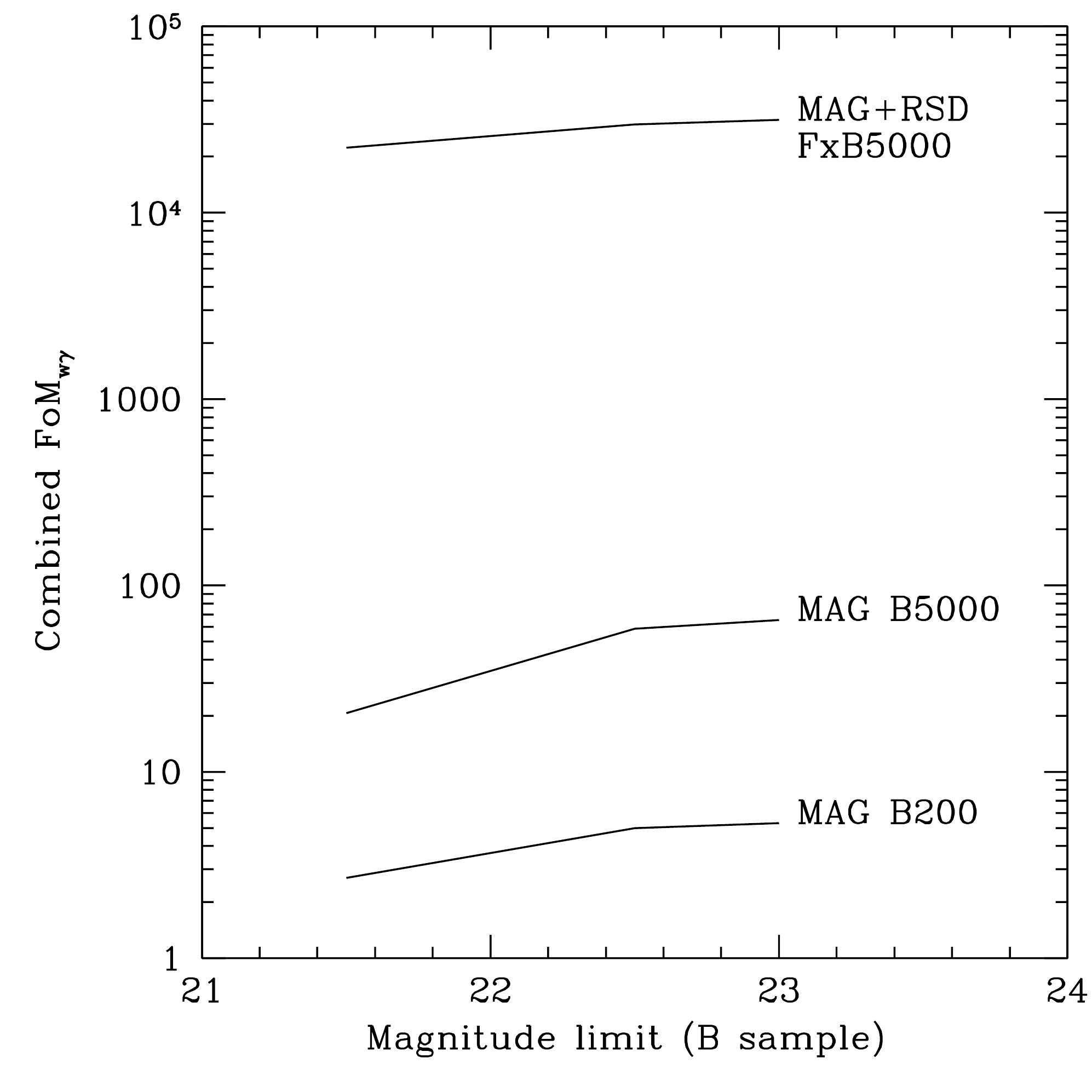}
\caption[]{
Lines show the combined FoM$_{w\gamma}$ as a function
of the magnitude limit of the bright sample for different
samples and probes, as labeled. The top line corresponds to the
FxB cross-correlation using both MAG and RSD. The lower lines 
correspond to MAG alone over the Bright sample with 5000 deg$^2$
or 200 deg$^2$.}
\label{fig:fomplotm}
\end{figure}

Figure \ref{fig:fomplotm} shows how the FoM changes as
we change the magnitude limit ($m_l$) of the
bright sample.  For $N(z)$ we use the distributions
shown in Fig.\ref{fig:nz}.
The combined FoM$_{w\gamma}$ for MAG in the Bright sample
depends strongly on $m_l$, as shown by the bottom lines
 in the figure, for both 200 and 5000 deg$^2$. When combined
with the Faint sample and RSD, the dependence is weaker.
For the latter case we find:
\beq
FoM_{w\gamma} \propto 1.26^{m_l}
\eeq

\subsection{Shot-Noise}
\label{sec:shotnoise}

\begin{figure}
\includegraphics[width=2.8in]{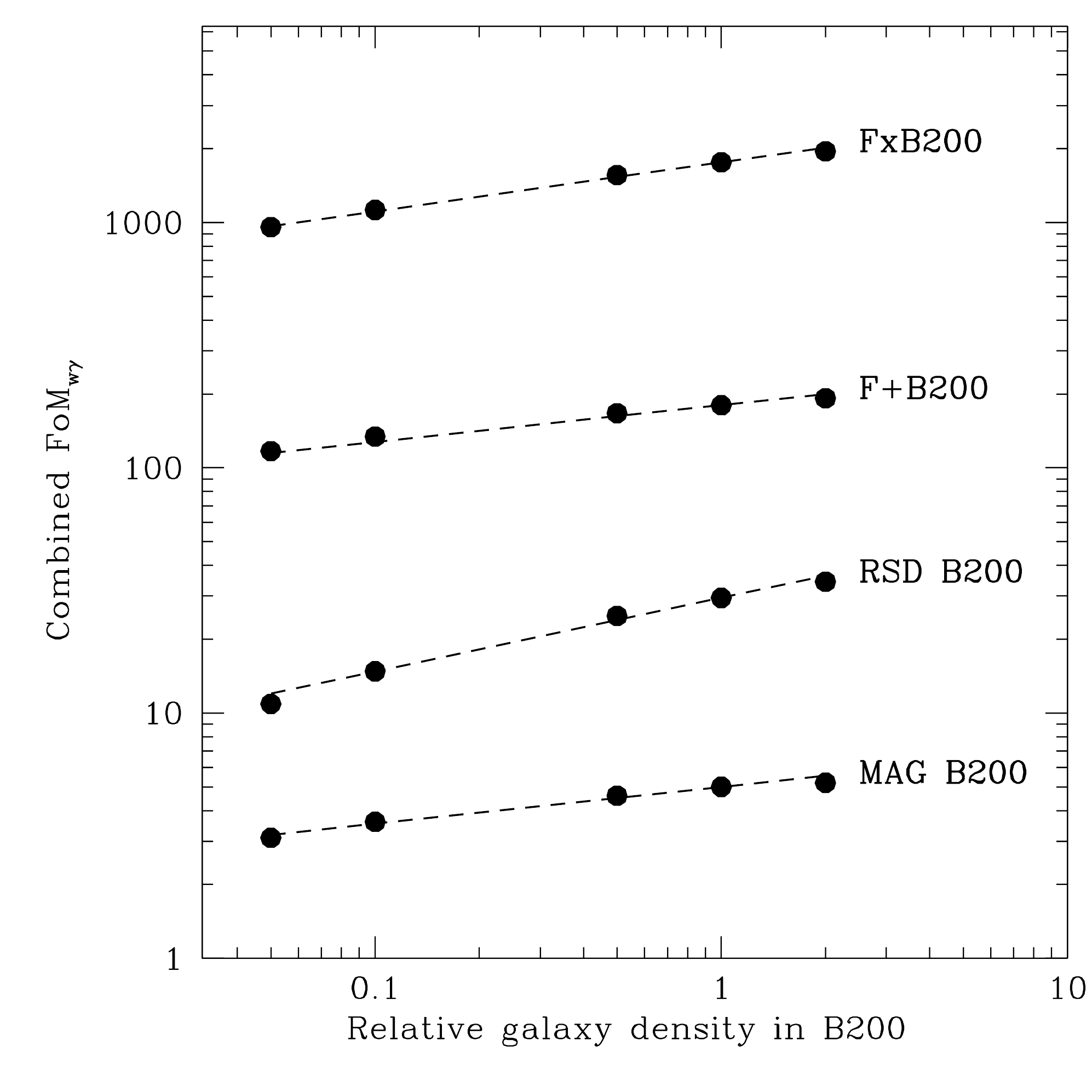}
\caption[]{FoM$_{w\gamma}$ as a function
of the relative galaxy density $\bar{n}$ in B200 with respect to the
fiducial values (in \S\ref{sec:fiducialsurveys}) for different 
probes, as labeled.
As usual, F+B adds MAG in the F200 sample
with RSD in the B200 sample, while FxB include 
all combinations and cross-correlations. The dashed
lines show a fit to $\bar{n}^\beta$.
}
\label{fig:fomplotD}
\end{figure}

What happens if we reduce or increase the galaxy density? 
The bright sample ($m_l<22.5$), for which we so far have 
assumed a PAU-like survey measuring
low resolution spectra for most of the galaxies, likely  has a lower
density for a truly spectroscopic survey. A different density can also come
from only considering a subset of the data or dividing the galaxies
in types.

Figure \ref{fig:fomplotD}
shows how the FoM from the different probes changes
as we reduce (or increase) the density in the bright sample
with 200 deg$^2$ (B200). The dilution is given in terms of
the default density in \S\ref{sec:fiducialsurveys}, which 
already corresponded to a 50\% completeness.
Both RSD and MAG are quite affected by sample dilution, especially when we
approach values lower than 10\% completeness. The data can be fitted
with a power-law in the relative density $\bar{n}^\beta$, with $\beta\simeq 0.15$
for MAG and $\beta\simeq 0.3$ for RSD (dashed lines in the figure).
It is therefore important to keep a high density if we want a high
FoM, specially for RSD.

The effect is less severe in the F+B200 combination ($\beta \simeq 0.15$)
because the density of the faint sample (F200) is kept
constant. For the FxB200 combination the effect is in between the one
for B200 and the one for F+B200. The effect of dilution in the FxB200
case can be  well described as:
\beq
FoM_{w\gamma}^{FxB} \propto \bar{n}^{0.2}
\eeq


\subsection{Redshift bin width}
\label{sec:bins}

\begin{figure}
\includegraphics[width=3in]{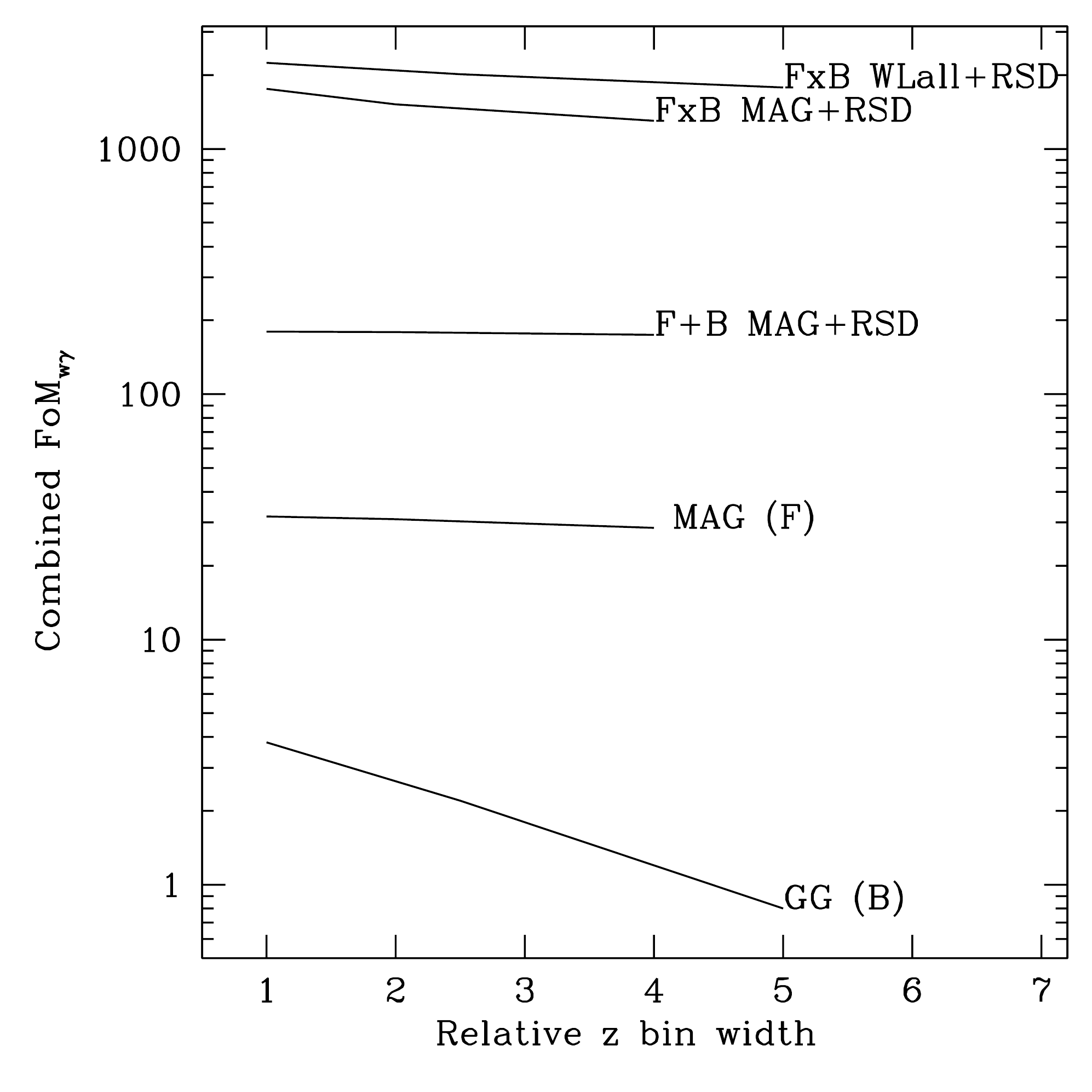}
\caption[]{Combined FoM$_{w\gamma}$ as a function
of the relative redshift bin width  (with respect to the
fiducial value of 0.014(1+z)) for the different
samples and probes. As usual, F+B adds MAG in the F sample
with RSD in the B sample, while FxB include all combinations and cross-correlations.}
\label{fig:fomplotb}
\end{figure}

Figure \ref{fig:fomplotb} shows how the 
 FoM$_{w\gamma}$ changes when we change the 
redshift bin width
from 4 to 20 times the photo-z error in the Bright
sample, i.e. 1 to 5 times the fiducial values.

Changing the bin width has little effect on the Faint sample for MAG. The
fiducial z-bin, $0.014(1+z)$, is already smaller
than the photo-z error
in the Faint sample,  $0.05(1+z)$, indicating that nearby bins
are strongly correlated.  Increasing the bin width by a factor of 4
only changes FoM$_{w\gamma}$ by 12\%.
This illustrates that the method to account for transition
probabilities in \S\ref{sec:trans} and their covariance in
\S\ref{sec:covar}, properly take into 
account the impact of photo-z errors in the MAG forecast.
We do expect some small degradation as we
increase the redshift bin width because of the lost of radial information.
This degradation is more significant for a bin width which is
larger than the photo-z error.


For the Bright sample different redshift bins
are independent as they are at least 4 times larger than
photo-z error. The impact of changing the bin widths 
is quite different for RSD and MAG. 
For RSD the results are independent of the bin
size,  once there are enough  bins. This is because we are
using 3D information within each bin to measure the distortions:
this is possible because we are in the linear regime and we assume
photo-z accuracy that is good enough for this task, see
\S\ref{sec:z-dist}.
The redshift bin slicing only affects  the radial resolution in the parameter
estimation, but it does not add new modes as happens when we do 2D clustering.
 So as long as there are enough bins to capture the slow cosmic evolution 
the results are unchanged in 3D.  The F+B case suffers little degradation
because it is just the sum of F with MAG and B with RSD, neither of
which changes with number of bins.

For MAG in the Bright sample the effect is more important
because the result is dominated by the 2D galaxy-galaxy (GG) correlations
(see Table \ref{table:bfix}), whose signal-to-noise depends linearly in the number
of redshift bins (see \S\ref{sec:s2n}).
If we reduce the number of independent bins,
we directly reduce the FoM. In the case of MAG for the Bright sample
a bin width that is twice as large reduces the FoM$_{w\gamma}$ by a
factor of two. This agrees with the reduction of the 2D
GG FoM for the B sample shown in Fig. \ref{fig:fomplotb}.
The degradation is smaller
in the GS because its signal-to-noise is independent
of the number of bins  (see \S\ref{sec:s2n}).

The influence of the 2D galaxy-galaxy contribution is also apparent in the top lines
of Fig. \ref{fig:fomplotb}, which shows the FxB combination for
MAG or for WL-all (which includes all galaxy and Shear correlations).
As we decrease the redshift bin width by a
factor of 4, the combine FoM of FxB in MAG+RSD increases by 75\%. 

Overall, this illustrates well one of the main points we
want to stress in this paper: that the benefit of the FxB combination
comes from the combination of many independent redshift bins.

\begin{figure}
\includegraphics[width=2.8in]{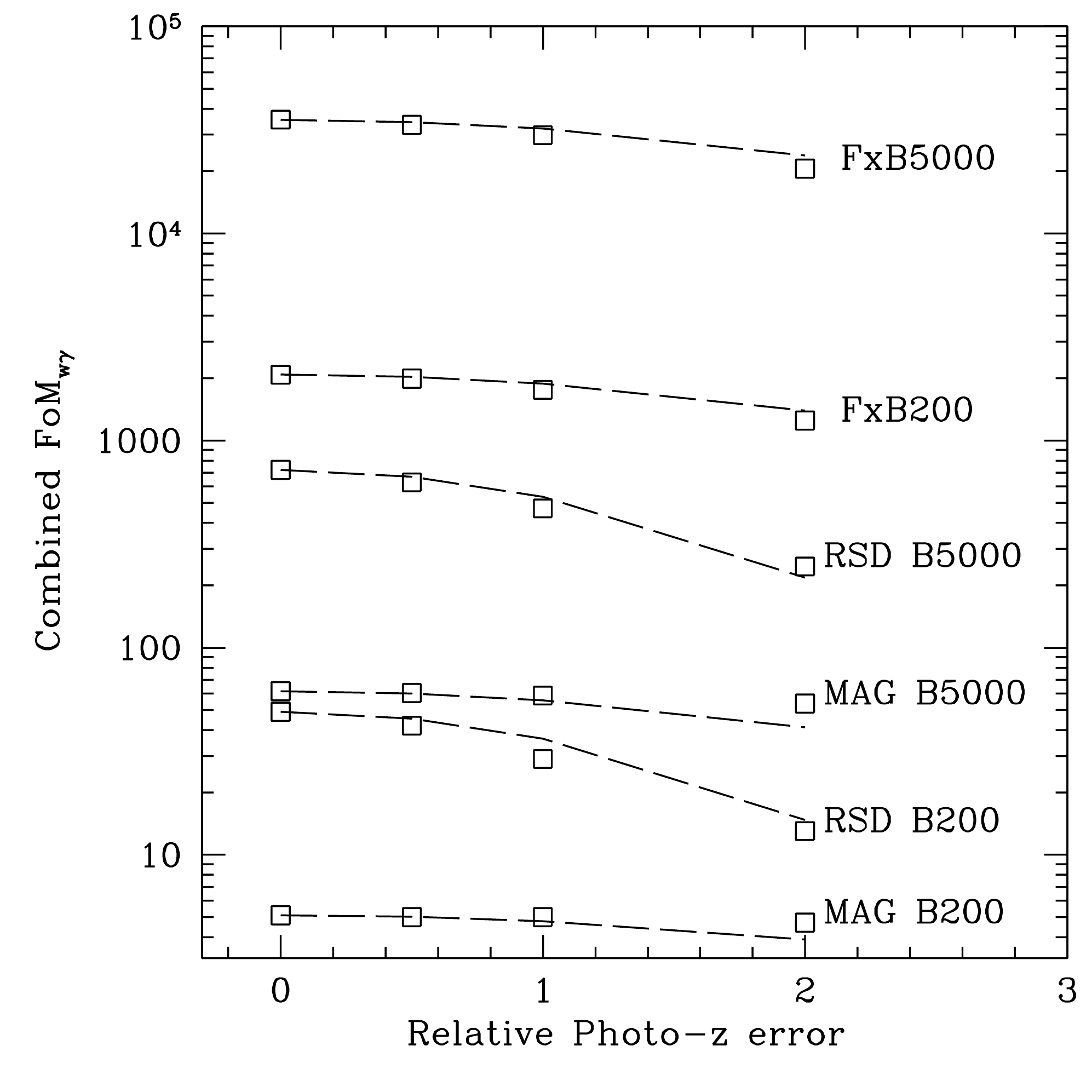}
\caption[]{Symbols show the combined FoM$_{w\gamma}$ as a function
of the relative photo-z of the bright sample (with respect to the
fiducial values in $\sigma_z=0.0035(1+z)$)  for the different
samples and probes, as labeled. Dashed lines show the best
 fit to $e^{-A\sigma_z^2}$.}
\label{fig:fomplotz}
\end{figure}

\subsection{Photo-z errors}
\label{sec:photoz}

\begin{figure*}
\includegraphics[width=3in]{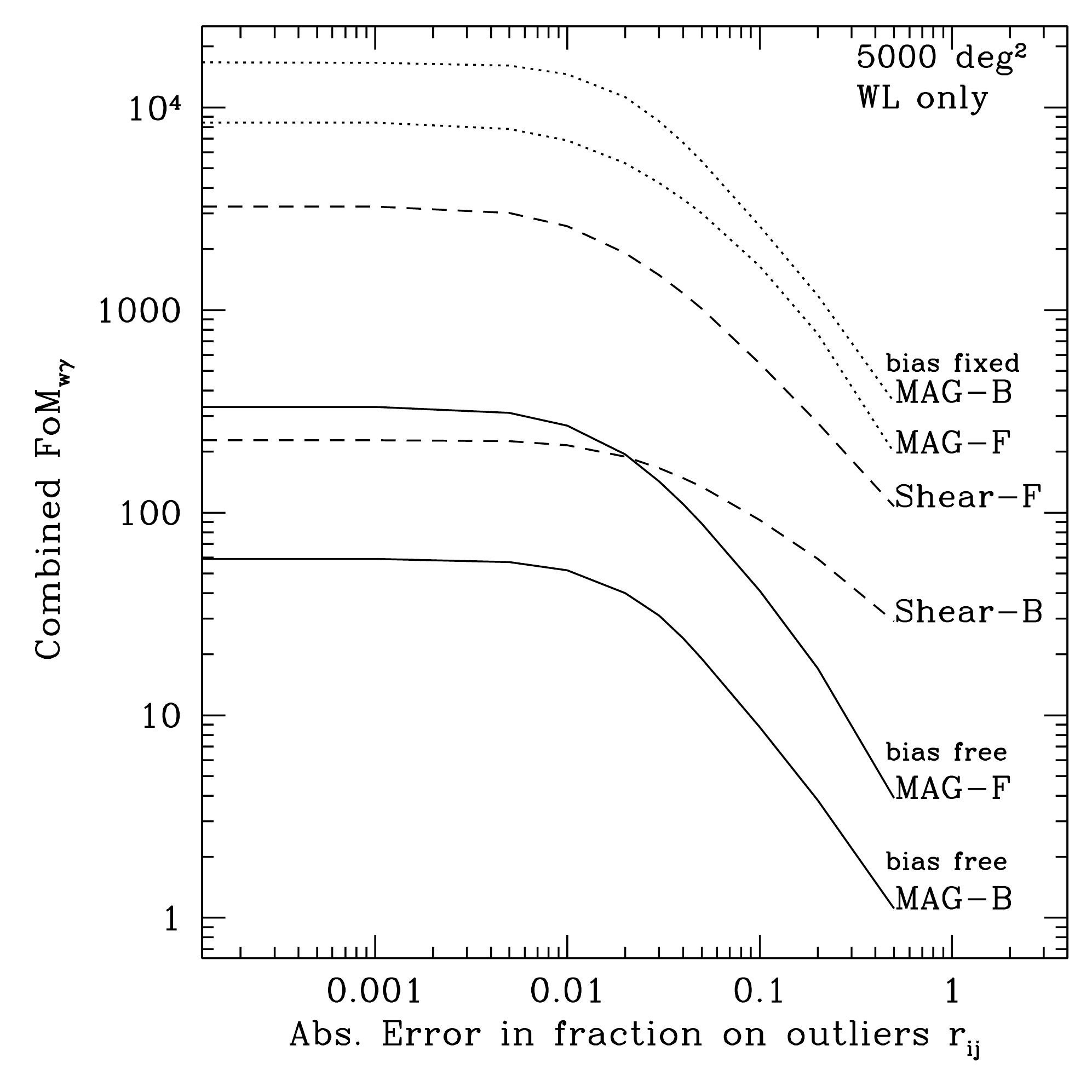}
\includegraphics[width=3in]{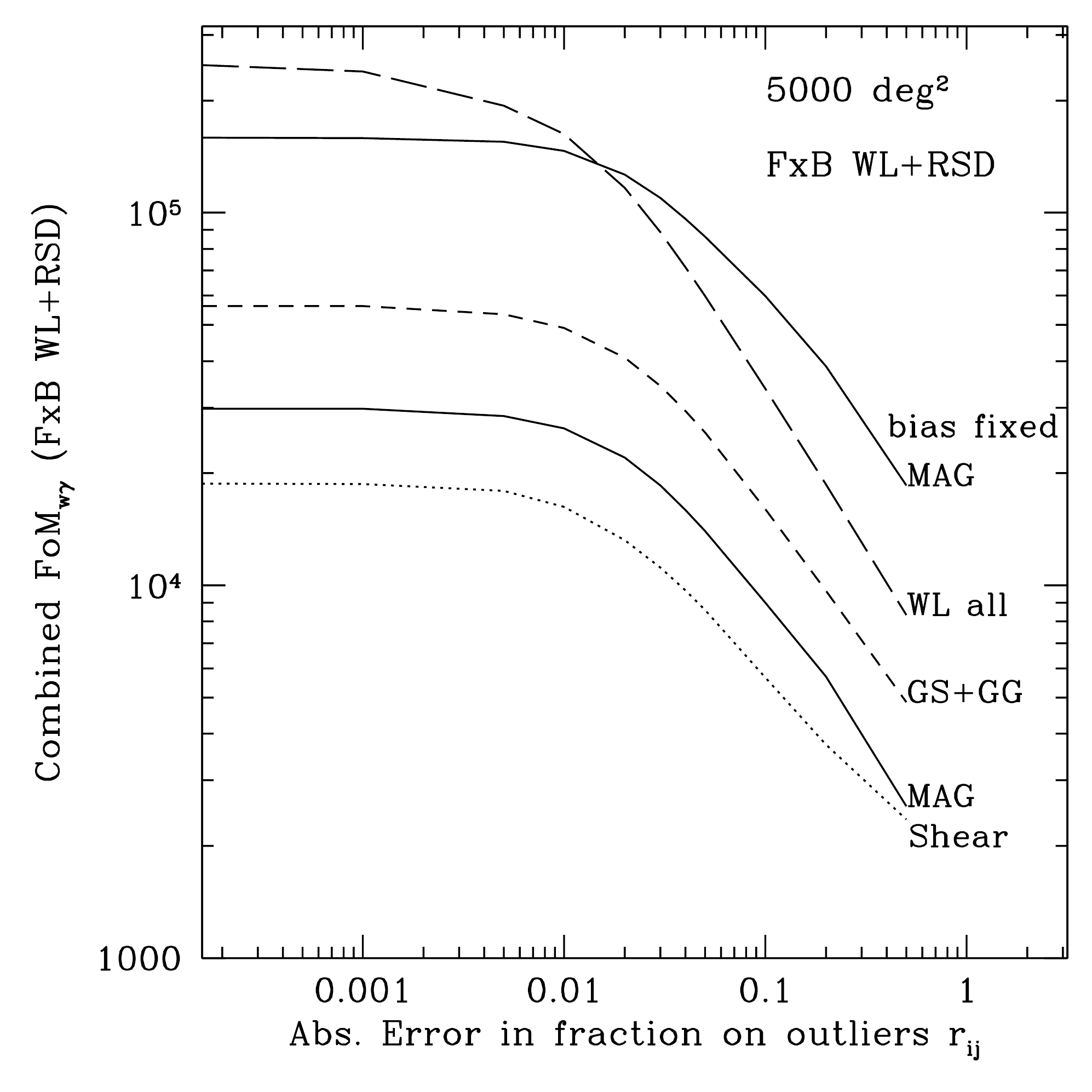}
\caption[]{Combined FoM$_{w\gamma}$ as a function
of the absolute error in our knowledge of the photo-z
transitions for  5000 deg$^2$ survey. {\sc Left}:
Results correspond to
cosmic magnification (MAG with free bias, continuous lines, or fixed
bias, dotted line) and  Shear-Shear cross-correlation
(Shear, dashed lines)  alone, i.e. without RSD, for both the Faint (F) and
Bright (B) samples. {\sc Right}:
  Results correspond to the FxB cross-correlation of the F and B sample,
including RSD with different WL probes: magnification
(MAG with free and fixed bias, top and bottom continuous line), Galaxy-Shear cross-correlations (GS,
short-dashed),  Shear-Shear cross-correlation (Shear, dotted line)
and all WL combined (long dashed lines).}
\label{fig:fomplotr}
\end{figure*}



Figure \ref{fig:fomplotz}
shows how the combined FoM$_{w\gamma}$
changes as we change the relative photo-z errors, $\sigma_z$, in the bright sample.
Here the photo-z error distribution is Gaussian.

The variation on FoM$_{w\gamma}$ is smaller for MAG than for RSD.
This makes sense as the bright MAG analysis uses relatively broad
redshift bins: $\Delta_z=0.014(1+z)$, i.e. 4 times the nominal photo-z error of
$\sigma_z=0.0035(1+z)$. This makes our result conservative but
quite insensitive to  photo-z errors. In the case of RSD the
degradation comes from the S/N reduction on modes affected 
by the photo-z error.

For the FxB results the effect is less severe as here we also include
results for the faint sample for which the photo-z error is fixed.
These tendencies can be fitted to an exponential decrease:
$e^{-A\sigma_z^2}$.
For both the 200 and 5000 deg$^2$ we find that to a good 
approximation:

\bea
FoM_{w\gamma}^{FxB} \propto e^{-0.1[\sigma_z/0.0035(1+z)]^2} \\
FoM_{w\gamma}^{RSD} \propto e^{-0.3[\sigma_z/0.0035(1+z)]^2}
\eea
These fits are shown as dashed lines in the figure. MAG in the B200 sample
falls a bit less steeply ($A\simeq 0.07$) than MAG in the B5000 ($A
\simeq 0.1$).

Note in Figure \ref{fig:fomplotz} 
how, despite having 25 less area,
the combined FxB200 result is better that the
separate MAG or RSD results over B5000. This is also true for 
results over the F5000 sample.

\subsubsection{Uncertainties in photo-z transitions} 

So far we have considered the case where we know perfectly well how
photo-z uncertainties change the observables. This corresponds
to the case $\Delta_r=0$, where $\Delta_r$ is our uncertainty in the 
fraction of galaxies $r_{ij}$ that are wrongly allocated to a given 
redshift bin, as defined by  Eq.\ref{eq:sigmar}.
The left panel in Fig.\ref{fig:fomplotr} shows how the
FoM for Magnification (MAG) and Shear-Shear cross-correlations (Shear)
degrades as we increase  the absolute error $\Delta_r$.
In the left panel we only include WL probes and no
RSD.

For MAG, the degradation in the FoM is similar  for the Bright and
Faint samples. In the limit of small errors in $r_{ij}$
the FoM of MAG is a 5.6 times larger for the Faint than for the
Bright sample (bias free), because the former is much deeper, has
more redshift range, volume and galaxies. Thus,
the better radial resolution in the Bright sample
does not quite compensate for its smaller volume when
we also have to fit for bias. But when
bias is known (dotted lines) the Bright sample produce 2 times
better results, which indicates that the larger volume in F is 
fully compensated with better radial resolution in B.. 
As we increase the uncertainties in the photo-z transitions the FoM 
degrades quickly, as we loose the capacity to
locate radial bins.

Recall that the Faint sample has a much larger photo-z error 
than the Bright sample and therefore the fraction of galaxies 
$r_{ij}$ will also be much larger. This means that for a fixed absolute error
in $r_{ij}$ the relative error  will be much smaller for the Faint
sample. Does it make sense then that the FoM is degraded 
equally for the Fain and Bright sample, as shown in Fig.\ref{fig:fomplotr}? 

As discussed in \S\ref{sec:bins} results for MAG in the Faint sample
depends very weekly on the redshift bin width as long 
as it is not much larger than the photo-z error.  To understand our new
results we can increase the bin width in the Faint sample to make it
similar  or slightly larger than its photo-z error. 
In this case we have that for both the Faint
and Bright sample $r_{ii} \simeq 1$ and $r_{ij} <1$ for $i\neq j$.
In this case, the variance in $C_{ij}$ caused by an
absolute error $\Delta_r$ in $r_{ij}$ is:
\beq
(\Delta C_{ij})^2 \simeq \Delta_r^2 (C_{ii}^2+C_{jj}^2)
\eeq
This should be compared to the variance in $C_{ij}$
in Eq.\ref{eq:covar}. 
In the limit $C_{ii} \simeq C_{jj} > C_{ij}$ we have:
\beq
{(\Delta C_{ij})^2 \over{Cov[C_{ij};C_{ij}]}} \simeq 2 \Delta_r^2
N(\ell)
\label{eq:s2nd}
\eeq
which is independent of $C_{ij}$ and therefore of
$\sigma_z$.\footnote{If we include shot-noise we have
to divide this by a factor $[1+1/(\bar{n}C_{ii})]^2$ which is
typically of order unity, but depends weakly on the sample and redshift bin width.}
 If we want this ratio to be smaller than $1/4$, so that the error in
$C_{ij}$ is not dominated by $\Delta_r$, we find that we need $\Delta_r<10^{-2}$
for $\ell \simeq 300$, which is in good agreement with 
 Fig.\ref{fig:fomplotr} and illustrates why this result is
the same for the F and B sample, and why its shape is also independent of
the redshift bin width.
We have checked this prediction by redoing
the results on the left panel of Fig.\ref{fig:fomplotr} for twice the 
bin width. For the Faint sample we find a very similar result with a
FoM$_{w\gamma}$ that is only 5\% lower for the larger bin width
 for all values of  $\Delta_r$.  For the Bright sample the results 
are a factor of 2 lower, as expected from \S\ref{sec:bins}, with 
a similar shape as function of $\Delta_r$.

For absolute errors larger than about $1\%$, 
the degradation for the shear-shear cross-correlation (dashed line
labeled Shear in left panel of Fig.\ref{fig:fomplotr})
is smaller than for MAG.  This is because MAG provides better
constraint than Shear alone (when bias is fixed).
When bias is known, MAG  has a larger FoM 
than Shear for $\Delta_r<1\%$ because of it's better radial resolution.
This is specially true for the Bright sample, which also has too small
a volume to take advantage of shear.  But because of the broad radial 
resolution, the FoM from Shear is less sensitive to  
the uncertainties in the photo-z transitions. 
Note that this formalism for degradation is more general (for narrow bins) than the
one based on the uncertainties in the centroid position
of the photo-z source distribution  (eg Huterer et al 2006).

For the MAG probe,  we need to know the transitions  to better than $4\%$ ($1\%$) if we want
the FoM to degrade less than $50\%$ ($11\%)$. This is the case
 for both the Faint and Bright 200 deg$^2$
samples.  For the 5000 deg$^2$ samples 
we need to know the transitions to better than about $2.5\%$ ($1\%$) if we want
the FoM of MAG+RSD to degrade less than $50\%$ ($13\%$). For WL+RSD
a $1\%$ uncertainty results in a $50\%$ degradation.
A bright sample with 5000deg$^2$ where we only know
transitions to $50\%$ gives similar FoM in MAG as a sample
with only 200deg$^2$ where transitions are known to be better than $2\%$.

The relative impact in the FoM 
is similar if we consider combined probes with
RSD, which we have assumed
is not affected by uncertainties in the photo-z transitions. This is illustrated
 in the right panel of Fig.\ref{fig:fomplotr}, which shows the corresponding 
results for the FxB combination of RSD with different WL probes.
Here we take $\Delta_r$ to be the same for the F and B sample, which is 
probably pessimistic for bright spectroscopic samples (but not for
PAU-like surveys).
The short-dashed line corresponds to the Galaxy-Shear (GS+GG) cross-correlation
(which also includes the galaxy-galaxy autocorrelation $C_{g_ig_i}$)
while the continuous lines correspond to MAG (i.e. $C_{g_ig_j}$). The
dotted line shows the shear-shear results. The long dashed line
combines (with appropriate covariance) all WL lensing probes (ie
MAG, GS, GG and SS). As
can be seen in the figure all probes are
 affected in similar way by the uncertainty in the photo-z
 transitions. When the uncertainty in the bin transition $\Delta_r$ are less than
0.1 we find:

\beq
FoM_{w\gamma}^{FxB} \propto e^{-(\Delta_r/0.18)(A/200)^{0.05}}
\label{eq:outliers}
\eeq
where $A$ is the survey area in deg$^2$. The $A$ dependence reflects 
the fact that to achieve a given FoM degradation we need to know the
transitions around 15\% better (i.e. $\Delta_r$ should be 15\% lower)  
for A=5000 deg$^2$ than for A=200 deg$^2$. This makes sense,
as the errors are lower for A=5000 deg$^2$, so the requirements
should be higher.
Accuracies of few percent in $r_{ij}$ seem within reach with current
or near future  photo-z codes with appropriate calibration, so 
this is not a critical limitation to measure magnification. 
 
Note how the FoM can increases by over a factor of 100 when
we cross-correlate the samples and include RSD, i.e. compare the values
of the FoM of FxB in the right panel with the FoM of F or B separately
in the left panel. Also note how magnification becomes better than
shear when combined with RSD (ie compare left and right panels).

\subsubsection{Non-Gaussian transitions} 

We have also explored the case of non-Gaussian transitions by
running a photo-z code over galaxy simulations and
exploring how this change the results with respect to the Gaussian case. 
We have estimated the photo-z transitions $r_{ij}$ using 
the photo-z code over  mock galaxy simulations. 
Details of this study will be presented elsewhere. 
Here we just want to point out that
in many cases we find that surprisingly the FoM  increases
for the Non-Gaussian transitions compared to the Gaussian case.
 The reason for this is that for the Gaussian comparison 
 we use a gaussian rms $\sigma_z$ corresponding to the 
the 68\% confident level of the photo-z distribution in simulations. The actual
(non-Gaussian)  photo-z
distribution is often much sharper in the center and has longer tails
than a Gaussian. If the distribution is known,
 this results in a better radial resolution and therefore better
FoM. In reality this improvement is limited by our uncertainties 
in the photo-z distribution (see \S\ref{sec:out} and below).

\begin{table}
\begin{tabular}{|c|ccc|}
\hline 
Probe/Sample &
FoM$_{w}$& FoM$_\gamma$ & FoM$_{w\gamma}$\\
\hline
FxB no wiggles &  78 (5.6) & 21 (6.3) & 1595 (16)\\
FxB  wiggles     &  86 (2.8) & 20 (5.5) & 1760 (35)\\
F+B no wiggles &  26 (0.4) & 6.5 (3.2) & 170 (0.6) \\
F+B  wiggles     &  28 (0.2) & 6.5 (3.2) & 180 (1.2) \\
\hline
\end{tabular}
\caption{Impact of having BAO wiggles in the MAG+RSD combination
 for 200deg$^2$ survey with
Planck and SN-II priors. Results without priors are in parentheses.
}
\label{table:wiggle}
\end{table}

\subsection{Impact of BAO wiggles}
\label{sec:wiggles}

The BAO wiggles in $P(k)$ can  be measured 
in the angular galaxy-galaxy correlations when we
use narrow redshift bins.  We test this by comparing results  
using the standard 
Eisenstein \& Hu (1998) $P(k)$ with and without the 
BAO feature for the same cosmology.

Results for MAG+RSD are shown in Table \ref{table:wiggle}. 
There is no impact on RSD as the shape is fixed in this case.
Without priors (in parentheses), the combined FoM in MAG+RSD (FxB)  increases by 
a factor of 2 due to  this effect. Of course this change mostly
affects $w(z)$ FoM as  $\gamma$ is measured by
the amplitude and not the shape of $P(k)$. 
We find very similar relative improvement in  FoM$_w$ 
for 5000deg$^2$ without priors.
Note how the F+B combination also increases by a factor of 2
with BAO wiggles.
This is despite  that in F+B we use MAG for the faint sample alone, 
which has low radial resolution, which translates into 
a lower resolution in $k$ for the shape measurement.

The increase in FoM$_w$ with priors is only 10\% for FxB for 200deg$^2$
and 31\% for 5000deg$^2$. Even with priors, these are quite substantial
gains and illustrate that the BAO wiggles are indeed well measured
with these probes. 

\subsection{Degeneracies}

We investigate here the gain in the MAG+RSD (FxB) combination for different
parameters. Table \ref{table:dege} shows how
the FoM$_{w\gamma}$ (with Planck and SN-II priors) 
degrades when we add a new parameter.
We see how the bias parameters
are the ones where the MAG+RSD combination degrades 
less than the separate RSD and MAG. So they are the ones that
benefit most from the MAG+RSD combination. For example note how
allowing $b_1$, the bias at the lower redshift of $z_i=0.25$, to vary
causes the FoM to degrade by 30-80\% for RSD or MAG separately
but only by 3\% for the FxB combination.

In some cases, like $n_s$, the degradation increases when we consider
the combined RSD+MAG probes. This is because the result 
for separate
probes is dominated by the priors. In this case adding a new parameter
does not degrade the FoM. The combination of probes provide better
constraints which are no longer dominated by priors. This results in
an increase in degradation. In the case of $\sigma_8$, the degradation
is always low as its value is dominated by priors and does not
affect much our FoM.

\begin{table}
{\center
\begin{tabular}{cccc}
Parameter & MAG & RSD & MAG+RSD \\
\hline 

$\Omega_m$  & 21.1 & 76.2 & 27.1 \\
$\Omega_{DE}$  & 48.7 & 69.2 & 50.9 \\
$\Omega_b$  & 19.3 & 75.8 & 18.6 \\
h & 19.8 & 76.0 & 21.2 \\
$n_s$  & 0.4 & 0.0 & 40.1 \\
$\sigma_8$  & 0.0 & 0.9 & 1.1 \\
$b_1^F$  & 77.6 & 28.4 & 2.6 \\
$~~b_1^B$  & 79.1 & 31.9 & 3.0 \\
$b_2^F$  & 81.5 & 40.5 & 5.0 \\
$~~b_2^B$  & 82.0 & 46.0 & 5.6 \\
$b_3^F$  & 92.0 & 66.7 & 47.3 \\
$~~b_3^B$  & 92.0 & 67.7 & 46.1 \\
$b_4^F$  & 93.9 & 78.3 & 62.8 \\
$~~b_4^B$  & 94.0 & 79.7 & 64.5 \\
\hline 
\end{tabular}
\caption{Degradation (per cent reduction) of the FoM$_{w\gamma}$ when we
add a new parameter to the FxB200 cross-correlation forecast. 
The change is relative to the case 
where we have all parameters varying except for the one
quoted (without priors). The 4 bias parameters for the Faint and Bright
populations $b_F$ and $b_B$ are the parameters
that gain most in the MAG+RSD combination.}
\label{table:dege}
}
\end{table}

\subsection{Marginalized errors}

Table  \ref{table:margi}  shows the marginalized
1-sigma error-bars of different parameters for each
probe. This corresponds to 200 deg$^2$ with Planck and
SN-II priors. The last column is for 5000 deg$^2$.
The table reflects in other parameters
what we have already found for the different FoMs: that the
FxB probe provides improvements in errors by factors of a few 
with respect to the  errors in separate MAG or RSD probes.
In particular note how the error in bias goes from $\simeq 5-8\%$ to
$\simeq 1-2\%$. The error in $\sigma_8$ does not change
as is dominated by priors. This is fine as our goal is
to measure the growth evolution and not 
the overall normalization.

\begin{table}
\begin{tabular}{c|cc|ccc|}
& MAG  & RSD   & \multicolumn{3}{c|}{MAG+RSD} \\\hline 
             & F200 & B200  & F+B200  & FxB200 & FxB5000\\
\hline 
$\Omega_m$  & 5.3\% & 47\% & 5.5\% & 3.3\% & 1.5\%\\
$\Omega_{DE}$  & 1.8\% & 9.2\% & 1.8\% & 1.2\% & 0.6\%\\
$h$  & 2.8\% & 24\% & 2.8\%& 1.8\%  & 0.83\%\\
$\sigma_8$  & 0.9\% & 0.9\% & 0.9\% & 0.9\% & 0.76\% \\
$\Omega_b$  & 5.8\% & 47\% & 5.7\% & 3.6\%  & 1.88\%\\ 
$w_0$ & 15\% & 17\% & 14\% & 5.8\% & 2.57\%\\
$w_a$ & 0.86 & 1.21 & 0.78 & 0.29 & 0.07\\
$n_s$  & 0.65\% & 0.66\% & 0.65\% & 0.62\% & 0.42\% \\
$\gamma$ & 142\% & 74\% & 29\% & 13.4\% & 4.6\%\\
$b_1^F$  & 11\% & - &  4.0\%& 1.7\% &0.95\%\\
$~~b_1^B$  & - & 5.1\% &  5.5\% & 1.8\% & 0.93\%\\
$b_2^F$  & 16\% & - &  4.9\% &2.1\% &1.07\%\\
$~~b_2^B$  & - & 4.6\% & 5.3\% & 2.7\% &1.18\% \\
$b_3^F$  & 22\% & -  & 4.5\% & 1.6\% & 0.92\%\\
$~~b_3^B$  & - & 5.9\%  & 3.5\% & 1.7\% & 0.92\% \\
$b_4^F$  & 26\% & - &  4.7\% & 2.0\% & 1.05\% \\
$~~b_4^B$  & - & 8.9\% & 9.6\% & 2.7\% & 1.20\% \\
\hline 
\end{tabular}
\caption{Marginalized  error-bars for each parameter
and different probes. For 200 deg$^2$, except the last column which is
for 5000 deg$^2$. All with Planck and SN-II priors.}
\label{table:margi}
\end{table}

\section{Survey Comparison}
\label{sec:survey}
 
Table \ref{table:surveys} shows a comparison of different surveys and
probes. In the upper section of Table \ref{table:surveys} we compare different
WL probes for the parent F5000 photometric sample.
The following entries correspond to probes for the brighter spectroscopic
samples, smaller area surveys and different probes either combined over 
different areas or cross correlated.

\subsection{Shear-Shear,  Galaxy-Shear and MAG}
\label{sec:wlprobes}

In the first three entries in Table \ref{table:surveys} we compare MAG with
Galaxy-Shear (GS+GG) and Shear-Shear (SS) using the same assumptions, fiducial
model and priors for a 5000 deg$^2$ survey.  Using SS, $C_{\kappa_i\kappa_j}$, is the most
conservative use of WL as it does not requiere any galaxy biasing modeling.
Note that  in GS+GG we also include 
the  galaxy-galaxy autocorrelations, i.e. it includes both $C_{g_ik_j}$ and $C_{g_ig_i}$,
but does not include SS, $C_{\kappa_i\kappa_j}$, or galaxy
cross-correlation, $C_{g_i,g_j}$, for $i\neq j$. 
Magnification (MAG) includes both auto and cross-correlations,
i.e.  $C_{g_i,g_j}$, for all $i$ and $j$.
As expected, the MAG result for F5000 is lower than
SS or the GS+GG, this is due to galaxy bias.
 When bias is fixed (entry \#2b in the table),
MAG is larger despite the  relatively poor radial resolution,
$\sigma_z=0.05(1+z)$. 
The galaxy-shear cross-correlation alone, i.e. $C_{gk}$, 
has lower noise than MAG, i.e. $C_{gg}$, as can be seen in the
 covariance of Eq.\ref{eq:covar}:
\bea
Cov(C_{gg},C_{gg}) &\propto& C_{gg}C_{gg} \\
Cov(C_{gk},C_{gk}) &\propto & C_{gg}C_{kk} 
\eea
where $C_{kk}$ is much smaller than $C_{gg}$. This means
that without noise GS performs better than MAG. But systematic effects
will be quite different for measuring accurate shapes versus accurate photometry. 
It is also likely that in practice there are higher densities of
galaxies  available for MAG than for Shear as we move to higher redshifts
and this has not been taken into account here.

An advantage of our approach is that we can include all
observables in the same FM (with appropriate covariances)
to provide a joint analysis of MAG+GS+GG+SS including all covariances (label WL-all
in the Table).  The WL-all case (entry \#4 in the
Table) includes all these weak lensing probes, and of course gives the
best results. The improvement of WL-all over GS+GG is over a factor of
6: compare entries \#4 and \#3. 
This is due to the direct measurement of the matter amplitude (ie
growth)  by SS which helps breaking degeneracies with bias
in GS+GG and MAG.

\begin{table*}
\begin{tabular}{llccccccc}
\hline
& Probe & Sample & $\eta$ &Volume  &$\sigma_z/(1+z)$ & FoM$_{w}$  & FoM$_{\gamma}$  & FoM$_{w\gamma}$   \\
\# & & & Complete & Gpc$^3/h^3$ &Photo-z error & $\gamma$ free & & $\times10^3$\\
\hline
&& \tt Photometric \\
&&  $22.5<i_{AB}<24$ \\
\hline
1& Shear-Shear (SS) &  F5000 & 1/2& 18 &0.05& 84 & 39 & 3.2 \\
2 & MAG  & F5000 & 1/2 & 18 & 0.05 & 54 & 6 & 0.33\\
2b & MAG (bias fixed) & F5000 & 1/2 & 18 & 0.05 & 142 & 59 & 8.4\\
3 &Galaxy-Shear (GS+GG)  & F5000 & 1/2 & 18 & 0.05 &  91 & 29  & 2.6\\
4 &Weak Lensing (WL-all)  & F5000 & 1/2 & 18 & 0.05 &  331 & 47 & 16 \\
\hline
&& \tt Spectroscopic \\
&&  $i_{AB}<22.5$  \\
\hline
5& MAG & B5000 & 1/2 & 7& 0.0035 &  64 & 0.9 & 0.06\\
5b& MAG (bias fixed) & B5000 & 1/2 & 7& 0.0035 &  196 & 85 & 17\\
6 &RSD &  B5000 &  1/2 &7 & 0.0035 &  27 & 17 & 0.47\\
7 & RSD  &  B5000spec &  1/20 &7 & 0 &  20 & 17 & 0.35\\
8 & RSD  &  B5000spec21.5& 1/26 & 3.5 & 0 & 10 & 16 & 0.2\\
9 & BAO &  B5000 & 1/2 & 7 & 0.0035 & 47 & - & -\\
10 &BAO&  B5000highz&  1/10 & 18 & 0& 78 & - & -\\
\hline
&& \tt Small: 200 deg$^2$ \\
\hline
11 & MAG+RSD & FxB200 &1/2 &0.7 & 0.05-0.0035 & 86 & 21 & 1.8\\
11b & MAG+RSD (bias fixed) & FxB200 &1/2 &0.7 & 0.05-0.0035 & 197 & 39& 7.7\\
11c & SS+RSD  & F+B200 & 1/2 & 0.7 & 0.05-0.0035 & 37 & 18 & 0.7\\
12& GS+GG+RSD & FxB200 &1/2 & 0.7 &  0.05-0.0035&105 & 21 & 2.2\\
13 & WL-all+RSD& FxB200 & 1/2 & 0.7 & 0.05-0.0035& 198 & 26 & 5.2\\
\hline
&& \tt Combine Independent \\
\hline
14 & BAO+RSD & B5000highz & 1/10& 18 & 0 &  100 & 27  & 2.7 \\
15 & BAO+WL-all & F+B5000highz &1/2-1/10& 7-18 & 0.05-0& 384 & 48  & 17 \\
16  & BAO+RSD+WL-all & F+B5000highz &1/2-1/10& 7-18 & 0.05-0& 597 & 66  & 40 \\
17 &MAG+RSD & F+B5000 &1/2 & 7-18 & 0.05-0.0035 & 82 & 29 & 2.4\\
17b &MAG+RSD (bias fixed) & F+B5000 &1/2 & 7-18 & 0.05-0.0035 & 686 & 112 & 77\\
17c & SS+RSD & F+B5000 &1/2 & 7-18 & 0.05-0.0035 & 262 & 58 & 15\\
18 &WL-all+RSD  & F+B5000 &1/2 & 7-18 & 0.05-0.0035 & 571 & 66  & 38\\
19 & SS-RSD  & F+B5000spec &1/2-1/20& 7-18 & 0.05-0&257 & 58  & 15\\
\hline
&& \tt Cross-correlation \\
\hline
20 & MAG+RSD & FxB5000 &1/2 & 7-18 & 0.05-0.0035 & 645 & 46 & 30\\
20b & MAG+RSD (bias fixed) & FxB5000 &1/2 & 7-18 & 0.05-0.0035 & 1141 & 139 & 159\\
20c & SS+RSD & FxB5000 &1/2 & 7-18 & 0.05-0.0035 & 316 & 60 & 19\\
21 &WL-all+RSD& FxB5000 &1/2 & 7-18 & 0.05-0.0035 & 2879 & 87 & 251\\
22 & WL-all+RSD & FxB5000spec  &1/2-1/20& 7-18 & 0.05-0& 2113 & 74 & 159\\
23 & WL-all+RSD & FxB5000spec21.5  &1/2-1/26 & 3.5-18 & 0.05-0& 1509 & 65 & 98\\
\hline
\end{tabular}
\caption{ FoM comparison for different surveys and probes: Shear-Shear
  (SS), Galaxy-Shear (GS+GG, which includes Galaxy-Galaxy
  autocorrelations),  Magnification (MAG),
 redshift space distortions (RSD) and Baryon Acoustic Oscillations
 (BAO). The WL-all case includes MAG, SS, GG and  GS. All cases
are marginalized over bias (except for \#2b, 5b, 11b, 17b and 20b, where bias is
fixed)  and 6 cosmological parameters (see Eq.\ref{eq:parameters}). 
All samples are
 5000deg$^2$ expect for \#11-13 which are 200deg$^2$. 
The 5th column gives the comoving volume to the highest
redshift used in each analysis. 
BAO alone (\#9-10) does not provide
FoM$_\gamma$ or FoM$_{w\gamma}$. }
\label{table:surveys}
\end{table*}

\subsection{Bright spectroscopic samples}

\begin{figure*}
\includegraphics[width=5.in]{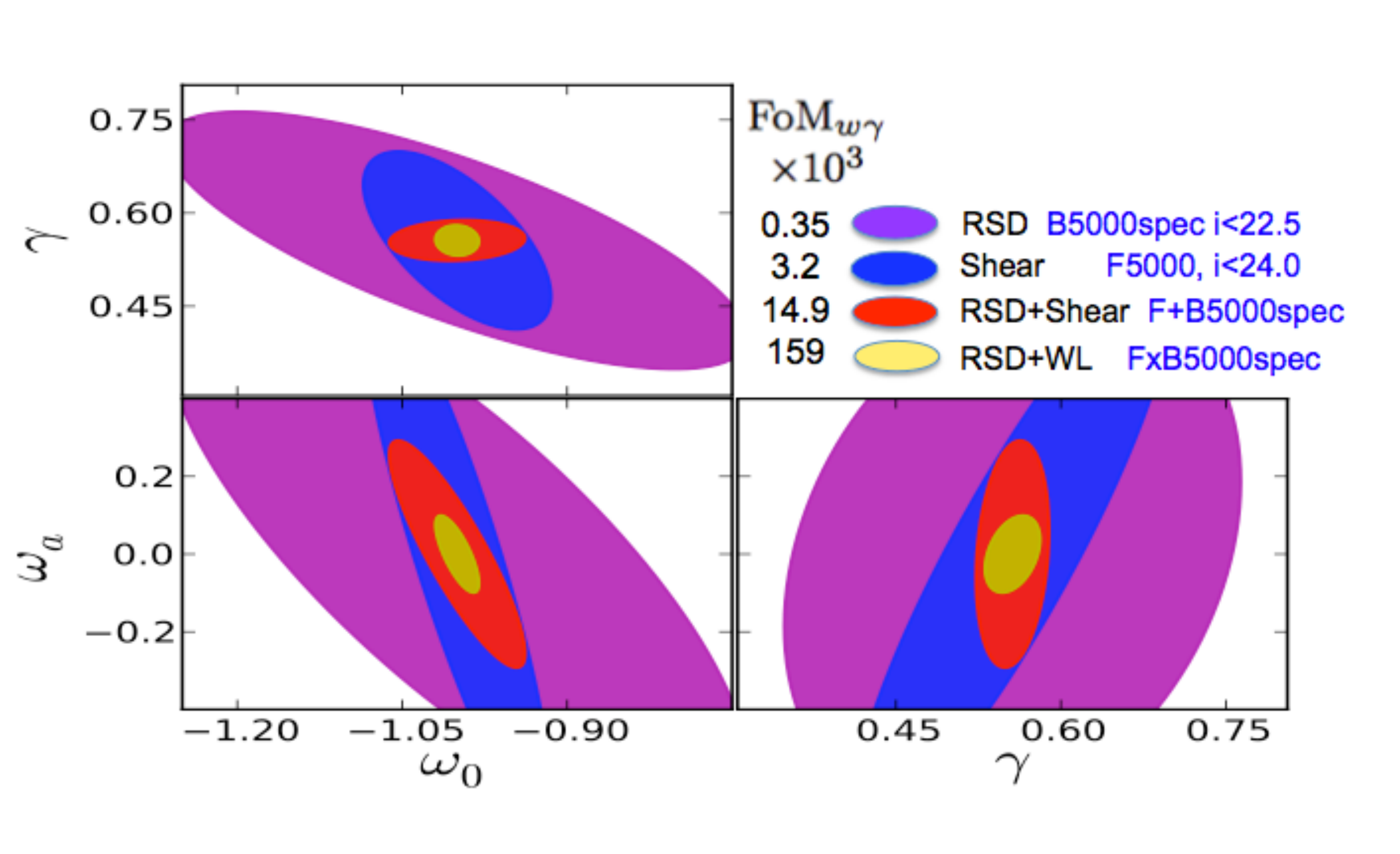}
\caption[]{Contours ($\Delta \chi^2=2.3$) 
of $w_0-w_a-\gamma$ in  weak lensing (only Shear-Shear)
over a Faint sample with $i_{AB}<24$ (F5000, blue) as compared with RSD over a bright
spectroscopic subset (B5000spec, purple) targeting 1000 galaxies per square degree
to  $i_{AB}<22.5$.   Bias evolution ($b(z)$)
and other cosmological parameters are marginalized over.
The red contours correspond to the combination of both
probes over different sky areas  (F+B), while the yellow contours
show the combination over the same area (FxB), including also
all cross-correlations between shear and galaxies in both samples.
This produces over a factor 10 larger combined FoM.}
\label{fig:degene}
\end{figure*}

As defined in \S\ref{sec:fiducialsurveys} the B5000 sample
is close to spectroscopic as it has very good photo-z,
$\sigma_z=0.0035(1+z)$, but a very high completeness
(of 10000 galaxies/deg$^2$ to $i_{AB}<22.5$). This is motivated by
the PAUCam concept (see \S1), but we also show results for more
standard (future)  spectroscopic flux limited samples. One 
of them is called B5000spec: it
has $\sigma_z=0$ and  1000 galaxies/deg$^2$.
Otherwise B5000spec is identical to B5000. A second sample
is called B5000spec21.5 is flux limited to $i_{AB}<21.5$ (instead
of $i_{AB}<22.5$ in B5000spec or B5000) with only 500
galaxies/deg$^2$. This gives a mean redshift of 0.4 and
about half the volume of $i_{AB}<22.5$ (see also long dashed line
in Fig.\ref{fig:nz}). As shown in Table \ref{table:surveys}
(comparing entries 6-7), the RSD results for B5000spec
are lower than for B5000, indicating that density is
more important than spectroscopic accuracy as found in
\S\ref{sec:shotnoise} and \S\ref{sec:photoz}. This is 
a good validation of the PAUCam concept. The result
for B5000spec21.5 is almost a factor of 2 lower, but still
quite competitive given that it will be much faster to get
spectra for galaxies which are 1 magnitude brighter.

When the B5000 sample is a subset of F5000 (i.e. in the FxB
case) we can also estimate MAG for B5000 as we will have
both good redshift accuracy and good photometry to do
shear or magnification. This is shown as entry \#5 in 
Table \ref{table:surveys}.
Note how MAG over B5000 provides slightly better FoM$_w$
(but lower FoM$_{w\gamma}$) than MAG over F5000. When bias is fixed
MAG over B5000 is always better than MAG over F5000 
(compare \#2b and \#5b), despite the larger volume and depth in F5000.
This of course is due to the better resolution in the radial bins. 

BAO is also included in our forecasts for comparison.
We follow the method of  Seo \& Eisenstein (2007)
using the same fiducial values and parameters as for the other
cases (see \S\ref{sec:FM}). We consider two different surveys for a BAO
follow-up of F5000. The usual B5000, which for FxB  can be 
considered a brighter
subset of F5000 and therefore lies in its foreground, with
most of the redshifts at  $z<0.5$. the other is  B5000highz 
which corresponds to a selection  (e.g. emission line galaxies) 
which is biased towards higher redshift targets
out of F5000, with around 2000 galaxies/deg$^2$, 
used to maximized the BAO probe. We use a constant number
of target galaxies per unit redshift in the range $0.2<z<1.7$
which has about 2.5 larger volume than B5000.
The BAO result for B5000highz is about a factor of 2 better than for
B5000, as expected. This is about the largest BAO FoM 
that one could get based on targets from the parent F5000 sample
as we have assumed it covers the same volume as F5000 and has quite a large density.
The result is almost as large
as the FoM$_w$ for MAG in B5000 (entry \#5), but note that with
MAG we can also measure $\gamma$, which is not possible with BAO.
As mentioned earlier, doing MAG over B5000 is in fact only a
subset of what we can do when we combine the good radial resolution
in B5000 with the photometrical (and WL) information in F5000, i.e. the
FxB combinations which will be discussed next.

\subsection{Smaller area surveys}

In the 11th-13th rows of Table \ref{table:surveys}, 
we see how a modest survey, with only 200 deg$^2$ can give a better FoM
based on cross-correlation (FxB200 in the table)
than much larger surveys (F5000 or B5000)  that consider only either MAG, SS,
GS, RSD or BAO separately, i.e. all entries above in Table \ref{table:surveys}, 
except entries \#3 and \#4  which combine several probes. 
In particular, the BAO PAU Survey proposed in Benitez et al. (2009) with 5000 deg$^2$ yields
a FoM$_w=47$ (entry \#9 in the Table) which is a factor of 2-3
times smaller than the combination of RSD with MAG (or WL probes) over
200 deg$^2$ (entries \#11-13). This represents a factor of 50-70 improvement
in the ratio of benefit over cost, if we assimilate cost with area and
benefit with FoM.  In addition, the RSD+WL-all
cross-correlation also provides constraints on $\gamma$
of about 5\% accuracy, which cannot be obtained from BAO.

If we combine the Fisher matrix of MAG and RSD from two large surveys (F+B5000)
we do  better than FxB200, but not by far. When we include
all WL  and RSD in the F+B5000 combination (row \#18-19) the improvement
is more significant, overcoming the FxB200 result by a factor of 3-7. As
the combined FoM scales roughly with the area (Eq.\ref{eq:area}),
doing a survey with $\sim 800$ deg$^2$ with the FxB cross-correlation will provide similar
combined FoM than the combination of the two larger but separate
surveys  (F5000+B5000), which correspond to a total of 10000 deg$^2$.
 This represents a cost saving of about a factor of
10, if we assume that cost scales with survey area.
 Also recall that F+B only uses RSD over B5000 and
lensing over F5000, while FxB include all probes and all cross-correlations.

In the case of the PAU Survey, we find from photo-z simulations that
the narrow-band  filters also improve the  photo-z accuracy of the Faint sample from
0.05(1+z)  to about 0.03(1+z). If we include this improvement,
the FoM$_{w\gamma}$ for FxB200 increases by another 30\%.

\subsection{Spectroscopic and Photometric combinations}

Entries \#20-21 of Table \ref{table:surveys} shows that the FxB5000 combination
are about 10 times larger FoM$_{w\gamma}$ than F+B (\#17-18), even when F+B
correspond to twice the area (as they come from independent surveys
and therefore have twice as many independent modes). Part of this gain
comes from the fact that in the FxB combination we have two probes (WL
and RSD) to measure the bias of the B population, while for the F+B
combination the bias in B is only measure with RSD. The improvement on 
going from F+B to FxB is larger in FoM$_w$ than in FoM$_\gamma$.
Note that the gain is smaller for SS+RSD combination (compare \#17c
to \#20c) because bias is already well measured with F+B with Shear.
This can also be seem comapring the cases with fixed bias  (\#17b and \#20b) .
When bias is fixed in the FxB case (\#20b) there is still a significant
gain with respect to the free bias case (\#20), indicating that there
is still room  for important benefits  if we can measure bias
in some other ways (\S\ref{sec:bias}).

In principle one can also infer the bias of the B spectroscopic population 
by identifying the same bright population within the photo-z 
($i_{AB}<24$) sample and doing   the cross-correlation of the faint
and bright galaxies. This can improve the F+B result by a factor of 5,
but is still a factor of 2 lower than FxB because of the larger
photo-z errors.

This gain is illustrated in Fig.\ref{fig:degene}, which shows the 3
projections $(w_0, w_a, \gamma)$ in the FoM$_{w\gamma}$ ellipsoid. Alone, the RSD results
(purple) in the spectroscopic sample (B5000spec) are less powerful in
constraining FoM$_{w\gamma}$ than Shear-Shear (blue),
but the improvement resides in the
complementarity of the combination to break degeneracies. 
This can be seen in the red
contours (F+B5000spec) which combine both RSD and Shear-Shear 
assuming they are independent.
When the B5000spec sample is a subset over the same area
of the F5000 sample, we can also
include all their cross-correlations  and provide a second route to measure 
bias evolution. This benefits both WL and RSD, and the combination
is over a factor of 10 larger (yellow FxB contours in the figure) than
F+B and a factor of 10 times larger than all WL in the Faint sample
alone (entry \#4 in Table \ref{table:surveys}).

Figure \ref{fig:degene2} shows the corresponding result using MAG
and RSD over 200 deg$^2$, instead of Shear and RSD over 5000 deg$^2$.
 In this case the gain in the combination FxB
(see Table \ref{table:bfix2}) is over a factor of 10 times larger than
F+B and over a factor 50 for the one in MAG in the faint sample alone. The
extra benefit here comes from the higher sampling rate in the PAUCam
strategy.

The BAO+RSD (B5000highz) combination (entry \#14) is comparable to the 
MAG+RSD in F+B5000 (\#17) or the FxB200 combinations (\#11).
BAO and RSD have also been used to constrain bias (e.g. see Amendola,
Quercellini and Giallongo 2005, Cinzia, Amendola and Branchini 2011). 
But this is a factor of 5 lower
than the WL-all+RSD independent combinations (\#18) and a factor of 90 lower
than the FxB cross-correlation result (\#21). 
The results for B5000spec21.5 (last entry in the Table) are about a
factor of 1.6 lower in the combined FoM than the corresponding $i_{AB}<22.5$
result (\#22).

\subsection{Spectroscopic follow-up strategy}

The WLxRSD (or MAGxRSD) strategy appears much better than following a BAO
approach over the same B5000 sample. The combination of WL
and RSD produces a  FoM$_w$ about 24 times larger that BAO alone 
over B5000. The reason is that in BAO
we are ignoring the $P(k)$ amplitude and WL distance ratios information.
 In fact, RSD alone can give comparable FoM than BAO despite the additional
 biasing parameters. It  also produces interesting constraints
in $\gamma$ which BAO cannot measure. 

 A potential advantage of
a BAO follow-up would be to 
sample larger volumes as in  B5000highz 
which corresponds to a selection of higher redshift
targets out of F5000, with around 2000 galaxies/deg$^2$ with 
constant density per unit redshift between 0.2 and 1.7.
This is shown in the \#10 entry of Table \ref{table:surveys}. The
improvement in FoM$_w$ is significant but still far from WL alone (\#4).
The combination of BAO and RSD produces only a modest
improvement over the BAO alone result, similar to MAG+RSD from
separate surveys.

To have a fair comparison or optimization of which could be the 
best strategy (going deep
as in B5000highz  or shallower as B5000spec) we would need to add BAO to 
WL and RSD as separate probes (F+B).
 This is not totally straight forward as these probes could
be  correlated. Our WL forecast already includes BAO wiggles
(see \S\ref{sec:wiggles}).  
What we can do here is to combine the BAO 
measurement with WL and RSD, as if they were from independent
samples. This is partially true as there is little volume overlap between
B5000highz and B5000spec (where the lenses to F5000 reside). 
In any case, this is an upper bound of the true result, 
because the covariance between WL
and BAO will lower the FoM. 
Results are shown as entries \#15-16 in Table
\ref{table:surveys}.  The BAO+WL-all combination (entry \#15) is similar to 
SS+RSD (entry \#19). When we combine everything BAO+WL-all+RSD (entry \#16) 
the outcome  is of course higher.
But it is still over a factor of 6 lower than the FxB WL-all+RSD 
cross-correlation (\#21). This indicates that a shallower
spectroscopic follow-up strategy provides better returns in terms
of FoM. The legacy value of such spectroscopic follow-up
could also be larger
as we will be able to relate dark matter, as traced by WL, to galaxy
formation, as traced by galaxies over the same structures.

\section{Conclusion}

In this paper we advocate the use of the cross-correlation of
a deep (faint) photometric sample with a
foreground brighter spectroscopic subset using narrow radial redshift bins. 
The foreground sample could also be photometric with
very good photo-z of $\sigma_z/(1+z) \simeq 0.0035$.
We consider three different types of probes in the analysis: 
1) angular clustering from galaxy-galaxy autocorrelation 
in narrow redshift bins, 2) weak lensing (WL)  from shear-shear,
galaxy-shear and magnification (i.e. galaxy-galaxy cross-correlation),
3) redshift space distortions (RSD), 
from the ratio of transverse to radial modes. 
The combination of such measurements
provides a significant improvement in the forecast for
the evolution of the dark energy equation of state, given by $w(z)$, 
and the cosmic growth evolution, given by  $\gamma$.

This improvement comes from the use of narrow redshift bins
and the measurement of galaxy bias,
which affects both RSD and WL cross-correlations,
but in  different ways. For transverse modes, biasing can be obtained by 
comparing  WL clustering with galaxy angular clustering, but this is
also degenerate with geometrical factors that depend on  cosmic expansion evolution.
In redshift space,  bias can be measured directly
by comparing line-of-sight to transverse clustering.
The combination of both methods results 
in a very accurate determination of bias evolution, also breaking the
degeneracy of growth with cosmic expansion evolution.  We have argued
that  this gain is only fully accomplished when both WL and RSD  measurements are 
done over the same area of the sky, which allows better measurement of
bias and the full understanding of cross-correlations. Once bias is
known, the foreground sample gives much better constraints 
than the deeper photometric sample, thanks to its better radial resolution.

A new figure of merit, FoM$_{w\gamma}$,  
that encodes both cosmic expansion $w(z)$ and
growth evolution $\gamma$ is introduced  to explore changes
with the sample area, $A$, the redshift completeness, $\eta$, the magnitude
limit of the bright sample, $m_l$, the photo-z or spectroscopic error,
$\sigma_z$ and the uncertainty in the fraction of photo-z transitions
$\Delta_r$ defined in \S\ref{sec:out}. For the magnification (MAG) and
RSD combination over the same area (FxB) we find

\beq
FoM_{w\gamma} \simeq 2080 ~\bar{A}^{0.89}~
\eta^{0.2}~1.26^{m_l-22.5}~
e^{-\bar{\sigma}_z^2-\bar{\Delta}_r \bar{A}^{0.05}}
\label{eq:FoM1}
\eeq
where $\bar{A}$ is the survey area in units of 200 deg$^2$,
$m_l$ is the magnitude limit of the bright sample, 
$\bar{\sigma}_z$ is the photo-z error of the bright sample
in units of $0.01(1+z)$ and $\bar{\Delta}_r$ is the 
absolute uncertainty in the fraction of photo-z transitions
in units of $0.18$.  
This equation can be used to extrapolate the predictions to different survey
parameters within the ranges we have explored, i.e.
$\bar{A} \in [1,50]$, $\eta>0.01$, $m_l \in [21.5,23.0]$,
$\bar{\sigma}_z<0.7$ and  $\bar{\Delta}_r<0.6$.
If we include all WL probes: shear-shear, galaxy-shear and
magnification,  the above
FoM$_{w\gamma}$ is about a factor of 2.9 (8.6)
 times larger than Eq.\ref{eq:FoM1}
for 200 (5000) deg$^2$. 
These results are robust and conservative. We only use linear scales
($k_{min}=0.1$) and
a relatively large redshift bin-width of $\Delta z\simeq 0.014(1+z)$.
Better results can in principle be obtained using smaller scales and
narrower redshift bins, but this requieres further assumptions and
modeling.
We use a conservative modeling of biasing, which uses four parameters for each
population (i.e. faint and bright subsamples) and no priors.
With two bias parameters per population the above FoM increases by about
40\%. If we fix bias the improvement is 4.3 (5.3) for 200 (5000) deg$^2$. 

Shear-Shear alone (excluding galaxy correlations)
produces smaller FoM than magnification alone with bias fixed, but
10 times lager when bias is not known.
Galaxy-Shear (including galaxy-galaxy autocorrelation and galaxy bias)
can  produce a factor 8 times better FoM$_w$ than MAG (i.e. galaxy-galaxy cross and
auto-correlation)  because the
latter is subject to larger sampling error from the background
galaxy-galaxy autocorrelation. All WL probes
benefit equally from the boost in the cross-correlation with RSD.
They provide similar cosmological information (see Van Waerbeke 2009),
so that the final FoM will depend on the size of systematic errors and 
the number density and volume that we can trace in each case.
As illustrated by our results in Eq.\ref{eq:FoM1},
 magnification alone, without shape information, can be used
to provide very competitive cosmological information.

We have not considered here photo-z bias or shear multiplicative
or additive biases as they are more survey dependent and
have been studied elsewhere (see Huterer et al. 2006, Bernstein 2009 and
references therein). But we have found new requirements on the
uncertainties of photo-z transitions for the magnification, galaxy-shear 
and shear-shear probes to work. 
For all cases, the transitions $r_{ij}$ (or photo-z contamination)
needs to be known to about 1\% (absolute error in $r_{ij}$) 
if we want only a small degradation in the FoM$_{w\gamma}$ 
(see Fig.\ref{fig:fomplotr} and Eq.\ref{eq:outliers}). These
values look within reach for future surveys. But note that
as indicated by Eq.\ref{eq:s2nd}, this requirement increases
as $\ell_{max}^{1/2}$ where $\ell_{max}$ is the maximum multipole
(or smaller angular scale) used in the fit. In our case we only use
linear scales (Eq.\ref{eq:lmax}). A fit using non-linear scales 
($\ell_{max} \simeq 6000$) needs to know photo-z transitions to 0.4\%
and also requieres understanding of how baryonic physics could
change the non-linear $P(k)$ (Van Daalen et al. 2011, Semboloni et
al. 2011).

Given current limitations on systematics of shear measurements,
specially for deeper ground based surveys, pushing for magnification as a
cosmological tool is a  good alternative to implement the 
WL+RSD cross-correlation advocated here.
A clear advantage for magnification would be to consider deeper
photometric samples (such as LSST which can reach $r\simeq 27.5$,
Ivezic, et al. 2011), where galaxy shapes are
dominated by atmospheric noise but we can still measure magnitude
and photo-z errors  (see Van Waerbeke et al. 2010, Namikawa, Okamura \&
Taruya 2011).  In this case, the magnification and RSD
cross-correlation  presented in this paper could play
a significant role, provided we can perform an appropriate deep spectroscopic 
follow-up of the foreground lenses and provided we can calibrate the
 fraction of photo-z transitions. The PAUCam concept, based on doing
 photometric redshifts from 40 narrow (100\AA) and 6 broad
 bands, seems well suited for such deep spectroscopic follow-up
as we have found that the characteristic PAU photo-z error of
$\sigma_z=0.0035(1+z)$ only degrades the FoM$_{w\gamma}$ by $\sim 10\%$
  (e.g.  Fig.\ref{fig:fomplotz} and Eq.\ref{eq:FoM1}).

To design a spectroscopic follow-up survey 
of a photometric (WL) survey,  we find  that we can obtain
6 times better FoM$_{w\gamma}$ by targeting the foreground lenses
 to do cross-correlations, as proposed here,
than trying to go deeper to measure BAO 
(compare entry \#21 to \#16 in Table 6), 
 as has been proposed in several future surveys,
 such as Euclid, BigBOSS, HETDEX or FOCAS (WFMOS).
Measuring BAO with a 5000 deg$^2$ PAU Survey (Benitez
 et al. 2009) provides  2-4 times smaller FoM$_w$
than doing RSD+MAG (or WL) cross-correlation with the same PAUcam 
instrument over a much smaller 200 deg$^2$ area (compare
\#11-13 to \#9 in Table \ref{table:surveys}). This represents
a factor 50 to 100 improvement in profit over cost,  defined as
FoM achieved per unit area. 

 In our analysis we have only
considered the gains in the photometric and
spectroscopic cross-correlation  using  the 2-point clustering statistics.
Higher order correlations will certainly add additional benefits both 
in the weak-lensing probes (e.g. see Gaztanaga \& Bernardeau 1998), in
clustering  and in galaxy biasing (Frieman \& Gaztanaga 1994,
Gaztanaga \& Scoccimarro 2004).
Besides the improvement in cosmological inference, the cross-correlations
of spectroscopic and photometric surveys
should also provide key information on the relation between galaxy evolution
and dark matter growth, which will bring some new light
into the problem of galaxy formation.

\section*{Acknowledgements}

We would like to thank Benasque Center for Science for
their great hospitality during the Cosmology 2010 workshop
where the ideas in this paper were presented and discussed. 
We thank Anne Bauer, Gary Bernstein, Yan-Chuan Cai,
Enrique Fernandez, 
Josh Frieman, Wayne Hu, Bhuv Jain, Antony Lewis and David Weinberg
for discussions and feedback on ideas in this paper.
Funding for this project was partially provided by the
Spanish Ministerio de Ciencia e Innovacion (MICINN), 
project AYA2009-13936,  Consolider-Ingenio CSD2007- 00060, 
European Commissions Marie Curie Initial 
Training Network CosmoComp (PITN-GA-2009-238356),
research project 2009-SGR-1398 from Generalitat de Catalunya 
and  the Juan de la Cierva MEC program.
M.E. was supported by the FI PhD grant from the government
of Catalunya (AGAUR).
The MICE simulations have been developed at the MareNostrum
supercomputer  (BSC-CNS) thanks to grants AECT-2006-2-0011
 through AECT-2010-1-0007. 
Data products are stored at the Port d'Informacio Científica (PIC).

\end{document}